%% file: main.tex
\documentclass[10pt]{article} 
\ifdefined\QEMScoreAnonymous
\usepackage{tmlr}
\else
\usepackage[preprint]{tmlr}
\fi

\input{math_commands.tex}

\usepackage{hyperref}
\usepackage{url}
\usepackage{booktabs}
\usepackage{amsmath}
\usepackage{xspace}
\usepackage{graphicx}
\usepackage{tikz}
\usetikzlibrary{arrows.meta,positioning}
\usepackage{array}     
\usepackage{colortbl}  

\definecolor{abGray}{HTML}{C9C9C9}
\definecolor{abMint}{HTML}{BFDFD2}
\definecolor{abCoral}{HTML}{ED8D5A}

\newcommand{\qemscore}{\texorpdfstring{\textsf{QEMScore}}{QEMScore}\xspace}

\title{\qemscore{}: How Much Does the Measurement Add to\\Learned Quantum Error Mitigation?}

\author{\name Yue Zhao \email yue.z@usc.edu \\
      \addr University of Southern California
      \AND
      \name Huayue Gu \email hgu2@kennesaw.edu \\
      \addr Kennesaw State University
      \AND
      \name Yushun Dong \email yd24f@fsu.edu \\
      \addr Florida State University
      \AND
      \name Xiyang Hu \email xiyang.hu@asu.edu \\
      \addr Arizona State University}

\newcommand{\findingslot}[1]{\noindent\textbf{Finding #1}}

\makeatletter
\newcommand{\releaselink}[2]{\if@accepted\url{#1}\else #2\fi}
\makeatother

\def\month{MM}  
\def\year{YYYY} 
\def\openreview{\url{https://openreview.net/forum?id=XXXX}} 

\begin{document}

\maketitle

\input{00_abstract}

\input{01_intro}
\input{02_related}
\input{03_benchmark}
\input{04_results}

\input{05_field}

\input{06_three_systems}
\input{07_discussion}
\input{08_statements}
\ifdefined\QEMScoreAnonymous
\clearpage
\fi
\bibliography{refs}
\bibliographystyle{tmlr}

\input{08_reproducibility}

\appendix

\input{09_appendix}
\input{A2_related}
\input{A3_benchmark}
\input{A4_results}

\input{A5_field}

\input{A6_three_systems}
\input{A7_discussion}

\end{document}

%% file: math_commands.tex
\usepackage{amsmath,amsfonts,bm}

\def\eqref#1{equation~\ref{#1}}

\def\1{\bm{1}}

\DeclareMathAlphabet{\mathsfit}{\encodingdefault}{\sfdefault}{m}{sl}
\SetMathAlphabet{\mathsfit}{bold}{\encodingdefault}{\sfdefault}{bx}{n}



%% file: 00_abstract.tex
\begin{abstract}
How much does the noisy measurement add to learned quantum error mitigation? An
accuracy table cannot say, because a model handed circuit structure can score
well without reading the measurement at all. \qemscore{} adds the comparison
that can. Each simulated circuit carries an exact ideal answer. The learned
mitigator is scored beside a capacity-matched control, a model just as flexible
that reads the same circuit description but never the measurement. Each
method's measurement spend is accounted and not equalized. We run a controlled
campaign on simulated circuits and reanalyze two published learned mitigators,
Q-LEAR and QRAFT, from their released hardware data.
Three findings stand out. First, under familiar within-family conditions (S0)
evaluated across two spin-chain families and three seeds, continuous couplings
identify the target, and the control that never reads the measurement matches
87.7 to 100.5 percent of the mitigator's gain over an affine fit to the circuit
description. A plain polynomial in the coupling parameters, fitted after the
campaign, beats the mitigator on all six evaluations, reflecting the selected
learners' capacity. Second, for these selected learners, matching most of the gain is not
matching the accuracy: on five of six evaluations the mitigator removes 19.5
to 74.5 percent of the error the capacity-matched control leaves, a
learner-specific gap rather than a measurement requirement. Third, on released hardware data where descriptors only
partially identify queries, the findings differ: flexible models of the
descriptors show negligible mean gain over affine fits in Q-LEAR, and measurement
inputs carry predictive gains in both Q-LEAR and QRAFT. These comparisons
reflect representation- and protocol-specific behavior rather than an isolated
cross-regime difference. A learned mitigator's accuracy should therefore be
reported beside such controls.
\end{abstract}

%% file: 01_intro.tex
\section{Introduction}
\label{sec:intro}

\input{figure-src/qemscore-figure1/figure1}

Before fault-tolerant error correction becomes practical, mitigation is
one route to more accurate expectation-value estimates
\citep{preskill2018nisq}. Learned mitigation trains a model to predict
ideal expectation values from noisy measurements and circuit features
\citep{czarnik2021clifford, lowe2021vncdr, strikis2021learningqem,
liao2024practicalqem}. A model given circuit structure can reconstruct
ideal values from parameters alone, without the measurement.
An accuracy table therefore leaves open how much measurement helps.

\qemscore{} makes that question testable (Figure~\ref{fig:construction}).
One circuit yields descriptors $x$, a noisy estimate $r$, and an exact ideal
value $y$ (a). The full method reads $(x, r)$; a capacity-matched control,
fitted from the same candidate family under the same selection rule, reads
$x$ alone (b). Both are scored against $y$, so the gap between them is the
measurement-associated part of the error (c). An accounted ledger records
what each method spent at its own declared total, without equalizing cost.
Table~\ref{tab:positioning} summarizes the label sources, shifts, and feature
controls described by prior efforts; our focus is the capacity-matched
comparison that withholds the measurement. Q-LEAR \citep{muqeet2024qlear} and QRAFT
\citep{patel2021qraft} group inputs by execution requirement without varying
it. In the primary evaluations the control reproduces most of the improvement
over an affine control. An evaluation of this shape can therefore reward
interpolation over circuit descriptors and report it as mitigation.

\paragraph{What \qemscore{} Contains.} As a software artifact, the suite
specifies five circuit families, six noise families, six evaluation settings,
and eight registered methods.
The completed campaign evaluates a focused subset: twelve campaign datasets
with exact labels, covering two spin-chain families (transverse-field Ising and
Heisenberg), one noise family (depolarizing plus readout at two strengths),
and the familiar setting (S0) alone. Six primary evaluations assess the two
families across three seeds at shipped steps and 640 training circuits.
Evaluation across the remaining three circuit families, five noise families, and
five changed settings is future work.
For field evidence, the benchmark reanalyzes two published systems under their
released hardware data: Q-LEAR \citep{muqeet2024qlear}, with 48 paired results on
eight IBM backends, and QRAFT \citep{patel2021qraft}, on five machines. A
secondary study scores three screened published systems under a separate protocol.

\paragraph{Contributions: Three Questions and Their Answers.}

\textbf{RQ1 (what the capacity-matched control reproduces).} \emph{In the
prespecified within-family evaluations, how much of the improvement over an
affine feature-only control does a capacity-matched feature-only control
reproduce?} It reproduces 87.7 to 100.5 percent of that improvement, in all
six primary evaluations (Section~\ref{sec:results-breaks}).
The full method reduces the control's remaining error by 19.5 to 74.5
percent on five rows. On the sixth the reduction is $-12.6$ percent, and the
gap's own interval spans zero.

\textbf{RQ2 (what the published data show).} \emph{Under the released
protocols of two published learned mitigators \citep{muqeet2024qlear,
patel2021qraft}, which inputs carry the ordered predictive gap in matched
retraining?} Measurement-derived inputs carry it in both. Only Q-LEAR's
archive carries an affine arm, so only that reanalysis estimates the capacity
contrast. Under the released Q-LEAR protocol, the nonlinear descriptor control
gives a ten-fit mean hardware error reduction of $+0.000055$ over the affine one
(Section~\ref{sec:field-qlear}). In QRAFT's released ablation, which includes
no affine arm, a descriptor-only arm is about six times worse than one adding
the forward observed probability (Section~\ref{sec:field-qraft}).

\textbf{RQ3 (what the inputs cost to collect).} \emph{How many shot-level
circuit evaluations does each feature group's collection step add, and what
signed error reduction per added evaluation follows?} On hardware, base
execution adds 1,024 evaluations for a 0.1969 gap; depth cut adds 3,072 for
0.0222, a 26.61 ratio. The design carries no shot-count sweep and no
reallocating arm, so nothing says what one further evaluation would buy
(Section~\ref{sec:field-qlear}, Appendix~\ref{app:field-cost}).

The primary within-family decomposition is the paper's one headline result,
fixed before the frozen runs (Appendix~\ref{app:protocol}).
A secondary study on three screened systems (ML-QEM, Synergy CNN, Q-Cluster)
supplements the paper's three questions
(Section~\ref{sec:results-third-brief}, Appendix~\ref{app:three-systems}).

\paragraph{Scope.} Three reader expectations are not met.
First, the comparison roster does not include Clifford data regression
or its variable-noise variant; \qemscore{} models their measurement
cost and does not run them. Second, the schema carries no
native-gate counts, angle-bin histogram, or sparse Pauli
encoding, making the restricted-feature arm an ablation rather than a
reproduction of the published method \citep{liao2024practicalqem}.
Third, there is no observable-shift setting; the per-observable forest
cannot predict an unseen observable, so generation refuses that split.
One caveat belongs beside the settings: the four strength levels are
not calibrated across noise families, so the unseen-noise setting
moves channel and effective severity together.

\qemscore{} ships no routing layer; this paper reports no routing
result and claims no quantum advantage and no hardware speedup.
The controlled conclusions are conditional on the tested simulator
regimes and budgets. The field conclusions are conditional on released
archived data from eight IBM backends, its simulator companion, and the
released protocols. Each of the secondary study's three readings is
conditional on that system's released material and declared evaluation
unit.

%% file: figure-src/qemscore-figure1/figure1.tex
\begin{figure}[t]
\centering
\includegraphics[width=\linewidth]{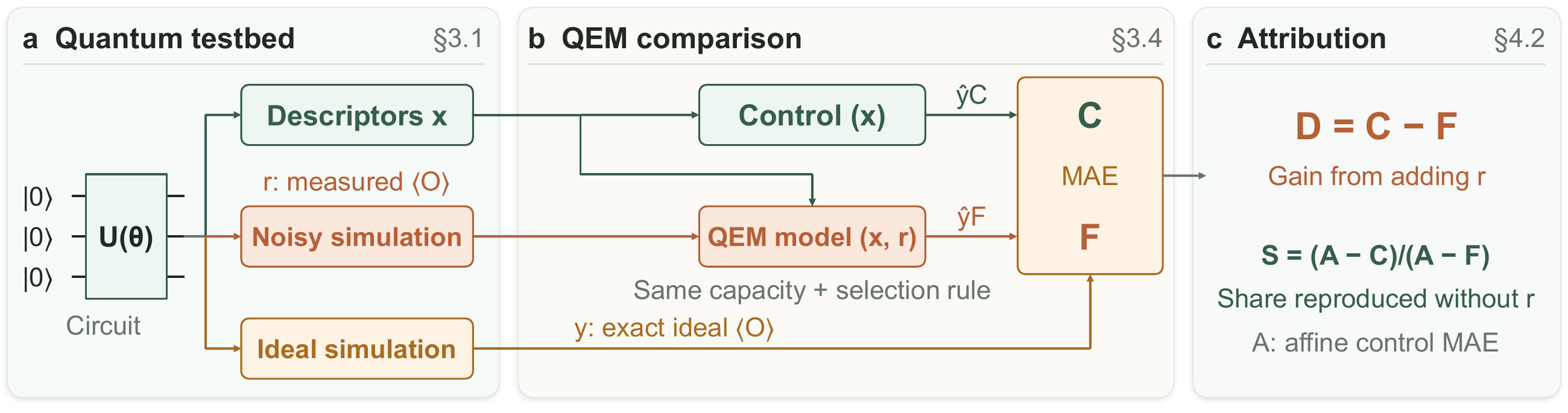}
\input{figure-src/qemscore-figure1/caption_v5}
\label{fig:construction}
\end{figure}

%% file: figure-src/qemscore-figure1/caption_v5.tex
\caption{\textbf{One circuit, one exact target, one input withheld: the two
models differ only in whether they read the measurement.}
(a) A circuit and observable $O$ yield descriptors $x$, a noisy estimate $r$,
and the exact ideal expectation $y$ from noiseless simulation.
(b) The learned QEM model reads $(x, r)$; a capacity-matched control reads $x$
alone. Both are fitted from the same candidate family under the same selection
rule and scored against $y$ by held-out macro mean absolute error, $F$ and
$C$; the diagram shows evaluation after fitting.
(c) The measurement-associated gap is $D = C - F$. Against the affine
feature-only control with error $A$, the share reproduced without measurement
is $S = (A - C)/(A - F)$; a large $S$ can coexist with a large remaining-error
reduction $D/C$.
The schematic depicts the controlled testbed (RQ1). Released-data reanalyses
(RQ2, RQ3) use each archive's own reference rather than exact $y$.}

%% file: 02_related.tex
\section{Background and Related Work}
\label{sec:related}

\subsection{From Noisy Measurements to Learned Mitigation}

A mitigation task starts from a circuit, an observable, a noise setting, and a shot
budget. Zero-noise extrapolation (ZNE) re-runs the circuit at amplified noise and
extrapolates back to zero \citep{temme2017error, li2017efficient}. Clifford data
regression (CDR) builds efficiently simulable training circuits related to the
target and learns a noisy-to-exact map \citep{czarnik2021clifford}; variable-noise
CDR (vnCDR) adds multiple noise levels \citep{lowe2021vncdr}. Learned mitigation
generalizes the recipe: a supervised model maps the noisy estimate and features of
the circuit and observable to a prediction of the ideal value
\citep{strikis2021learningqem, liao2024practicalqem}.\looseness=-1

This paper registers ZNE and defines its bill. Mitiq \citep{larose2022mitiq} serves
one purpose in this work: continuous integration cross-checks the benchmark's own
ZNE folding and Richardson extrapolation against it on shared cases. Table~\ref{tab:roster-errors} reports untouched-test ZNE errors. The paper does
not evaluate CDR or vnCDR.
Both build a fresh training set for every target circuit, and the construction rule that
fixes what those circuits are is not settled here.\looseness=-1

\subsection{Positioning and Controlled Shift}

\begin{table}[t]
\centering
\footnotesize
\setlength{\tabcolsep}{3.5pt}
\renewcommand{\arraystretch}{1.05}
\caption{Evaluation-design properties of the prior efforts this manuscript discusses, as stated in their releases. An en-dash indicates this manuscript makes no statement, rather than that the property is absent. Takeaway: the rows differ on labels, declared shifts, and feature controls rather than along a single axis.}
\label{tab:positioning}
\vspace{1pt}
\begin{tabular}{@{}
  >{\raggedright\arraybackslash}p{3.6cm}
  >{\raggedright\arraybackslash}p{3.6cm}
  >{\raggedright\arraybackslash}p{4.1cm}
  >{\raggedright\arraybackslash}p{4.1cm}
  @{}}
\toprule
\textbf{Prior effort}
  & \textbf{Label source}\newline\scriptsize Target for predictions
  & \textbf{Declared shifts}\newline\scriptsize Conditions held out or extended
  & \textbf{Feature control}\newline\scriptsize Comparison withholding input \\
\midrule
\rowcolor{abGray!35}\multicolumn{4}{@{}l}{\textit{Learned mitigators}}\\
\citet{liao2024practicalqem}
  & noiseless simulation; QEM-mitigated hardware & depth, coupling, noise drift & -- \\
Q-LEAR \citep{muqeet2024qlear}
  & ideal distribution & -- & drops observed probability \\
QRAFT \citep{patel2021qraft}
  & output distribution & -- & static vs.\ full feature sets \\
\rowcolor{abGray!35}\multicolumn{4}{@{}l}{\textit{Classical-ML shift benchmarks}}\\
WILDS \citep{koh2021wilds}
  & -- & curated distribution shifts & -- \\
ADBench \citep{han2022adbench}
  & -- & regimes, corruptions & -- \\
\rowcolor{abGray!35}\multicolumn{4}{@{}l}{\textit{Quantum mitigation benchmarks}}\\
QEM-Bench \citep{bao2025qembench}
  & ideal expectations & Trotter steps, circuit sizes, observables & -- \\
\rowcolor{abCoral!25}\qemscore{} (ours)
  & exact simulation labels & S0 evaluated; 5 single-axis changes specified & capacity-matched + 4 surrogates \\
\bottomrule
\end{tabular}
\vspace{-0.15in}
\end{table}

Table~\ref{tab:positioning} places \qemscore{} among the prior efforts closest to
this one.
\citet{bao2025qembench} standardize what learned mitigators are trained and scored
on; \qemscore{} is a separate artifact that standardizes what a score is allowed to
mean. The two are complementary rather than competing.
The evaluation pattern follows controlled shift benchmarks in classical machine
learning: WILDS curates distribution shifts \citep{koh2021wilds}; ADBench
stress-tests anomaly detectors \citep{han2022adbench}; and
\citet{bowles2024better} shows how benchmarking choices change conclusions in
quantum machine learning.
Appendix~\ref{app:related-extended} gives the full survey.\looseness=-1

%% file: 03_benchmark.tex
\section{\qemscore{}: Benchmark Design}
\label{sec:benchmark}

This section summarizes the benchmark's design. An evaluation item is a
four-part tuple: a circuit, an observable, a noise configuration, and a shot
count. \qemscore{} defines six evaluation settings, the familiar one (S0) and
five single-axis changes (S1 to S4 and S6). The completed campaign evaluates the
familiar setting (S0) alone across twelve datasets; the five changed settings
are future work on this artifact (Section~\ref{sec:results}). Its second
instrument reads released
data from published systems (Section~\ref{sec:field-protocol}). Measurement cost
is accounted at each method's own declared total, with no method subsampled to a
common budget (Appendix~\ref{app:ledger}). Full construction, composition,
setting, and scoring details appear in
Appendix~\ref{app:benchmark-task}
through~\ref{app:benchmark-field}.

\subsection{Task Definition and Testbed Construction}
\label{sec:task}

An item carries a noisy estimate $r$, the exact ideal value $y$, and a
structural description of the circuit and observable, including the sampled
physical parameters. A model can therefore predict the ideal label from
circuit parameters alone. The surrogate comparisons estimate the noisy
input's predictive benefit for the specified models and representation
(Section~\ref{sec:results}).

Every testbed is constructed by one deterministic pipeline in five steps
with the shared-circuit paths summarized in Figure~\ref{fig:construction}(a):
one seeded circuit sample feeds a noisy
path (compile, execute, producing $r$) and an exact path (noiseless
simulation, producing $y$). The two paths converge into one verified item.
Exact labels are the reason the benchmark runs on simulators: with $y$ in
hand it can measure when a correction helps and when it lands farther from
the truth than the raw estimate it replaced. Construction details, the five
steps, the two dataset schemas, and the assembly verification are in
Appendix~\ref{app:benchmark-task}.

\subsection{Benchmark Composition: Circuit Families and Noise Models}
\label{sec:inventory}

Five circuit families span structured to unstructured workloads: transverse-field
Ising, Heisenberg chain, QAOA-MaxCut (four graph classes), near-Clifford random,
and random Clifford. Every family has exactly computable ideal values, either by
statevector or stabilizer simulation. Clifford ideal values are discrete, so the
benchmark separates them at generation time: each dataset declares one stratum,
and no Clifford cell enters a headline statistic
(Appendix~\ref{app:clifford}). The first campaign evaluates two spin-chain
families (transverse-field Ising and Heisenberg) under depolarizing plus readout
noise alone across twelve datasets; the remaining three circuit families and five
noise families are software capabilities for future transfer studies.
The campaign uses two family pools per
role, with training totals of 320 or 1280 circuits, 640 validation and 320
test circuits. Three observables ship, all Z-type Pauli strings. No aggregate
MaxCut cost and no whole-energy observable is generated; this is also why the
observable axis contributes no evaluation setting. The generator requires every
observable predicted in validation or test to appear in training, and refuses to
write a dataset that breaks that closure. Observable transfer is declared future
work in Appendix~\ref{app:discussion-extended}.

Six noise families each sit on a four-level strength grid; the completed campaign
evaluates depolarizing plus readout at two strengths. The four strength levels are not calibrated
across noise families, so the unseen-noise setting moves the channel and its
effective severity together rather than isolating the channel alone. Severity
calibration was not applied to the shipped registry before the campaign ran, and
the cross-family agreement test remains a standing expected failure. Full family
descriptions, the noise-trio design, and the strength grids are in
Appendix~\ref{app:benchmark-composition}.

\subsection{Artifact Capabilities: Six Evaluation Settings}
\label{sec:settings}

Table~\ref{tab:settings} summarizes the six predeclared settings.
No observable setting exists: the random-forest candidate
fits one independent forest per physical observable, and generation refuses to
write a dataset whose prediction observables never appear in training. Observable
transfer is future work (Appendix~\ref{app:discussion-extended}). Full setting recipes and
interpretation caveats are in Appendix~\ref{app:benchmark-settings}.

\begin{table}[t]
\centering
\footnotesize
\setlength{\tabcolsep}{4pt}
\renewcommand{\arraystretch}{1.15}
\caption{Six predeclared evaluation settings. \emph{Setting}: the artifact's
identifier and a plain name; S5 is reserved for an unseen-observable setting,
which generation refuses.
\emph{What changes}: the held-out or extended part of the evaluation item.
\emph{Train domain}: what training items contain.
\emph{Test domain}: what test items contain.
\emph{What it tests}: the question each row answers.
\emph{Status}: whether the campaign has run the setting or only specified it.
The completed campaign evaluates S0 alone (mint); S1 to S4 and S6 are
future work on this artifact. Each row changes one
declared part of the familiar setting; rows do not share a severity scale
and are never ranked against each other.
Takeaway: each row specifies one conditional stress test under one
declared intervention; coupled physical properties may still move within that
intervention.}
\label{tab:settings}
\vspace{3pt}
\begin{tabular}{@{}
  >{\raggedright\arraybackslash}p{1.02in}
  >{\raggedright\arraybackslash}p{0.92in}
  >{\raggedright\arraybackslash}p{0.95in}
  >{\raggedright\arraybackslash}p{0.98in}
  >{\raggedright\arraybackslash}p{1.30in}
  >{\raggedright\arraybackslash}p{0.62in}@{}}
\toprule
Setting & What changes & Train domain & Test domain & What it tests & Status \\
\midrule
\rowcolor{abMint!35}
S0~Familiar conditions & Circuit instances only & Same families, noise, strengths & New instances, same conditions & Accuracy where training looks like deployment & Campaign run \\[3pt]
S1~Unseen noise type & Held-out noise family & Source noise trio & Held-out noise trio, same index & Transfer across mechanism and severity & Specified \\[3pt]
S2~Stronger noise & Top strength held out & Lower strengths & Top strength of seen families & Extrapolation above trained range & Specified \\[3pt]
S3~Unseen circuit family & One family held out & Remaining families & Held-out family & Transfer to new circuit structure & Specified \\[3pt]
S4~Deeper circuits & Depth extended above training & Standard depth range & Extended depths, fresh seeds & Whether the fit holds on longer circuits & Specified \\[3pt]
S6~Fewer shots & Test-time budget reduced & Standard shot count & Reduced shot count & Behavior under higher estimator variance & Specified \\
\bottomrule
\end{tabular}
\vspace{-0.1in}
\end{table}

\subsection{Methods, Budgets, and Scoring}
\label{sec:scoring}

Eight registered methods cover four roles. The raw estimate is the uncorrected
measurement. Zero-noise extrapolation uses global digital folding at scale
factors 1, 3, and 5 with three-point Richardson extrapolation. Two learned arms
follow: a ridge regressor and a restricted-feature ablation of the published
learned mitigator \citep{liao2024practicalqem}. Both consume one versioned
feature vector of thirty columns carrying no noise family, no channel parameter,
and no strength index. The ablation is not a reproduction: the dataset schema
carries no native-gate counts, no angle-bin histogram, and no sparse Pauli
encoding. Four surrogate controls fix what a learned score means: the
feature-only control receives every feature except the noisy estimate, the
noisy-only control sees the noisy estimate and shot count alone, a shrinkage
control predicts the source training mean, and the shuffled-noisy control
permutes the training noisy-estimate column.

Clifford data regression and its variable-noise extension appear in
Section~\ref{sec:related} as published methods and are absent from the roster.
The release ships a closed-form cost model and a training-circuit viability
probe for them, and no executable mitigator.

Budgets are accounted in circuit evaluations, split into training, extra
test-time, and base-measurement buckets. Feasibility tiers cap spending at
$2.5\times10^{6}$, $2.5\times10^{7}$, and $2.5\times10^{8}$ circuit
evaluations. The runner refuses any configuration whose bill exceeds its tier.
The ledger accounts for measurement spend at each method's own declared total;
no method spends a cheap allowance down to a common total. This campaign
runs at the highest tier, 250,000,000 circuit evaluations. While the campaign
roster artifacts record declared per-arm totals, the audit does not
independently verify them as executed measurement counts, and this version
reports no empirical campaign cost comparison (Appendix~\ref{app:ledger}).

Scoring uses macro mean absolute error against the exact label. Errors are
averaged within each declared evaluation cell, then across cells with equal
weight, so large easy cells cannot dominate. Intervals come from a bootstrap
that resamples whole physical circuits, keeping observables that share one
circuit together. This campaign reports no harm endpoint.
Predictions are never clipped before scoring. Full method descriptions,
the surrogate-control design, and the scoring protocol are in
Appendix~\ref{app:benchmark-scoring}.

%% file: 04_results.tex
\section{Experiments}
\label{sec:results}

The campaign evaluates transverse-field Ising and anisotropic Heisenberg
spin-chain circuits at ten qubits and three first-order Trotter steps.
In these families, instances are parameterized by fixed low-dimensional global
couplings: two coupling parameters ($J, h$) for transverse-field Ising and three
($J_x, J_y, J_z$) for Heisenberg, each sampled uniformly over fixed intervals
at a fixed step duration $dt$. Shipped step sizes are $dt = 0.2$ and $0.15$,
respectively; the large step is $dt = 0.6$ for both families. Each family has 160 or
640 training circuits, 320 source-validation circuits, and 160 untouched test
circuits. Experiments use depolarizing-plus-readout noise at the first and
third strength levels, middle-site and middle-pair observables, and 2048 shots.
Two step regimes by two training sizes by three seeds (101, 211, and 307) yield
twelve campaign datasets, all audited without failure.
The six primary evaluations comprise the two families across the three seeds at
the shipped step sizes and 640 training circuits per family ($n=640$).

Appendix~\ref{app:results-full} holds the extended protocols and auxiliary
untouched-test errors (Table~\ref{tab:roster-errors}). Cellwise signal attenuation
is reported in Appendix~\ref{app:controlled-diagnostics}.

\subsection{Validation Check: Do Models Beat the Controls?}
\label{sec:results-controls}

\textit{Which learned arms pass the source-validation gate?}

\begin{figure}[t]
\centering
\includegraphics[width=6.5in]{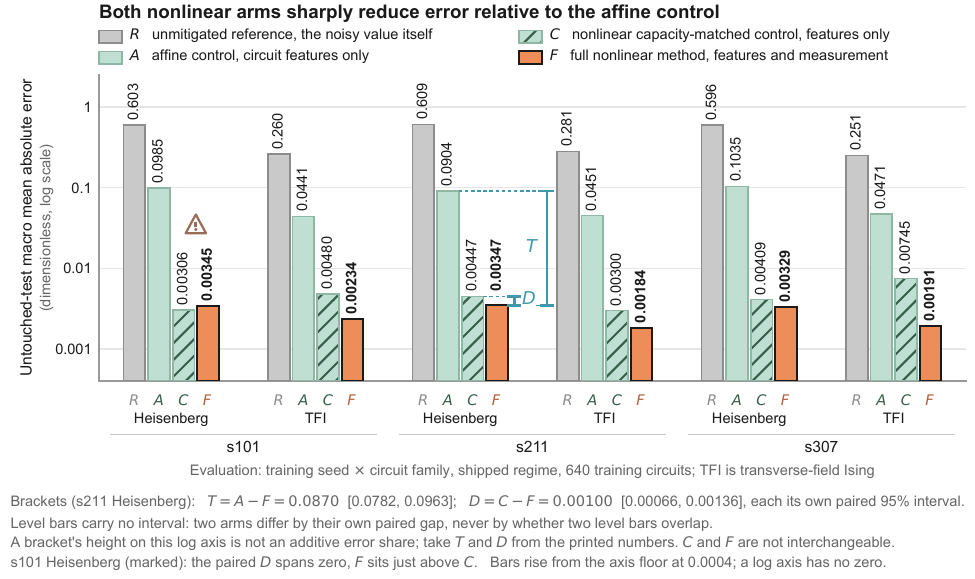}
\caption{\textbf{Both nonlinear arms sharply reduce error relative to the affine control on all six primary evaluations.}
Each bar is a macro mean absolute error against the exact label on untouched test rows, equally weighted over four strength-by-observable cells, on a logarithmic axis. $R$: unmitigated noisy reference. $A$: affine feature-only control. $C$: capacity-matched feature-only control, fitted from the ablation's own candidate family. $F$: the full nonlinear method, this campaign's restricted-feature
ablation (features and measurement). Both feature-only arms read no measurement. Bars are point estimates with no interval; two arms differ by their paired gap, not by whether bars overlap. Brackets on seed~211 Heisenberg define $T{=}A{-}F$ and $D{=}C{-}F$; the note prints their point estimates and paired 95 percent circuit-bootstrap intervals, conditional on the fitted and selected models. Coverage is not joint across evaluations. Bracket heights on the log axis are not additive error shares.}
\label{fig:share-ladder}
\end{figure}

Passing requires the macro mean absolute error to beat both the feature-only
and the noisy-only control by a factor of 1.05. The margin must hold in at
least two circuit families with continuous ideal values, and in each of them
both paired circuit-blocked intervals must lie wholly above zero.
\textbf{The restricted-feature ablation passes on all twelve campaign
datasets; ridge passes on the six large-step datasets and fails on all six at
the shipped step sizes.}
Across the twelve family results at the shipped step sizes and both training
sizes, the affine feature-only control's error is 1.0004 to 1.0145 times
ridge's, inside the 1.05 margin.
The noisy-only control's error is 1.60 to 4.52 times ridge's on those rows.
These model-selection diagnostics supply neither an independent replication
nor an attribution decision.
Figure~\ref{fig:share-ladder} shows the four arms of that decomposition
without measurement totals.

\subsection{RQ1: How Much Improvement Can Circuit Features Alone Reproduce?}
\label{sec:results-breaks}

\textit{What gain does the nonlinear arm show against a feature-only
control with comparable model capacity?}

The capacity-matched control uses the ablation's random-forest and
multilayer-perceptron candidates with the noisy-estimate column deleted at
fitting, validation, and prediction.
Full and feature-only pipelines select independently under the
one-standard-error rule, using the same searches, preprocessing, and training
schedules.
The feature-only counterpart has zero measurement cost.

\begin{figure}[t]
\centering
\begin{minipage}[t]{3.1in}
\centering
\raisebox{\dimexpr3.55in-\height\relax}{\includegraphics[width=3.1in]{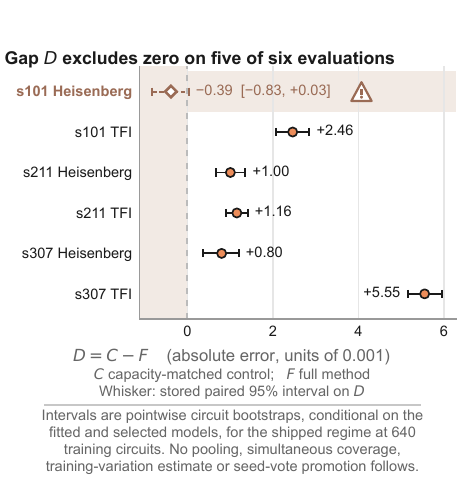}}\\[2pt]
{\small (a)}
\end{minipage}\hfill
\begin{minipage}[t]{3.1in}
\centering
\raisebox{\dimexpr3.55in-\height\relax}{\includegraphics[width=3.1in]{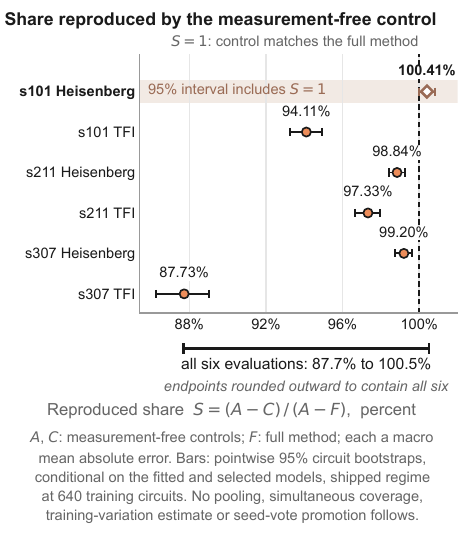}}\\[2pt]
{\small (b)}
\end{minipage}
\caption{\textbf{Five of six gap intervals lie wholly above zero; seed~101 Heisenberg spans zero with $S{=}1.0041$.}
(a)~Primary gap $D{=}C{-}F$ on untouched test rows, in units of~0.001.
(b)~Reproduced share $S{=}(A{-}C)/(A{-}F)$ as a percentage; $S{=}1$ means the control matches the full method.
$A$, $C$, $F$: affine, capacity-matched, and full nonlinear macro mean absolute errors of Figure~\ref{fig:share-ladder}.
Whiskers: pointwise 95 percent circuit-bootstrap intervals, conditional on the fitted and selected models, at 640 training circuits per family. Coverage is not joint across rows; three seeds estimate no replication rate.
Bracket below~(b): outward-rounded range of the six point estimates, not a confidence interval.
The ablation cuts the control's remaining error by 19.5 to 74.5 percent on the five rows whose gap excludes zero; on the sixth the reduction is $-12.6$ percent. The row with the lowest share here carries the largest such reduction (Figure~\ref{fig:share-vs-reduction} in Appendix~\ref{app:results-full}).}
\label{fig:gap-and-share}
\end{figure}

Two ratios follow, with different numerators and different denominators:
\begin{equation}
\label{eq:two-ratios}
T \;=\; A - F \;=\; (A - C) + D,
\qquad
S \;=\; \frac{A - C}{T} \;=\; 1 - \frac{D}{T},
\qquad
\frac{D}{C} \;=\; \frac{C - F}{C}.
\end{equation}
In Equation~\ref{eq:two-ratios}, the share $S$ divides the part the
capacity-matched control reproduces by the ablation's total improvement $T$
over the affine control. Its companion, the
remaining-error reduction, divides the gap $D=C-F$ by the control's own error
$C$. A share near one makes $D/T$ small and says nothing about $D/C$, so it can
accompany a large reduction in remaining error. The raw estimate $R$ sets the
ladder's scale and enters neither ratio. Table~\ref{tab:controls} in
Appendix~\ref{app:figure-tables} prints all four levels for every row.
Neither ratio settles whether reading the measurement is necessary.
At fixed width, step count, step size, and observable, the exact label is a
deterministic function of the sampled couplings. An unrestricted descriptor-only
predictor therefore has zero Bayes risk on this item distribution. A positive
$D$ compares finite-sample, model-class, fitting, and selection effects of the
evaluated pipelines, not information available only in the noisy estimate.
This is an information statement, not a claim of efficient classical simulation.
Our own
post-campaign diagnostic supplies a sharper reference: an ordinary least-squares
degree-five polynomial in the couplings alone (21 coefficients per observable for
transverse-field Ising, 56 per observable for Heisenberg), reading no
measurement, beats the full method on all six rows, reaching test macro mean
absolute errors between 0.0000424 and 0.000292
(Appendix~\ref{app:controlled-diagnostics}, Table~\ref{tab:poly-degree-5}).
These comparisons characterize the selected learners; they do not establish the
best achievable accuracy with or without measurements.

\findingslot{1 (primary within-family decomposition). In the six primary
evaluations, the capacity-matched feature-only control reproduces 87.7 to
100.5 percent of the improvement the restricted-feature ablation shows over
the affine feature-only control.}
Five of the six shares have intervals entirely below one; seed 101 with
Heisenberg runs from 0.9997 to 1.0084 (Figure~\ref{fig:gap-and-share}).
Holding the learner class fixed, the ablation reduces the control's remaining
error by 19.5 to 74.5 percent on those five rows. The endpoints are rounded
outward to contain all five estimates.
On seed 101 with Heisenberg that reduction is $-12.6$ percent, and the gap's
own interval spans zero.
Table~\ref{tab:primary-gap} in Appendix~\ref{app:figure-tables} reports the six
gaps beside each gap as a fraction of $T=A-F$.

The secondary evolution-step diagnostic compares the capacity-matched gap $D$
across two step regimes at 640 training circuits per family.
Its intervals are pointwise and conditional on the fitted and selected models;
three of six lie below zero, two above, and one spans zero
(both positive contrasts belong to seed 211; Table~\ref{tab:secondary-step} in
Appendix~\ref{app:results-secondary}).

%% file: 05_field.tex
\subsection{Reanalyzing Released Q-LEAR and QRAFT Data}
\label{sec:field-protocol}
\textit{How is each published system's ladder built and scored, and does the
replay reproduce the published numbers?}

The controlled half shows what an evaluation of this shape can reward. It cannot
show how often that shape occurs. We therefore ran ordered input comparisons on
the released data of two published learned mitigators, Q-LEAR
\citep{muqeet2024qlear} and QRAFT \citep{patel2021qraft}. Q-LEAR's archive
covers six application circuits on eight IBM backends, giving 48 paired results,
with a simulator companion. QRAFT's archive covers five machines and ships both
its processed inputs and its predictions.

Four rungs make the Q-LEAR ladder. An affine and then a nonlinear descriptor
control read the same four circuit descriptors and the output state's weight,
so the step between them moves model capacity alone. The base-statistics rung
adds two statistics computed across the base executions, and the full released
method adds three features collected from three depth-cut copies of the
circuit. The two declared primary comparisons are the nonlinear descriptor
control against the full method, and the base-statistics rung against the full
method. Every rung is fitted from scratch under one declared recipe, scored by
the standard Hellinger distance specified here, and accounted at its own
declared total.

QRAFT's three arms read descriptors, add the forward observed probability, then
add the reverse-execution statistics and machine identity. One reading scores
the released predictions and the other retrains at ten seeds, both by per-row
mean absolute error because the released output carries no circuit identifier. The QRAFT
panel specification was committed before its scoring script was run against any
data. Neither reading reproduces QRAFT's published results, and the panel reads
out mean absolute error while QRAFT's own models minimize a weighted
classification cost. Both readings split rows rather than circuits, so neither
says anything about unseen application circuits, and the Q-LEAR files retain no
circuit identifier either. Nothing on the field half carries an exact ideal
value, and nothing binds the Q-LEAR specification to a time before fitting.
Appendix~\ref{app:benchmark-field} holds the full ladder construction, the
QRAFT panel, the reproduction gate, and the execution counts.

All six published Q-LEAR application values fall inside our ten-fit range, with
a largest application mean deviation of 0.0134. That is range coverage rather
than an equivalence test, and it does not hold at every backend.
Appendix~\ref{app:field-replay} gives the full reproduction panel with
Table~\ref{tab:field-repro}.

\subsection{RQ2 and RQ3: Measurement Gains and Costs in Q-LEAR}
\label{sec:results-field}
\label{sec:field-qlear}

\textit{Under the released Q-LEAR protocol, how much of the ordered improvement
does a nonlinear descriptor control reproduce, and how much needs the
measurement-derived inputs?}

\begin{figure}[t]
\centering
\includegraphics[width=6.5in]{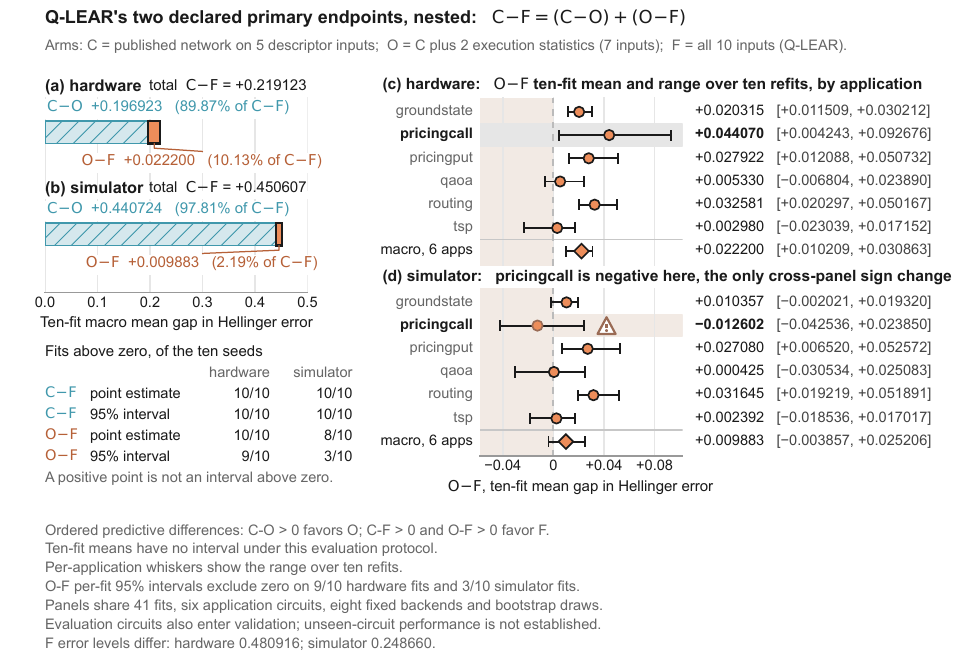}
\caption{\textbf{The base-statistics step carries 89.87 percent of the total gap on hardware and 97.81 percent on the simulator.}
Endpoints are nested: $C{-}F=(C{-}O)+(O{-}F)$. $C$: nonlinear descriptor control (five inputs). $O$: adds base execution statistics (seven inputs). $F$: full ten-input released method. Positive gap means the fuller set has lower error.
(a,~b) Ten-fit macro mean $C{-}F$ split into its two additive pieces, hardware and simulator.
(c,~d) $O{-}F$ ten-fit mean per application and macro row; whiskers show the range over ten refits, not a confidence interval.
The lower-left block counts fits with positive point estimates and fits whose 95 percent intervals lie wholly above zero. Both panels score the same 41 fits on six circuits and eight backends with identical bootstrap draws, so the simulator column is not an independent replication. The depth-cut step has an interval above zero on 9 of 10 hardware fits and 3 of 10 simulator fits.}
\label{fig:qlear-endpoints}
\end{figure}

\textbf{The nonlinear descriptor control adds a ten-fit hardware macro mean of
$+0.000055$ over the affine one, and every per-fit interval spans zero.}
The anticipated large capacity contribution is therefore not observed, which is
weaker than absence and is the strongest statement these fits support.
Measurement-derived inputs carry the ordered gap on hardware. There the
full released method improves on the nonlinear descriptor control by a ten-fit
macro mean of 0.2191, with per-fit intervals above zero on all ten fits. The
base-statistics rung opens 0.1969 of that gap and the three depth-cut circuits
0.0222. That depth-cut interval is above zero on nine of ten hardware fits and
three of ten on the simulator.
Figure~\ref{fig:qlear-endpoints} reports the two measurement steps on both
panels.

\textit{RQ3: what collection requirements accompany these ordered gains?}
\nopagebreak

\nopagebreak
\textbf{On hardware the base-execution step adds 1,024 shot-level circuit
evaluations for a ten-fit mean gap of 0.1969, and the depth-cut step adds
3,072 for 0.0222.}
Per added evaluation those are $1.923 \times 10^{-4}$ and
$7.227 \times 10^{-6}$, a descriptive ratio of 26.61 with no interval
reported. The simulator companion costs 44,843 and 134,528 evaluations for
the same two steps, and seven of its ten depth-cut intervals cross zero
(Appendix~\ref{app:field-cost}).
One absence governs every number here. The design carries no shot-count sweep
and no arm that reallocates a fixed number of evaluations across methods, so
nothing here says what one further evaluation would buy.

\begin{figure}[t]
\centering
\includegraphics[width=6.5in]{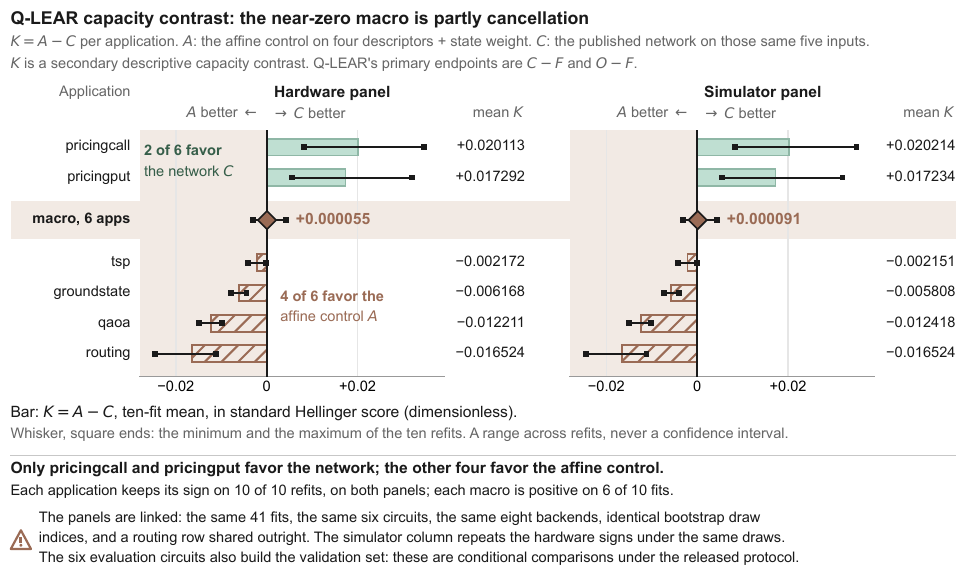}
\caption{\textbf{The near-zero macro mean ($+0.000055$ hardware, $+0.000091$ simulator) is partly cancellation: two pricing applications favor the network and the other four favor the affine control on all ten refits of both panels.}
$K{=}A{-}C$ per application. $A$: affine control on four circuit descriptors and the state weight. $C$: published network on the same five inputs; the two arms differ only in learner capacity. Positive~$K$ means the network has lower error.
Bars: ten-fit means in Hellinger score. Whiskers with square ends: minimum and maximum over ten refits, a range and not a confidence interval. Per-application rows carry no interval because one application leaves no resample unit. $K$ is a secondary descriptive capacity contrast; Q-LEAR's primary endpoints are $C{-}F$ and $O{-}F$ (Figure~\ref{fig:qlear-endpoints}). Both panels share 41~fits, six circuits, eight backends, and identical bootstrap draws, so the simulator column is not independent confirmation.}
\label{fig:qlear-cancellation}
\end{figure}

\textbf{The near-zero average is partly cancellation: the nonlinear descriptor
control improves on the affine one for both pricing applications on all ten
fits, by hardware means of $+0.020113$ and $+0.017292$.} It worsens the other
four applications on all ten fits, and the simulator carries the same signs.
The gap from the nonlinear descriptor
control to the full released method varies by an order of magnitude across
applications, from 0.053 on QAOA to 0.565 on routing. It stays positive on
every one of the six. Figure~\ref{fig:qlear-cancellation} gives the six
capacity contrasts on both panels.
Appendix~\ref{app:field-qlear} gives the full ladder specification, scoring
protocol, and descriptor-information discussion.

\subsection{RQ2: How Much Do Measurements Improve QRAFT?}
\label{sec:field-qraft}

\textit{Does a descriptor set richer than Q-LEAR's four circuit numbers and a
state weight close the gap in a second published system?}

\textbf{Descriptors alone score roughly six times worse than descriptors plus the
forward observed probability: 12.4838 against 2.1074 in the released predictions,
and 13.5903 against 2.2685 under matched retraining.}
Conditions favor
the descriptor-only arm, because the released split is shuffled by row and a
circuit's output states therefore appear in both training and test. It loses
anyway, on every machine and every seed.
The final feature group adds reverse-execution statistics and machine identity.
It improves matched retraining by about six percent, from 2.2685 to 2.1246, and
loses in the released predictions, 2.4467 against 2.1074. That bounded statement
concerns this score, and it is not a claim that QRAFT's reverse-circuit method
fails under the objective QRAFT chose, which this work does not evaluate.
Table~\ref{tab:field-qraft} in Appendix~\ref{app:field-qraft} gives both panels,
the full three-arm specification, the per-machine and per-seed ratios, and the
panel's three limits.

\subsection{Scope of the Q-LEAR and QRAFT Findings}
\label{sec:field-scope}

\textit{Taken together, what do the two reanalyses support, and what stays
outside them?}

The Q-LEAR ladder estimates the capacity contrast the controlled campaign
anticipated, and does not show that pattern occurring. The QRAFT panel compares
three feature sets instead, with no affine arm, no reproduced share and no
interval, so its error ratio does not test the controlled share.
Two instances are not prevalence, and
this design cannot support prevalence in either direction. Both archives come
from one release, so the field half holds two feature designs rather than two
independent environments. Neither reanalysis claims execution-free prediction; every rung conditions on a support that was chosen by
executing the circuit. Grouping information sources by origin predates this work: Q-LEAR's own RQ3
retrains after individual feature deletions, and QRAFT already compares static,
static-plus-forward, and full forward-and-reverse feature sets
\citep{muqeet2024qlear, patel2021qraft}. What these panels add is the ordered
group comparison, under one retraining recipe and a single score per panel, with
the controls this paper specifies. Appendix~\ref{app:field-cost} places the released
collection counts beside each step.

%% file: 06_three_systems.tex
\subsection{Additional Case Studies: ML-QEM, Synergy CNN, and Q-Cluster}
\label{sec:results-third-brief}

\textit{What does a descriptor control's reproduced share reveal in screened
published systems?}

Beyond the three primary research questions, a six-criterion structural screen
admitted ML-QEM
\citep{liao2024practicalqem}, Synergy CNN \citep{cantori2024synergy}, and
Q-Cluster \citep{patil2025qcluster}. Each uses its own loss and a scoring rule
committed before fitting. Affine $A$ and capacity-matched $C$ use the same
execution-independent descriptors; $C$ follows the system's model family and
selection rule, while $F$ uses its full features.
For $S=(A-C)/(A-F)$, a predeclared interval wholly above 0.5 reads present, wholly below
absent, and spanning 0.5 indeterminate (Appendix~\ref{app:three-systems}).

\textbf{ML-QEM reads present at $S = 0.5985$ $[0.5808, 0.6172]$; Synergy CNN
reads absent, with all six upper interval limits at or below $+0.2755$.}
Of ML-QEM's 2,500 four-qubit test rows, 277 reuse a training descriptor block.
Mean per-row expectation-vector errors are $A = 0.4419$, $C = 0.2376$,
$R = 0.2097$, and $F = 0.1005$. Thus $S = 0.5985$ coexists with $C$
performing worse than raw $R$: $(R-C)/(R-F) = -0.2552$.
Synergy CNN's controls use reduced data, so no reading holds at the published
budget.

\textbf{Q-Cluster reads present at $S = 0.8795$ $[0.7723, 0.9658]$ with
calibration and $0.9211$ $[0.8480, 0.9716]$ without.}
It scores intermediate regression on a circuit-overlapping split, not final
mitigation accuracy. None establishes a clean independent occurrence of the
controlled pattern in final mitigation accuracy, or its absence.
Appendix~\ref{app:three-systems} gives the full analysis and
Figure~\ref{fig:three-systems}.

%% file: 07_discussion.tex
\section{Discussion and Limitations}
\label{sec:discussion}
\label{sec:limits}

In the controlled campaign, continuous couplings identify the exact target at
fixed family, size, observable, and evolution time. The residual gap $D$ is therefore
learner-specific; it measures no irreducible information benefit of the measurement.
Released field descriptors only partially identify queries, and the archives
differ in targets, scores, labels, and splits. QRAFT tests no capacity contrast;
Q-LEAR's nets to near zero across opposite-signed applications, with measurement-derived inputs carrying the predictive
gap (Section~\ref{sec:results-field}). These case studies establish representation- and
protocol-specific comparisons, without isolating a cause of cross-regime
differences. The restricted-feature ablation uses \qemscore{}'s own features
and establishes nothing about \citet{liao2024practicalqem}'s published method.
Appendix~\ref{app:discussion-extended} expands the interpretation.

The descriptor-only control is an evaluation tool, not a replacement for
quantum execution. In the controlled campaign it learns from exact simulated
targets; its low error does not supply those targets for arbitrary circuits.
In the released archives, even descriptor-only arms condition on output
support obtained by executing the circuit. The input comparisons therefore
diagnose what a specified learning pipeline gains from additional measurement
features. They do not establish that an equally accurate, execution-free
estimator is available for a new circuit.

\paragraph{Limitations.} Controlled conclusions depend on the tested simulator
regimes, circuit families, and budgets; hardware transfer remains untested.
Field conclusions depend on archives from eight IBM backends, their simulator
companion, and released protocols. Each screened-system reading depends on its
released material, evaluation unit, and split (Section~\ref{sec:results-third-brief}).
Cost is accounted, not equalized (Appendix~\ref{app:ledger}).

Clifford data regression and its variable-noise variant have cost formulas but
no executable comparators. The four strength levels lack cross-family severity
calibration. A pre-label reject-and-fallback layer is absent; checking decoded
run artifacts cannot establish label-free routing (Appendix~\ref{app:discussion-extended}).
Validation neither replays sampled counts nor recomputes stored structure
features; Appendix~\ref{app:protocol} details the separate campaign audit.
This paper claims no quantum advantage or hardware speedup.
The five changed settings, S1 to S4 and S6 (Table~\ref{tab:settings}), remain
unevaluated. Appendices~\ref{app:limitations-full} and~\ref{app:discussion-extended}
detail the limitations and future evaluations.

\section{Conclusion}
\label{sec:conclusion}

\qemscore{} evaluates how much the measurement adds to learned quantum error
mitigation by comparing each learner with a capacity-matched control that
reads the same circuit descriptors without the measurement. This comparison
separates the improvement reproduced by descriptors from the remaining
measurement-associated gap. Exact simulated targets, released-data reanalyses,
and measurement-cost accounting make these quantities inspectable alongside
prediction error.

Across the six primary within-family evaluations, descriptor-only controls
reproduce 87.7 to 100.5 percent of the gain over an affine control. Yet the
full method reduces their remaining error by 19.5 to 74.5 percent on five
evaluations; the sixth gap's interval spans zero. Recovering most of the gain
therefore leaves room for meaningful accuracy differences within the selected
learners. In the released Q-LEAR and QRAFT archives, measurement-derived inputs
carry predictive gains. Together, these protocol-specific comparisons show why
accuracy alone cannot establish how much a learned mitigator benefits from
measurement inputs.

The practical recommendation is to report the full learner beside
descriptor-only controls, with same-unit errors, uncertainty, circuit overlap,
and each method's measurement spend. A reproduced share needs its affine
reference and remaining-error reduction to be interpretable. This reporting
practice makes the contribution and collection cost of measurement inputs
visible, giving subsequent evaluations a clearer basis for judging progress
in learned quantum error mitigation.

%% file: 08_statements.tex
\subsubsection*{Broader Impact Statement}

This simulation and archived-data study informs learned-mitigation reporting.
We identify no direct dual-use concern. Aggregate comparisons do not identify
individual harmful corrections.

\makeatletter
\if@accepted
\subsubsection*{Acknowledgments}

This work builds on Qiskit and Qiskit Aer, Stim \citep{gidney2021stim} for
stabilizer labels, Mitiq \citep{larose2022mitiq} as the zero-noise extrapolation
reference, and scikit-learn.
\fi
\makeatother

%% file: 08_reproducibility.tex
\subsubsection*{Reproducibility Statement}

Controlled datasets regenerate from recorded seeds. Validation rebuilds
circuits, recomputes exact labels, and checks hashes and ledger conservation.
The campaign archive and per-file checksums are at
\releaselink{https://github.com/yzhao062/QEMScore/releases/tag/campaign-archive-v1}%
{a public release tagged \texttt{campaign-archive-v1}, whose host is withheld
for double-blind review; the anonymous supplement contains only the compact
controlled-diagnostics package behind Tables~\ref{tab:roster-errors},
\ref{tab:signal-difficulty} and~\ref{tab:poly-degree-5}, not the campaign
archive or the field reanalyses}.
Appendices~\ref{app:repro-full} and~\ref{app:repro} detail reproducibility.

%% file: 09_appendix.tex
\section{Formal Protocol and Code Crosswalk}
\label{app:protocol}

Apart from the setting identifiers described in Table~\ref{tab:settings}, this
appendix is the only place internal identifiers appear; the artifact's manifests,
presets, and result files use these names.

\begin{table}[ht]
\caption{Crosswalk from the plain names of the main text to the identifiers in
the artifact. The budget-tier letters L, M, and H are unrelated to the
severity labels L1 to L4. Takeaway: every main-text name resolves to exactly
one identifier that appears in a shipped manifest or result file.}
\label{tab:crosswalk}
\begin{center}
\small
\begin{tabular}{lll}
\toprule
\bf Main-text name & \bf Artifact identifier & \bf Kind \\
\midrule
Familiar conditions & S0 & Evaluation setting \\
Unseen noise type & S1 & Evaluation setting \\
Stronger noise & S2 & Evaluation setting \\
Unseen circuit family & S3 & Evaluation setting \\
Deeper circuits & S4 & Evaluation setting \\
Fewer shots & S6 & Evaluation setting \\
Raw estimate & \texttt{raw} & Method \\
Zero-noise extrapolation & \texttt{zne} & Method \\
Ridge regression & \texttt{ridge} & Method \\
Restricted-feature ablation & \texttt{liao} & Method \\
Feature-only control & \texttt{feat-only} & Surrogate control \\
Noisy-only control & \texttt{noisy-only} & Surrogate control \\
Shrinkage floor & \texttt{shrinkage} & Surrogate control \\
Shuffled-noisy control & \texttt{shuf-noisy} & Surrogate control \\
Noise strengths (weakest to strongest) & L1 to L4 & Severity levels \\
Budget tiers (2.5M, 25M, 250M evaluations) & L / M / H & Circuit-evaluation caps \\
Smoke data tier & T0 & Dataset tier \\
\bottomrule
\end{tabular}
\end{center}
\end{table}

The grammar also reserves the identifier S5 for an unseen-observable setting,
and generation refuses to emit it. A split whose prediction observables never
appear in training violates the source-closure rule that the fixed
per-observable forest needs. That refusal is a regression test rather than an
accident, and the setting is out of scope here (Appendix~\ref{app:limitations-full}). Of
the dataset tiers, only the smoke tier ships; the paper core is frozen at
protocol freeze, after the pilot cost and power audit.

The runner refuses overlapping training, validation, and test item IDs, and the
run-artifact validator rechecks the recorded item overlaps. These checks do not
certify physical-circuit disjointness across roles. The campaign checks that
stronger property on realized circuit hashes before fitting. Validation rows
come from the source domain. For future transfer experiments, the five
changed-setting recipes use source-domain validation. Their source noises are
the stochastic trio at the three lower strengths. The unseen-noise recipe uses
the held-out trio on its uncalibrated grids (Appendix~\ref{app:noise}); the
stronger-noise recipe uses the source families' top strength. These recipes are
outside the first campaign, whose numeric design and selection protocol are
specified in Section~\ref{sec:results}. Hyperparameter grids are frozen before
any training run, selection uses validation error only, and test labels never
touch tuning.

The primary result is the six within-family decompositions at the shipped
step sizes and 640 training circuits per family. Let $A$, $C$, and $F$
denote the affine feature-only, capacity-matched feature-only, and full
nonlinear errors. We report $T=A-F$, $K=A-C$, $D=C-F$, and the share
$S=K/T$ separately for every primary family-and-seed row, with pointwise
circuit-bootstrap intervals conditional on the fitted and selected models.
No share is selected or averaged. Evolution-step and training-size
contrasts are secondary and cannot promote the primary claim.
Section~\ref{sec:results-breaks} states the numeric design. The
decomposition and its primary setting were declared after inspecting a
rehearsal at generator seed 4243, which is not a campaign seed and whose
scores are not reported here. Before final
generation, the campaign freezes the resolved specifications, the code revision,
the bootstrap stream mapping, the audit rules, and the requirements for recording
models and predictions. After generation and before fitting, it checks realized
counts, physical role separation, and the intended cross-size identity
relationships. Selected configurations and item-keyed predictions are retained
after fitting and bound to the dataset and code revision. This campaign does not require a
second noise parameterization or multiple feasibility tiers.

Role counts in a split specification are per-role totals, divided across the
resolved family, width, and depth pools, so a per-configuration target must be
authored pool by pool. All randomness derives from one master seed through named
streams, and every per-item seed is recoverable from the manifest.

The campaign carries its own audit, because the released validator does not
replay sampled counts or recompute compiled structure (Section~\ref{sec:limits}).
Before any campaign number enters a table, an external audit replays each
family's seeded parameter draws for every pool instance, recompiles every
distinct circuit to compare stored depth and two-qubit gate count, and replays
complete seeded histograms and both observable statistics on a sample selected
by a fixed index rule declared before any outcome is inspected. That sample
covers both circuit families, both evolution-step regimes, both training-size
artifacts, every generator seed, all three roles, and both
severities, with independent ideal-label calculations for each family. For the
descriptive zero-noise extrapolation results it re-executes the folded circuits
at the declared scales and compares the final extrapolated predictions, because
the shipped implementation discards its folded histograms after use and no stored
counts exist to compare against. The
audit code, environment, selected identifiers, comparisons, and failures are
released beside the campaign artifacts. An audit failure blocks use of the
affected results until its cause is resolved and the resolution is recorded.
Exact-equality and numerical-tolerance rules are frozen before outcomes are
inspected. This is a stratified campaign audit, and
the paper describes it as one; it is not an independent replay of every campaign
execution. The index rule is fixed: per family, regime, and seed pool, training
instances 0, 79, 159, 160, 399, and 639; validation 0, 159, and 319; and test 0,
79, and 159, skipping the indices a training size does not contain. Across both
families, both regimes, and all three seeds that is 144 distinct circuits, 288
base measurement groups, and 144 folded executions.

\section{Measurement Bill: Formulas and Tiers}
\label{app:ledger}

The cost unit is the circuit evaluation: one backend call times its shots.
The ledger defines three buckets per method: $B_{\mathrm{train}}$ for
generating every circuit a method fits or selects on, which is the training split
and the source-validation split together, $B_{\mathrm{extra}}$ for
mitigation-specific circuits at test
time, and $B_{\mathrm{pred}}$ for the base noisy measurement. Exact ideal
labels from simulation are logged separately; they are not noisy backend
calls, and burying them would flatter supervised methods. Compatible
observables of one circuit, noise, and shot configuration share a single
measurement, which is executed once and charged once.

The artifact offers three feasibility caps: 2,500,000, 25,000,000, and
250,000,000 circuit evaluations. The first campaign uses one declared cap,
recorded in its resolved manifest. A cap applies to
$B_{\mathrm{train}} + B_{\mathrm{extra}} + B_{\mathrm{pred}}$ per method
and per paired budget cell. If a method exceeds it, the runner refuses the
whole run before execution and reports each offending method's shortfall.

The tier caps spending; it does not equalize it. No allocator spends a cheap
method's remaining allowance on extra base-measurement shots. Under this accounting rule, methods at the same tier can incur different
realized totals. The micro-preset example in Appendix~\ref{app:limitations-full}
illustrates that difference. The nominal total is
training plus base prediction spend; ZNE's extra folded executions are what make
its realized total larger. Section~\ref{sec:scoring} states the campaign's
cost-reporting limits in this version.
Appendix~\ref{app:field-cost} separately reports source-implied field
collection requirements.
Wall-clock time is secondary engineering data; simulator speed and queue
behavior distort it, so the fairness unit is the circuit-evaluation count.

\section{Noise Families and Strength Grids}
\label{app:noise}

Six named noise families each have a four-level strength grid. The first campaign
uses depolarizing plus readout at the first and third strengths. For future
transfer recipes, the source trio is depolarizing, dephasing, and amplitude-phase
damping, each with readout. The held-out trio is coherent overrotation,
correlated two-qubit crosstalk, and a mixed heterogeneous model. Channel
constructions, grid values, and infidelity diagnostics ship with the artifact.

The shipped grid values are monotone placeholders. No calibration has matched
them to a common severity across families, and the release says so in the
module that defines them. A calibration proposal ships beside them, naming the
operational scalar (average gate infidelity per layer at a reference shape),
the depolarizing anchor, and a solver that computes matched grids. That
proposal is marked inactive, and generation binds the placeholder registry. A
standing expected failure in the test suite asserts that the shipped families
disagree with the anchor, and it is the artifact's own record that the gap is
open. The consequence for future transfer experiments is direct. Comparing an unseen noise
family against a trained one mixes a change of channel with an undeclared
change of strength. The result is a transfer measurement rather than an
isolation of the mechanism change. Matched severity grids and an
alternative-matching sensitivity check are left to a future transfer campaign.
Cross-family severity
comparisons stay out of scope in either case.

\section{Statistical Protocol}
\label{app:stats}

Accuracy uses mean absolute error against the exact label, computed on the raw
expectation scale. All observables in this release are single-basis Z or ZZ
parities, so no per-observable normalization applies. Root-mean-square error
is a tail diagnostic, and signed bias is reported by stratum. The artifact can also compute
$h=\max(0,|m-y|-|r-y|)$ and summarize its total, mean, maximum, and
positive fraction. These are artifact capabilities, not endpoints of
the reported campaign, and no harm results are reported here. Predictions are never clipped before scoring, and leaving the
physically valid range is reported as its own rate. A future transfer campaign must predeclare its degradation summary before use.

Aggregation uses macro means and interquartile ranges across predeclared
cells, so large easy cells cannot dominate. Macro aggregation gives cells equal
weight; we retain cell-level signed biases to expose cancellation in their
aggregate. The cell key for split datasets is method, role, split identity,
circuit family, noise family, severity, and observable, so records from two
evaluation settings cannot merge under equal remaining fields. Legacy
single-domain fixtures keep the earlier key without split identity, and the
runner declares which of the two an artifact uses. The report separates strata
as well: the continuous-regression headline, the Clifford control, and one
section per declared split are computed and rendered separately, and Clifford
cells enter no headline statistic.

\textbf{Provisional protocol for future transfer comparisons.} Such comparisons
would use paired cell differences, Wilcoxon signed-rank tests
\citep{wilcoxon1945individual}, Holm correction within each metric family,
median paired differences, and rank-biserial effect sizes. Mean ranks with
critical-difference diagrams \citep{demsar2006statistical} would be secondary.
Circuit-blocked bootstrap intervals keep all rows of a sampled circuit together.
The proposed win, tie, loss, and harm categories, their practical margins, and
the degradation summary remain to be finalized before a transfer campaign.
They do not govern this campaign's findings. Its primary decomposition is
specified in Section~\ref{sec:results-breaks} and Appendix~\ref{app:protocol};
the evolution-step and training-size contrasts are secondary.

Two properties of the intervals bound what they support. They are pointwise and
conditional on the fitted and selected models, so they carry neither training
nor model-selection uncertainty. And a difference between two settings is
estimated directly, as a contrast with its own interval, rather than inferred
from one setting passing a threshold and another failing it.

\section{Field Statistical Protocol}
\label{app:field-stats}

This appendix fixes the score, the aggregation, the fits, and the intervals
behind the field reanalysis. The Q-LEAR ladder, recipe, seeds, score, weights
and resample unit were fixed before the controlled campaign ran, and the
analysis follows that text. QRAFT's panel carries its own separately documented commitment, made before
its own scoring rather than before the controlled campaign. Neither commitment
covers the descriptive efficiency presentation, which was formed after both
analyses completed. One binding
artifact is absent, and it belongs here rather than in Limitations: no pre-fit
freeze file was written for the field half. Checksums taken before review show
that the inputs did not change afterwards, and they cannot establish when the
specification was frozen.

The affine control is fitted once and reused. C, O, F, and the off-ladder C4
diagnostic are each fitted at seeds 0 through 9, giving forty-one fits. An
affine descriptor control and a nonlinear descriptor control both read the four
released circuit descriptors together with the state weight, and they differ
only in learner capacity. A third descriptor arm drops the state weight, so it
predicts one value for every output state of a circuit. The base-statistics rung
adds two statistics taken across the base runs, a percentile of the observed
probability and a maximum-over-minimum odds ratio. On top of that, the full
released method adds the three depth-cut features.

Scores are the standard Hellinger distance between the predicted and the target
distribution, taken on the released observed-state support. Released files
record percentages, which the scorer converts to probabilities. Negative entries
are clipped at zero, and each vector is then normalized by its own sum before
the distance is taken. A nonfinite prediction is caught ahead of the clip and
refused, rather than being read as zero probability.
The distance is the standard Hellinger form, $H(p,q) = \frac{1}{\sqrt{2}}
\lVert \sqrt{p} - \sqrt{q} \rVert_2$, with the $1/\sqrt{2}$ factor carried, so
values lie in $[0,1]$. The release's own second research question renormalizes
differently, and analyses computed under the two normalizations are never mixed
into one decomposition.

Aggregation is equal weight in two stages. One distribution error is produced
per application and backend, the eight backends are averaged equally within an
application, and the six applications are then averaged equally. These 48 paired
field results and their fits are not included in the \texttt{campaign-archive-v1} release. The six
applications are \texttt{groundstate}, \texttt{pricingcall}, \texttt{pricingput},
\texttt{qaoa}, \texttt{routing}, and \texttt{tsp}.

Spread across the ten fits of each nonlinear arm describes refitting variation and is
not a confidence interval.

Intervals are pointwise application bootstraps at the 95\% level. Each one
conditions on a single fit, on the eight backends, and on the released labels,
and resamples applications. Refitting and model-selection uncertainty therefore
sit outside every field interval, and the ten-fit spread describes that variation
separately. Counts of intervals above zero are descriptive counts over the same
six applications rather than independent replication rates, and the hardware and
simulator panels reuse those same applications.
Each interval comes from 10,000 resamples at root seed 20260906, taking the
2.5th and 97.5th percentiles of the resampled statistic for a 95 percent
interval. These three settings are read from the field scorer's own constants.
They are separate from the controlled campaign's physical-circuit resampling,
which uses its own root seed, and the two are never mixed.

Two arithmetic checks run on every seed and panel. The two ordered steps sum to
the full gap, with residuals of 0 or $5.6\times10^{-17}$. Re-scoring the
unchanged predictions under the hardened scorer reproduces every reported field
number. That scorer declares the cohort, the backends, and the seeds as
constants instead of deriving them from whichever records survive. It refuses a
duplicate fit record, stops the run on a missing fit, and marks a seed as not
estimable when a prediction vector is nonfinite.

The evaluation protocol requires a verified reproduction of the released hardware table
before any ablation is interpreted, and the run followed that order. No
reproduction tolerance was declared beforehand, so calling the replay faithful is
a post-hoc judgement rather than a test against a prespecified bound.

The released generator builds its validation set from the same circuit directory
that both evaluation panels read, and that directory holds the six application
files. Released CSV files carry no circuit identifiers, so neither a row-level
mapping nor a circuit-disjoint split can be reconstructed from the archive.
Field results are therefore conditional predictive comparisons under the released
protocol, and they are not evidence about unseen application circuits.

\textbf{The QRAFT panel, scored separately.} QRAFT's released data
support a different score, and the two scores are never compared or combined.
Rows there are scored by mean absolute error in probability points, weighted
equally within a machine and equally across the five machines. Hellinger distance
is not computable on that archive, because it needs a circuit grouping the
released output does not carry.

Two readings of that archive run side by side. One reads the three arms as
shipped in the released prediction file, 1,524 rows, and fits nothing. The other
retrains all three arms from the released processed inputs, 10,155 rows, at ten
seeds under one recipe. Arms are QRAFT's own: seven circuit descriptors; ten
inputs adding the forward observed probability at three severities; and seventeen
inputs adding the reverse-execution statistics and machine identity. Spread
reported for the retrained panel is the ten-seed range rather than a confidence
interval.

No interval is reported for either panel, and that was decided before the run.
Released output carries no circuit identifier and no state label, so a circuit's
states cannot be reassembled and no resample respecting within-circuit
correlation can be constructed. Grouping by machine and descriptor tuple yields
735 groups over 1,524 rows at a median size of two, which are partial circuits
rather than circuits. Five machine-level clusters are too few for a stable 95\%
interval, and a wrong interval is not an acceptable substitute for none.

Neither panel reproduces QRAFT's published results. The released-prediction panel
reads their shipped predictions without checking them against the numbers printed
in their paper. Retraining uses a histogram gradient-boosting regressor in place
of their MATLAB classification ensembles with Bayesian hyperparameter search.
Their models minimize a weighted classification cost that
penalizes errors on high-probability states, while the retrained panel minimizes
squared error, and both are read out here under mean absolute error. A model
fitted for one objective can score worse under another without being worse at its
own, so nothing here evaluates QRAFT's method under the objective QRAFT chose
\citep{patel2021qraft}. Absolute errors from the retrained panel carry no weight;
only the ordering across arms does. Both panels split rows rather than circuits,
so neither says anything about unseen application circuits.

\section{Field Execution Accounting}
\label{app:field-cost}

Counts in this appendix are read from the released generators rather than from
execution records. The unit matches Appendix~\ref{app:ledger}: one backend call
times its shots, reported here as shot-level circuit evaluations. These are
source-implied intended counts, and they establish nothing about historical
billing, retries, or wall-clock time.

Per application circuit, the released hardware generator builds two copies of the
transpiled circuit, one full inverse, and three depth-cut circuits, then advances
its index by six. Every one of those circuits runs at 1,024 shots. The first base
run supplies the observed support that the descriptor arms are conditioned on.
Both base runs are required by the base-statistics rung, because the percentile
and the maximum-over-minimum ratio are taken across them. Released generation also
executes the full inverse, which no retained feature reads.

\begin{table}[h]
\caption{Required shot-level circuit evaluations per rung, read from the released
generators. Hardware counts are constant across applications; simulator counts
are equal-weight means over applications, whose mean output-state count is
21.8958. Ideal supervision is one run per circuit on both paths, reported
separately here and excluded from every efficiency in this appendix. Takeaway:
every rung conditions on at least one 1,024-shot execution, so no rung of this
ladder is execution-free.}
\label{tab:field-rungs}
\begin{center}
\small
\begin{tabular}{lrr}
\toprule
\bf Rung or path & \bf Hardware, per circuit & \bf Simulator, mean per circuit \\
\midrule
Descriptor arms, support enumeration only & 1,024 & 1,024 \\
Base-statistics rung, adds the base statistics & 2,048 & 45,866.67 \\
Full released method, adds the depth-cut circuits & 5,120 & 180,394.67 \\
Released generation, adds the unused full inverse & 6,144 & 225,237.33 \\
Ideal supervision, counted separately & 1,024 & 1,024 \\
\bottomrule
\end{tabular}
\end{center}
\end{table}

Simulator counts are derived rather than inherited from the hardware path. That
generator places feature collection inside its per-state loop, at two repetitions
per call and 1,024 shots. Five calls run per output state: the base circuit, the
full inverse, and the three depth-cut circuits. Per-circuit counts there follow
$1024 + k \times 1024 \times s$, where $s$ is the circuit's number of output
states. The multiplier $k$ is 0, 2, 8, and 10 for the first four rows of
Table~\ref{tab:field-rungs}. Simulator figures in that table are equal-weight
means over applications, and no individual circuit has a fractional count. A
feature group therefore costs once per application circuit on the hardware path,
and once per output state on the simulator path. That difference is a property of
the released collection code rather than of the features.

\begin{table}[h]
\caption{Each ordered step, the shot-level circuit evaluations it adds, its
ten-fit mean Hellinger reduction, and that reduction divided by the added
evaluations. The last column counts per-fit application-bootstrap intervals that
cross zero, out of ten fits. Takeaway: on hardware the base statistics return
26.61 times more error reduction per added evaluation than the depth-cut
circuits, descriptively and with no interval on the ratio.}
\label{tab:field-steps}
\begin{center}
\small
\setlength{\tabcolsep}{4pt}
\begin{tabular}{llrrrc}
\toprule
 & \bf Added feature & \bf Added & \bf Mean & \bf Reduction per & \bf Intervals \\
\bf Panel & \bf group & \bf evaluations & \bf gap & \bf evaluation & \bf crossing zero \\
\midrule
Hardware & Base statistics & 1,024 & 0.1969 & $1.923\times10^{-4}$ & 0 of 10 \\
Hardware & Depth-cut circuits & 3,072 & 0.0222 & $7.227\times10^{-6}$ & 1 of 10 \\
Simulator & Base statistics & 44,843 & 0.4407 & $9.828\times10^{-6}$ & 0 of 10 \\
Simulator & Depth-cut circuits & 134,528 & 0.0099 & $7.346\times10^{-8}$ & 7 of 10 \\
\bottomrule
\end{tabular}
\end{center}
\end{table}

Each efficiency in Table~\ref{tab:field-steps} divides an equally weighted macro
gap by an equally weighted mean collection count, with the same weights in
numerator and denominator. It is not an average of application-specific
efficiencies, and the two quantities are not interchangeable. Dividing a per-fit
interval by a known positive constant carries its uncertainty through unchanged
in shape, so zero crossings survive the scaling rather than being hidden by it.
Table~\ref{tab:field-eff-intervals} prints all forty scaled intervals. They are
the uncertainty this design does supply, and they are separate from the ratio
below, for which no uncertainty is reported.
\begin{table}[t]
\caption{Signed efficiency intervals per fit, formed by dividing each per-fit
gap interval by its step's fixed evaluation count. Hardware columns are in units
of $10^{-5}$ and $10^{-6}$; simulator columns are in units of $10^{-6}$ and
$10^{-7}$. These are pointwise intervals conditional on each fit, the eight
fixed backends and the released labels. Takeaway: the base-statistics step
excludes zero on every fit of both panels, while the depth-cut step crosses zero
on one hardware fit and seven simulator fits.}
\label{tab:field-eff-intervals}
\begin{center}
\small
\setlength{\tabcolsep}{4pt}
\begin{tabular}{lrrrr}
\toprule
 & \multicolumn{2}{c}{\bf Hardware} & \multicolumn{2}{c}{\bf Simulator} \\
\cmidrule(lr){2-3}\cmidrule(lr){4-5}
\bf Fit & \bf Base stats & \bf Depth-cut & \bf Base stats & \bf Depth-cut \\
 & $\times10^{-5}$ & $\times10^{-6}$ & $\times10^{-6}$ & $\times10^{-7}$ \\
\midrule
0 & [9.57, 33.91] & [1.77, 5.02] & [8.88, 11.39] & [-1.17, 0.85] \\
1 & [9.24, 33.54] & [3.69, 10.62] & [8.68, 11.38] & [-0.73, 1.39] \\
2 & [8.39, 33.25] & [2.56, 18.91] & [8.78, 11.37] & [1.14, 2.71] \\
3 & [8.71, 33.59] & [4.96, 10.77] & [8.80, 11.48] & [0.25, 1.68] \\
4 & [7.36, 32.34] & [4.87, 13.88] & [8.49, 11.01] & [0.31, 2.05] \\
5 & [9.96, 34.24] & [1.81, 9.99] & [8.88, 11.51] & [-0.19, 1.48] \\
6 & [9.78, 33.51] & [2.39, 8.60] & [9.12, 11.48] & [-1.68, 1.04] \\
7 & [8.54, 32.95] & [4.43, 12.65] & [8.69, 11.18] & [-0.33, 2.15] \\
8 & [8.02, 33.36] & [3.84, 13.35] & [8.61, 11.28] & [-0.59, 2.40] \\
9 & [8.47, 33.89] & [-0.41, 13.08] & [8.61, 11.27] & [-0.51, 2.40] \\
\bottomrule
\end{tabular}
\end{center}
\end{table}

On hardware, the ratio of the two ten-fit mean efficiencies is 26.61. That is a
descriptive ratio for this collection and retraining protocol, and no uncertainty
interval is reported for it; the component intervals stay conditional on each
fit. The simulator panel gets no efficiency multiple, because seven of ten
per-fit intervals for its depth-cut step cross zero and two of those point
estimates are negative. Any comparison whose improvement interval crosses zero is
left without an inverse efficiency.

One asymmetry belongs beside the first ordered step. The descriptor control
already conditions on support from the first base batch, and the base-statistics
rung uses numerical statistics from both base batches. That contrast therefore
includes access to information from a batch that had already been run, rather
than only the effect of running a second one. Collection paths also differ by a
large constant: the same base-statistics group needs 1,024 shots on the hardware
path and 44,843 on the simulator path, a factor of 43.79. Within each panel the
depth-cut step costs exactly three times the base step, so that path factor
cancels and each ratio reduces to three times the ratio of mean gaps. One more
constant belongs beside the top rung. The unread full inverse adds what the
base-statistics step adds: 1,024 evaluations on hardware and 44,843 on the
simulator. The ladder's last rung therefore costs as much as the step that
returns the larger mean reduction on both panels, and no retained feature
reads it.

Scaled intervals hold each fit, the eight backends, and the released labels
fixed, and they normalize to this corpus's fixed mean cost. An interval for an
aggregate efficiency over a changing application mixture would additionally have
to recompute the mean cost on each drawn set of applications.

Grouping information sources by origin predates this work, and two prior results
carry that credit. Q-LEAR's own third research question retrains after individual
feature deletions and reports a 32 percent median error increase from deleting
the observed probability \citep{muqeet2024qlear}. QRAFT compares static,
static-plus-forward, and full forward-and-reverse feature groups, and documents
their execution requirements \citep{patel2021qraft}. What this appendix adds is
the source-implied count placed beside each ordered step of one released ladder,
under one retraining recipe.

Ideal supervision is one 1,024-shot run per circuit on both paths, and it is an
ideal simulation at that shot count rather than an exact statevector label. It is
counted in the corpus totals below and excluded from every efficiency above.
Released corpus sizes are 9,486 training rows over 56 circuits and eight
backends, and 1,050 validation rows over six circuits and eight backends. At
10,240 shots per state plus 2,048 per circuit, the generator implies 98,054,144
simulator shots for training and 10,850,304 for validation, or 108,904,448
combined. Those totals include 507,904 ideal-simulator supervision shots. The
corpus is shared across the refitted controls and the seeds, and no deployment
volume is assumed, so the total is reported without a per-use amortization or a
time estimate.

Read these counts under three restrictions. This design has no shot-count sweep
and no arm that holds evaluations fixed across methods. Nothing here therefore
supports a reading about what one further evaluation would buy at the margin, or
about how evaluations should be allocated. Counts describe generating the
released data rather than reanalyzing it, and reanalyzing files that already
exist costs zero new quantum executions, which is a different accounting
question. Neither panel
is execution-free, because every rung conditions on a support that was chosen by
executing the circuit.

\section{The Clifford Control Stratum}
\label{app:clifford}

Under any Pauli observable, a random Clifford circuit's ideal expectation lies
in $\{-1, 0, 1\}$. Pooling this family into a continuous-regression headline
would conflate a discrete target with a regression target. The family is here
as a scalable stabilizer-labeled control (exact labels at up to 20 qubits via
\citealp{gidney2021stim}), and its results are reported as a separate stratum.

Separation is enforced at three places. The runner refuses a dataset containing
more than one stratum, so a Clifford run and a workload run are always separate
artifacts. The metric cell key carries the stratum through aggregation. And the
report computes the continuous-regression headline over continuous runs alone,
renders the Clifford stratum in its own section, and excludes Clifford cells
from every headline statistic and cost. A reader who runs the shipped report
command on a mixed manifest therefore sees two sections rather than one pooled
table.

\section{Model Selection and Feature Diagnostics}
\label{app:modelsel}

Selection applies a one-standard-error rule on mean circuit-block validation
error. Ties
break toward the larger signed improvement over the raw estimate on
validation, then lower cost, then the simpler model. For changed settings the
validation data stay in the familiar domain. That tie-break quantity is the
signed sum of per-item improvements, without the positive part. It is
therefore a different quantity from the unused harm measure defined in
Appendix~\ref{app:stats}: a
candidate that helps a lot on some items and harms a lot on others scores well
on it. The two are named separately throughout for that reason.
This campaign retains the existing signed-improvement tie-break.

One versioned feature specification fixes the model input as an ordered vector.
It carries the noisy estimate, the shot count, the qubit count, family
indicators, the sampled couplings, angles and evolution step size, depth and
insertion counts, graph-class indicators, compiled two-qubit gate count,
compiled depth, and observable locality. It carries no noise family, no channel
parameter, and no strength index, and every learned arm here consumes only that
vector. The item record itself still carries the declared noise family and
strength, which is how zero-noise extrapolation rebuilds the execution. The
specification is the single source of truth for model inputs, and every row
records its identifier. Carrying the sampled parameters is what makes a
simulator surrogate possible, and it is why the controls below are
load-bearing.

Four surrogate controls run beside the main methods. The feature-only control
drops the noisy estimate. The shrinkage floor predicts the source training mean
for the item's family, using the pooled source mean for a family absent from
training. Both have zero circuit-evaluation cost. The noisy-value-only
predictor and the shuffled-noisy-value test carry the same source and test
measurement charge as ridge; the shuffle permutes the training noisy-value
column deterministically and leaves prediction inputs alone. Accuracy that
survives the shuffle, or a competitive feature-only score, is reported as a
diagnostic. It motivates the direct gain estimate and the capacity-matched
comparison; neither observation alone establishes that the original model
ignores measurements.

Two comparisons sit on top of those controls, and they are not
interchangeable. The runner's automatic alarm triggers when the feature-only
control's test mean absolute error is at most five percent above ridge's. It is
a post-hoc comparison of two reported scores, it uses no uncertainty, and it
reads one slice. The released source-validation diagnostic requires an error ratio above
1.05 against each of the feature-only and noisy-only controls. Both paired
circuit-blocked intervals must lie strictly above zero in at least two
continuous families, at a declared resample count and seed. It refuses to evaluate a family whose circuit pool
cannot support the blocked resampling, and reports that family as not evaluable
rather than as a pass. Its stated limits are in Section~\ref{sec:results-controls}. The diagnostic
assigns no attribution decision and is not an equivalence test.
Untouched-test interpretation uses the six primary decompositions in
Section~\ref{sec:results-breaks}.

\section{Reproducibility Details}
\label{app:repro}

Dataset artifacts are write-once: generation refuses an output path that
already exists, run artifacts never modify datasets, and reports are views
over result manifests. Every item row carries per-item seeds, structure
features, noise configuration, shot counts, the noisy estimate with its
standard error, and the exact label with its label method. Split artifacts add
component hashes and split identity, which the legacy fixtures do not carry.
Package and backend versions live in the manifest, together with the seed
derivation, the active severity grids, the feature specification, and the
dataset hash.

The command-line surface is two preset-oriented subcommands, \texttt{generate}
and \texttt{run}. Reporting is the separate module entry point
\texttt{python -m qemscore.reports}, and validation is a Python interface
(Appendix~\ref{sec:artifact}). A split specification other than a shipped
preset is authored in Python. Methods are added through a phased fit, select,
and predict registration interface rather than the scikit-learn estimator
contract; that interface is tested but not declared stable across releases.
The dataset validator checks schema and hash integrity, declared roles, exact
labels, and measurement-bill conservation. The runner checks item-ID overlaps
between roles; physical-circuit separation requires the campaign check in
Appendix~\ref{app:protocol}. Together, the manifest chain and one master seed with
named streams make any table's inputs identifiable and any dataset regenerable.
Reproducing a table takes a short sequence of commands rather than one, and
the numerical trace beside each report names the source runs and cells behind
every printed number.

\section{Detailed Tables Behind the Result Figures}
\label{app:figure-tables}

These tables give the numerical values behind the result figures in
Section~\ref{sec:results} and Appendix~\ref{app:three-systems}, including the
reported errors, contrasts, and uncertainty intervals.

Table~\ref{tab:controls} is the record behind Figure~\ref{fig:share-ladder}:
four error levels per evaluation, and the rounded pointwise share interval
beside them.
Panel (a) of Figure~\ref{fig:gap-and-share} is recorded in
Table~\ref{tab:primary-gap}, which adds the gap as a fraction of the total
improvement over the affine control. Table~\ref{tab:field-ladder} is the record
behind Figure~\ref{fig:qlear-endpoints}, and keeps the capacity rung that opens
the ladder. Its per-fit intervals lie above zero on 0 of 10 fits on both panels,
where the base-statistics rung is above zero on 10 of 10.
Table~\ref{tab:field-apps} is the record behind
Figure~\ref{fig:qlear-cancellation}. It reports the hardware application
contrasts, and keeps the per-application gap of the full method over the
descriptor control, which stays positive on every application. Table~\ref{tab:third-instance} is the record behind
Figure~\ref{fig:three-systems}, and summarises the three system readings
with selected sensitivities and scope limits.

\begin{table}[htbp]
\caption{The four-arm ladder on the untouched test rows, for the six
primary evaluations of the campaign. Entries are macro mean absolute
error against the exact label, equally weighted over the four
strength-by-observable cells. Both feature-only arms read no measurement,
and the capacity-matched one is fitted from the ablation's own candidate
family under the same selection rule. The share is the part of the
ablation's improvement over the affine control that the capacity-matched
control reproduces, formed inside each circuit-blocked bootstrap draw. Five
of the six shares have intervals entirely below one, and the gap between the
capacity-matched control and the ablation excludes zero on the same five
rows. Seed 101 with Heisenberg is the exception in both. Takeaway: a control
that reads no measurement reproduces 87.7 to 100.5 percent of that
improvement, and the six shares disagree rather than reducing to one
number.}
\label{tab:controls}
\begin{center}
\small
\setlength{\tabcolsep}{4pt}
\begin{tabular}{llrrrrl}
\toprule
 & & \bf Raw & \bf Affine & \bf Capacity- & \bf Restricted- & \bf Share \\
 & & \bf estimate & \bf feature-only & \bf matched & \bf feature & \bf reproduced \\
\bf Seed & \bf Family & & \bf control & \bf control & \bf ablation & \bf [95\% CI] \\
\midrule
101 & Heisenberg & 0.6026 & 0.09850 & 0.00306 & 0.00345 & 1.0041 [0.9997, 1.0084] \\
101 & Transverse-field Ising & 0.2601 & 0.04412 & 0.00480 & 0.00234 & 0.9411 [0.9327, 0.9494] \\
211 & Heisenberg & 0.6092 & 0.09045 & 0.00447 & 0.00347 & 0.9884 [0.9841, 0.9925] \\
211 & Transverse-field Ising & 0.2813 & 0.04509 & 0.00300 & 0.00184 & 0.9733 [0.9666, 0.9795] \\
307 & Heisenberg & 0.5965 & 0.10348 & 0.00409 & 0.00329 & 0.9920 [0.9875, 0.9964] \\
307 & Transverse-field Ising & 0.2509 & 0.04712 & 0.00745 & 0.00191 & 0.8773 [0.8628, 0.8903] \\
\bottomrule
\end{tabular}
\end{center}
\end{table}

\begin{table}[htbp]
\caption{The primary gap $D=C-F$ on the untouched test rows: the
capacity-matched feature-only error minus the restricted-feature ablation's
error, for the six primary evaluations. Intervals are pointwise
circuit-bootstrap intervals conditional on the fitted and selected models, so
they omit retraining and model-selection variation. $D/T$ expresses the gap as
a fraction of the ablation's total improvement over the affine control.
Takeaway: five of the six intervals exclude zero, and the gap is a small
fraction of $T$ on every row.}
\label{tab:primary-gap}
\begin{center}
\small
\begin{tabular}{llrlr}
\toprule
\bf Seed & \bf Family & \bf $D$ & \bf Pointwise 95\% CI & \bf $D/T$ \\
\midrule
101 & Heisenberg & $-0.000385$ & $[-0.000821,\ 0.000028]$ & $-0.004$ \\
101 & Transverse-field Ising & $0.002461$ & $[0.002065,\ 0.002851]$ & $+0.059$ \\
211 & Heisenberg & $0.001005$ & $[0.000664,\ 0.001352]$ & $+0.012$ \\
211 & Transverse-field Ising & $0.001156$ & $[0.000911,\ 0.001409]$ & $+0.027$ \\
307 & Heisenberg & $0.000800$ & $[0.000373,\ 0.001198]$ & $+0.008$ \\
307 & Transverse-field Ising & $0.005547$ & $[0.005148,\ 0.005955]$ & $+0.123$ \\
\bottomrule
\end{tabular}
\end{center}
\end{table}

\begin{table}[htbp]
\caption{Ordered macro contrasts under the released Q-LEAR protocol, ten fits per
panel. The first step changes capacity at fixed descriptors; the steps after it
retain the nonlinear recipe and add measurement-derived feature groups. Reported are the mean gap that
step opens and how many per-fit application-bootstrap intervals lie above zero.
Takeaway: the estimated macro capacity gain is small, and every
per-fit interval spans zero. The base-statistics step has a positive
interval on both panels for every fit; the panels disagree about the
depth-cut step.}
\label{tab:field-ladder}
\begin{center}
\small
\begin{tabular}{lrrrr}
\toprule
 & \multicolumn{2}{c}{\bf Hardware} & \multicolumn{2}{c}{\bf Simulator} \\
\cmidrule(lr){2-3} \cmidrule(lr){4-5}
\bf Step added & \bf Ten-fit mean & \bf Above zero & \bf Ten-fit mean & \bf Above zero \\
\midrule
Nonlinear capacity over affine descriptors & $+0.000055$ & 0 of 10 & $+0.000091$ & 0 of 10 \\
Base statistics & 0.1969 & 10 of 10 & 0.4407 & 10 of 10 \\
Three depth-cut circuits & 0.0222 & 9 of 10 & 0.0099 & 3 of 10 \\
\bottomrule
\end{tabular}
\end{center}
\end{table}

\begin{table}[htbp]
\caption{Hardware contrasts by application, ten fits each. The first two columns
give the mean and the ten-fit range of the nonlinear descriptor control against
the affine one. A positive value there means the nonlinear control has the lower
error. The last two columns give the full released method against the nonlinear
descriptor control and against the base-statistics rung. Takeaway: the near-zero
descriptor average is partly cancellation, between two pricing applications that gain
and four that lose. The gap the full method holds over the descriptor control
stays positive on every application.}
\label{tab:field-apps}
\begin{center}
\small
\begin{tabular}{lrrrr}
\toprule
\bf Application & \bf Nonlinear over & \bf Range over & \bf Full over & \bf Full over base \\
 & \bf affine, mean & \bf ten fits & \bf nonlinear & \bf statistics \\
\midrule
Ground state & $-0.006168$ & $[-0.007797, -0.004441]$ & 0.193404 & 0.020315 \\
Pricing call & $+0.020113$ & $[+0.008253, +0.034651]$ & 0.123003 & 0.044070 \\
Pricing put  & $+0.017292$ & $[+0.005528, +0.031953]$ & 0.140534 & 0.027922 \\
QAOA         & $-0.012211$ & $[-0.014959, -0.009906]$ & 0.053112 & 0.005330 \\
Routing      & $-0.016524$ & $[-0.024532, -0.011166]$ & 0.565461 & 0.032581 \\
TSP          & $-0.002172$ & $[-0.004164, -0.000123]$ & 0.239225 & 0.002980 \\
\bottomrule
\end{tabular}
\end{center}
\end{table}

\begin{table}[htbp]
\caption{The three admitted systems under the scoring rule committed before any
arm was fitted, each beside the qualification its reading carries. Intervals
resample each system's own declared evaluation unit and are conditional on the
fitted and selected models. No share is averaged or pooled across systems.
Takeaway: the three readings disagree, and each is bounded by a property of the
release it was read from.}
\label{tab:third-instance}
\begin{center}
\small
\setlength{\tabcolsep}{4pt}
\begin{tabular}{p{1.8cm}p{1.5cm}p{3.6cm}p{5.8cm}}
\toprule
\bf System & \bf Reading & \bf Reproduced share $S$ & \bf What the reading rests on \\
\midrule
ML-QEM
 & Present
 & $0.5985$ $[0.5808, 0.6172]$, 2,000 row-bootstrap draws
 & Declared four-qubit experiment; 277 of 2,500 test rows share a descriptor
   block with training, favoring the descriptor control, and removing those rows
   alone still reads present at $0.5534$ $[0.5348, 0.5730]$; joint substitutions
   can move the reading; calibration group empty \\
\addlinespace
Synergy CNN
 & Absent
 & Six panel estimates from $-0.3498$ to $+0.0633$; every upper limit at or
   below $+0.2755$; 10,000 draws each
 & Reduced-data reanalysis fitting the controls on 72 to 90 rows per fold;
   calibration group empty; applicability at the original published budget is
   not established, and exploratory evidence there indicates budget sensitivity
   toward a larger share \\
\addlinespace
Q-Cluster
 & Present
 & $0.8795$ $[0.7723, 0.9658]$ with calibration and $0.9211$ $[0.8480, 0.9716]$
   without, 10,000 draws
 & Declared primary-seed row bootstrap; the scored object is the effective
   bit-flip regression stage, not the final output distribution the scoring plan
   named; every circuit appears in training and test; the exploratory circuit
   bootstrap is borderline at declared secondary seed 42 with calibration, where
   20 of 40 replicate seeds put the lower limit above 0.5 \\
\bottomrule
\end{tabular}
\end{center}
\end{table}

%% file: A2_related.tex
\section{Extended Related Work}
\label{app:related-extended}
\subsection{From Noisy Measurements to Learned Mitigation}

A mitigation task starts from a circuit, an observable, a noise setting, and a shot
budget. The backend returns a noisy estimate $r$ of the ideal expectation value
$y$. Zero-noise extrapolation (ZNE) re-runs the circuit at amplified noise and
extrapolates back to zero \citep{temme2017error, li2017efficient}. Clifford data
regression (CDR) builds efficiently simulable training circuits related to the
target, learns a noisy-to-exact map on them, and applies it to the target
\citep{czarnik2021clifford}; variable-noise CDR (vnCDR) adds multiple noise levels
\citep{lowe2021vncdr}. Learned mitigation generalizes the recipe: a supervised
model maps the noisy estimate and features of the circuit and observable to a
prediction $m$ of $y$ \citep{strikis2021learningqem, liao2024practicalqem}.

This paper registers ZNE, defines its bill, and reports its untouched-test
errors in Table~\ref{tab:roster-errors}. It does not evaluate CDR or vnCDR. Both build a
fresh training set for every target circuit, and the construction rule that fixes
what those circuits are is not settled here. A released probe over 31 targets
records the construction spaces, duplicate rates, and design-matrix conditioning
that the decision needs. The code ships that probe and the two cost formulas,
without a fit-and-predict path for either method. Mitiq \citep{larose2022mitiq}
serves one purpose in this work: continuous integration cross-checks the
benchmark's own ZNE folding and Richardson extrapolation against it on shared
cases.

\subsection{What Existing Learned-QEM Studies Already Cover}

Four prior efforts sit closest to this one, and they are close in different ways.
\citet{bao2025qembench} published QEM-Bench at ICML 2025, a benchmark for the
same task: twenty-two datasets spanning circuit types and noise
profiles, with predefined generalization settings that include held-out Trotter
steps, random-circuit sizes, and Pauli observables, together with a transformer
baseline. That resource standardizes what learned mitigators are trained and scored
on. \qemscore{} is a separate artifact, developed independently, and it standardizes
something else: what a score is allowed to mean. Its additions are one measurement ledger that charges every method in
circuit evaluations without equalizing them, datasets whose circuits and
exact labels a released validator rebuilds before accepting them, and
surrogate controls for assessing the noisy input's predictive benefit. The two are
complementary rather than competing.

The closest prior work on the model side is the machine-learning QEM study of
\citet{liao2024practicalqem}, and its coverage is wider than in-distribution
accuracy. That study evaluates linear regression, random forests, multilayer
perceptrons, and graph neural networks against digital ZNE, in simulation and on
IBM hardware up to 100 qubits, where it trains the model to mimic ZNE. It tests interpolation and extrapolation over
circuit depth and coupling strength, mitigates Pauli observables unseen in
training, includes a variational-eigensolver application, adapts to drifted noise
by fine-tuning on a few hundred extra circuits, and reports resource overhead in
circuit counts, both overall with training circuits included and at runtime alone.

On one of those axes the prior study reaches further than \qemscore{} does. Every
dataset here must be closed over its observables: each observable measured in a
prediction role has to appear in training. A method that fits one regressor per
observable cannot predict an observable it never fitted. Generation enforces the
rule for every dataset, refusing to write a dataset with a training-unseen test
observable. \qemscore{} therefore has no unseen-observable setting, and transfer
across measured quantities stays future work.

What \qemscore{} adds is protocol. The prior shift tests are case studies with
per-experiment evaluation choices. \qemscore{} predeclares one familiar setting
and five single-change settings: an unseen noise type, stronger noise, an unseen
circuit family, deeper circuits, and fewer shots. The grammar is closed.
Generation refuses a dataset in which a second component of the item tuple differs
between the training and test domains. Training and test values of the declared
component must themselves be disjoint. Item identities are disjoint across
training, validation, and test, and validation data are drawn from the training
domain alone, so model selection never sees the shifted side. The prior
noise-shift result is an adaptation demonstration, with fine-tuning on the
target. The five changed-setting recipes specify zero-shot transfer instead,
with no fine-tuning, and the completed campaign has not evaluated that
protocol. One qualification travels with those
recipes. The strength index is a grid label, and the released grids are not
calibrated to a common physical scale across noise families. In the unseen-noise
setting the error mechanism and its severity therefore move together.

Two further differences are in what gets reported and what gets charged.
\citet{liao2024practicalqem} report no harm endpoint. Others report aggregate
signs that can go negative: \citet{muqeet2024qlear} print per-application and
per-backend percentage changes against the unmitigated output, some of them
negative. We are not aware of a learned-QEM study that reports a per-item excess
against an exact label. The artifact can compute per-item excess absolute loss against exact
labels, but this campaign did not adopt that endpoint and reports no harm
quantity. On cost, the prior study reports overhead
per experiment against one reference method. \qemscore{} charges every method in
one unit, the circuit evaluation. Every bill carries three declared buckets: the
measurements that build the training data, method-specific extra measurements at
test time, and the base measurement of the test item. A feasibility cap at a
declared tier is checked per method before a run starts. These totals are
accounted at declared values, and they are not equalized. Inside one tier, the raw
estimate, ZNE, and the trained methods spend different amounts, and no method
spends a cheap allowance down to a common total. Nothing in this paper is budget
matched.

The Liao-style model family enters \qemscore{} as a restricted-feature ablation
rather than a reproduction. This benchmark's item schema carries no native-gate
count vector, no angle-bin histogram, and no sparse Pauli-observable encoding, and
every run artifact records those three omissions by name. The arm fits the
published architecture to the same feature vector the other learned methods here
receive: one random forest per observable, with a companion multilayer perceptron.
Source validation data select between the two. The arm is charged the same
training and test measurement groups as ridge, which holds data volume fixed and
isolates the architecture inside this benchmark. That comparison does not recover
the missing encoding. A win or a loss for this arm therefore says nothing about
the published method. The direction of any difference is not predictable either,
because the feature vector carries circuit parameters that may help a model
reproduce the simulator rather than use the noisy measurement.

Two of those four efforts are the published learned mitigators this paper
reanalyzes. Both already sort their inputs by what collecting an input requires.
Q-LEAR \citep{muqeet2024qlear} trains a neural network to predict a value for each
output state of a program. Those predictions are normalized into a distribution and
compared against the ideal one. Its release ships the generation notebooks, the
trained model, and the evaluation code. The corpora cover 56 training circuits and
six application circuits for validation, on eight IBM backends, with a simulator
companion. Inputs mix four circuit descriptors and a state weight with statistics
that require executing the circuit. Those statistics are a percentile of the
observed probability and an odds ratio formed as a maximum over a minimum, both
taken across two base runs. Three depth-cut variants of the circuit supply the rest.
Q-LEAR's own RQ3 already retrains after deleting individual features, and it reports
a 32 percent median error increase when the observed probability is dropped. The
qualitative importance of base-execution information in that system is its finding
rather than ours. QRAFT \citep{patel2021qraft} executes a program and its reverse,
then uses the reverse-execution statistics to correct the measured output
distribution. Its released \texttt{train.m} fits three models on nested feature
sets. The smallest reads seven circuit descriptors alone: width, depth, separate U1,
U2, U3 and CX gate counts, and the output state's Hamming weight. A ten-feature set
adds the forward observed probability at three severities, and a seventeen-feature
set adds the reverse-execution statistics and the machine identity on top of that.
Every arm predicts the same target, the true probability of the output state. QRAFT
describes the three as static, static-plus-forward, and full forward-and-reverse
comparisons, and it documents what collecting each requires. Separating features by
whether they require executing the circuit therefore predates this work in both
systems. The two are already connected, since Q-LEAR's published evaluation includes
a comparison against QRAFT. That evaluation's Table~4a scores the comparison as a
relative reduction in Hellinger-distance output error for Q-LEAR's multilayer
perceptron, with QRAFT's error as the denominator \citep{muqeet2024qlear}. Each of
the six application circuits contributes one reduction, computed after averaging that
circuit's errors across the eight IBM backends or their noisy simulators. The table's
Average row prints 25.2 percent on the simulators and 25.0 percent on the hardware,
and its text rounds both to 25 percent. Only the simulator figure follows from the
six displayed circuit rows. On the hardware those rows average to 24.55 percent
instead, a gap that rounding the average to a whole percent would close, though the
paper states no such convention.

\citet{muqeet2024qlear} and \citet{patel2021qraft} already group features by
execution requirement, and this paper does not claim that idea. What this paper adds
is an instrument and a normalization. The instrument is the \qemscore{} generator of
Section~\ref{sec:benchmark}, which produces circuits whose ideal answers are exactly
known. We can therefore fit surrogate controls and score them beside the full
method, including one matched in capacity that reads no measurement. Applying those
same controls to Q-LEAR's archive gives an ordered group comparison under
retraining, conditional on the released support: base-execution statistics against
the three depth-cut circuits. The rows there are noisy observed states and that
support was chosen by executing the circuit, so nothing here is execution-free
prediction. We then divide each macro gap by the shot-level circuit evaluations that
its feature-collection step adds, reading those counts from the released generation
notebooks rather than assuming them. The result is an ordered predictive comparison
normalized by collection counts. Nothing here measures the marginal benefit of
buying shots, because the design contains no shot-count sweep and no alternative
allocation. Counts implied by released source establish nothing about billing,
retries, or wall-clock time. The QRAFT arms enter the same way, transcribed from the
released \texttt{train.m} and read out under per-row mean absolute error. One reading
scores the released predictions without fitting anything, and the other retrains all
three arms under one fixed recipe. QRAFT's own models minimize a weighted
classification cost. This readout is therefore neither a reproduction of their
published results nor a verdict on their method under the objective they chose. The
code and data read for it are the copy carried in the Q-LEAR archive, which makes it
a second feature design rather than a second environment. Two instances are two
instances. This design cannot say how often either pattern occurs in published work,
in either direction, and nothing here establishes that Q-LEAR's published comparison
against QRAFT overstates a measurement contribution.

\qemscore{} does not evaluate selective mitigation. \citet{liao2024practicalqem}
name predictive uncertainty as future work, and it stays future work here, for a
reason worth stating in plain terms. No local check
on a decoded run artifact can establish that a prediction was produced without
access to the test labels. A reject score fitted from such an artifact therefore
cannot be shown to be label-free, whether it abstains on low-confidence inputs
\citep{geifman2017selective} or reports a conformal interval
\citep{shafer2008conformal}. Such a layer becomes measurable when fitting runs
inside a stage that has not yet seen labels, or when artifacts carry a verifiable
producer signature. This release has neither, and reports no routed, selective, or
abstaining result.

\subsection{Reliability Benchmarks and Controlled Shift}

The evaluation pattern follows controlled shift benchmarks in classical machine
learning. WILDS curates distribution shifts and reports the gap between familiar
and shifted performance as its central finding \citep{koh2021wilds}; ADBench
stress-tests anomaly detectors across supervision regimes and corruptions
\citep{han2022adbench}; and \citet{bowles2024better} shows how benchmarking
choices change conclusions in quantum machine learning. \qemscore{} differs in
what its changes mean. Noise type, noise strength, circuit family, depth, and shot
budget are experiment-level controls with exact labels, so a measured degradation
attaches to the named change rather than to an unknown mixture of shifts. Two
things still travel inside a named change. Coupled physical properties move with
it, since an unseen circuit family also changes gate mix and depth statistics. And
the unseen-noise setting carries the uncalibrated severity difference described
above, so it measures transfer to a different mechanism at a different strength.

%% file: A3_benchmark.tex
\section{Benchmark Design in Full}
\label{app:benchmark-full}
Section~\ref{sec:intro} posed three questions. This section gives a common
rationale for those questions and for the six primary evaluation settings. An
evaluation item is a four-part tuple: a circuit, an observable, a noise
configuration, and a shot count. The definition makes the scope traceable; it
does not by itself fix the number of questions or settings, and we say where each
count comes from.

\emph{Exact labels organize the questions.} A learned correction is harmful when it
lands farther from the ideal value than the raw estimate it replaced, and that
ideal value is exact in simulation yet unavailable at deployment. Exact simulator
labels also let us estimate the noisy input's predictive benefit for specified
learners and features. A competitive feature-only control shows that those
features support accurate prediction within its model class. The signed test
gains and their intervals bound the benefit established by the comparison. The
first question estimates the full method's gain against a control with
comparable model capacity (Section~\ref{sec:results-breaks}); the
evolution-step comparison is a secondary panel beside it
(Appendix~\ref{sec:results-help}).

\emph{The controlled changes are benchmark capability, not this campaign.} The six
settings below are what the artifact can generate and validate. Only the familiar
setting runs in the first campaign, so degradation under the five changed settings is
future work on this artifact rather than a result of this paper
(Section~\ref{sec:results}).

\emph{A further question, failure predictability, is deferred.} An earlier design
asked whether signals available before the ideal value is known can flag harmful
corrections, and answered it with a fixed routing rule evaluated as a
diagnostic. That layer is not in this release. A reject option is worth reporting
only if a reader can confirm that the abstention decision used no label
information, and we found no check over a released prediction artifact that
establishes it. The artifact records what a method returned; it does not record
what the method read while deciding. We removed the layer rather than publish a
property the benchmark cannot verify. Appendix~\ref{app:discussion-extended} states what a
later version would need to make the question answerable.

\emph{The tuple bounds the conditions.} Its four parts provide the axes on which
we construct controlled item-level changes. We select one familiar setting and
five predeclared changes along those axes; Section~\ref{sec:settings} explains
the selection rule and Appendix~\ref{app:limitations-full} records the omitted alternatives.
The observable contributes no setting in this release, for a reason internal to
the generator that Section~\ref{sec:inventory} gives with the observable
inventory.

\emph{Measurement cost is accounted, not equalized.} Every method's bill is
recorded in one unit, the circuit evaluation, split into three declared buckets:
training measurements, mitigation-specific measurements at test time, and the
base measurement of the test item. A configuration whose bill exceeds its tier
cap is refused before execution rather than silently shrunk. No method spends a cheap allowance down to a common total, so realized
totals can differ by method. Section~\ref{sec:scoring} and
Appendix~\ref{app:ledger} describe the library's accounting rule.
The campaign records each arm's buckets and realized totals in its
roster artifacts, as counts each method declares rather than audited
measurements, and this paper draws no cost comparison from them.

The rest of the section builds the pieces: the item and how each testbed is
constructed (Section~\ref{sec:task}); the circuits and noise models that populate
it (Section~\ref{sec:inventory}); the six primary settings as train/test recipes
(Section~\ref{sec:settings}); and the methods, budgets, and scoring
(Section~\ref{sec:scoring}).

\subsection{Task Definition and Testbed Construction}
\label{app:benchmark-task}

An item carries a noisy estimate $r$, the exact ideal value $y$, and a structural
description of the circuit and observable. That description includes the sampled
physical parameters of the instance: couplings and step size for the spin-chain
families, graph class, edge structure and variational angles for QAOA, insertion
count and angle for near-Clifford circuits, alongside two-qubit gate count,
compiled depth, and observable locality. A model can therefore predict the ideal
label from circuit parameters alone. The surrogate comparisons estimate the noisy
input's predictive benefit for the specified models and representation
(Section~\ref{sec:results}).

Every testbed is constructed by one deterministic pipeline in five steps
with shared-circuit paths summarized in Figure~\ref{fig:construction}(a).
One seeded circuit sample feeds a noisy path,
which compiles the circuit and executes it to produce $r$, and an exact path,
which evaluates the ideal value $y$ on the same sampled circuit by noiseless
simulation. The two paths converge into one verified item.

Figure~\ref{fig:construction} connects these paths to the model comparison
and attribution; the five construction steps are detailed below.

\emph{Step 1: circuit sampling.} Each instance is drawn from its family's
parameter distribution under a named seed stream: for the spin-chain families,
couplings are drawn uniformly from a fixed interval and the number of evolution
steps from a small integer set; for QAOA, a graph class, edge structure, and
per-layer angles; for the random families, the layer structure and the placement
of non-Clifford insertions. The distributions and the seed scheme are fixed in
the code and recorded in every dataset manifest. Appendix~\ref{sec:results-help}
states the first campaign's numeric design. Its resolved manifests, audit
settings, and code revision were fixed after the separate rehearsal and
before final generation (Appendix~\ref{app:protocol}).

\emph{Step 2: compilation.} Before execution, every circuit is transpiled with a
seeded compiler to one fixed basis (CX, RZ, SX, X), so that noise attaches to the
gate set actually executed and the structural features (two-qubit gate count,
compiled depth) describe the executed circuit. Compilation belongs to the noisy
path alone. The exact label is computed on the sampled circuit, so no compiler
choice can move it.

\emph{Step 3: noisy execution.} The selected noise family and strength are
attached to the compiled circuit, and the circuit is executed once per (circuit,
noise, shots) configuration. Every observable in this release is a Z-type Pauli
string, so one computational-basis measurement estimates all observables of a
configuration; they read their estimates from that single shared execution, and
the cost ledger charges it once. Basis-rotated observables are not implemented.

\emph{Step 4: exact labels.} The ideal value of every observable is computed on
the sampled circuit by noiseless simulation: dense statevector simulation for the
general families and stabilizer-tableau simulation for Clifford circuits. The two
label routes are independent implementations, and on Clifford circuits, where
both apply, the test suite requires them to agree over every Z-type observable to
the twelve-digit precision the datasets store. Label computation is logged
separately from the measurement bill.

\emph{Step 5: assembly and verification.} Items are serialized in a canonical
order and hashed; circuits, counts, observables, and noise configurations are
written as content-addressed sidecars; the manifest records every seed,
parameter, dependency version, and count; and a dataset directory is written
once and never overwritten. Circuit identity is a hash of the canonical
instruction stream, so two instances count as one circuit exactly when they
execute the same instructions. The released validator rebuilds each circuit from
its canonical descriptor, recomputes the exact label, and refuses the dataset if
any row disagrees. It also checks the schema, the sidecar hashes, the role
assignments, the observable semantics, the recorded seeds, and the conservation
of the measurement bill. It does not re-execute the noisy sampling: the stored
counts are bound to their row by content hash rather than reproduced. Within the
recorded environment, one master seed regenerates a testbed bit for bit.
Manifests record the expected lock digest, the observed digest when supplied, and
whether they match.
Continuous integration reproduces two benchmark dataset hashes from seeds alone.

Two dataset schemas exist in the repository. The benchmark's datasets use the
split schema described here, with source and target domains, a validation role,
and content-addressed sidecars. Six smaller fixtures in an earlier single-domain
schema remain as continuous-integration data and carry no benchmark results.

Exact labels are the reason the benchmark runs on simulators, and they are the
point: with $y$ in hand, the benchmark can measure when a correction helps and
when it lands farther from the truth than the raw estimate it replaced.

\subsection{Benchmark Composition: Circuit Families and Noise Models}
\label{app:benchmark-composition}

Three criteria selected the circuit families. Each family must represent a
workload class where mitigation is actually applied; its ideal values must be
exactly computable at benchmark scale; and together the families must span
structured to unstructured circuits, so that transfer between them is a
meaningful test. Table~\ref{tab:families} lists the result.

\begin{table}[t]
\caption{Circuit families, the qubit range each sampler admits, the exact-label
route, and why each family is included. Takeaway: every family is either a
mitigation workload or a control, and every ideal value is exact.}
\label{tab:families}
\begin{center}
\small
\begin{tabular}{llll}
\toprule
\bf Family & \bf Qubit range & \bf Labels & \bf Why it is here \\
\midrule
Transverse-field Ising (Trotter) & 1--12$^{\dagger}$ & Statevector & Canonical dynamics workload \\
Heisenberg chain (Trotter) & 2--12 & Statevector & Richer two-body dynamics \\
QAOA-MaxCut, four graph classes & 2--12 & Statevector & Variational workload, parameter structure \\
Near-Clifford random & 1--14 & Statevector & Between Clifford and fully general \\
Random Clifford & 1--20 & Stabilizer & Exact labels at the largest width; control \\
\bottomrule
\end{tabular}

\vspace{4pt}
\parbox{\textwidth}{\footnotesize $^{\dagger}$~Four of the five bounds are
enforced by their samplers. The 12-qubit transverse-field Ising bound is a
campaign constraint declared in the protocol manifest, and the sampler accepts
any positive width. The column gives admissible widths, which the released data
do not exhaust: the current fixtures run at 3 to 6 qubits, and the campaign
widths are declared in Appendix~\ref{app:protocol}.}
\end{center}
\end{table}

The first three families are the workloads, and each has a documented mitigation
record. Trotterized transverse-field Ising evolution is the benchmark circuit of
the 127-qubit utility experiment, whose headline expectation values rest on
zero-noise extrapolation \citep{kim2023utility}, and the same circuit class is a
structured testbed in the closest learned-mitigation study
\citep{liao2024practicalqem}. Heisenberg-chain dynamics has a direct mitigated
hardware precedent, at three-spin scale, across two hardware platforms
\citep{lotstedt2024comparison}. QAOA-MaxCut \citep{farhi2014qaoa} has published
mitigated runs on trapped-ion and superconducting processors, the latter with
learned mitigation on up to 40 qubits \citep{kakkar2022symmetryqaoa,
sack2024largescaleqaoa}.

The other two families are structural. Near-Clifford circuits are the class
data-driven mitigation trains on: CDR builds circuits of this kind, related to
the target circuit, as its training set \citep{czarnik2021clifford}. The
benchmark's near-Clifford family is a different object. It decorates a random
Clifford circuit with a small number of seeded T or $R_Z(\theta)$ insertions,
drawn independently of any target circuit. That makes it a generic class between
Clifford and fully general circuits, with exact statevector labels, and not a
per-target training set for anything the benchmark runs. Random circuits are
the standard characterization workload
\citep{magesan2012randomizedbenchmarking}, and Clifford circuits are exactly
simulable at scale \citep{aaronson2004stabilizer, gidney2021stim}; the random
Clifford family therefore earns its place twice, as the family whose exact labels
stay tractable at the largest admissible width and as a control.

Clifford ideal values are discrete ($-1$, $0$, or $1$ under any Pauli
observable), so pooling them with continuous targets would average two different
regression problems. The benchmark separates them at generation time: each
dataset declares one stratum, and the runner refuses a dataset that mixes strata,
so a Clifford run and a workload run are always separate artifacts. The reporting
code carries the separation through: the continuous-regression stratum is the
headline, the Clifford stratum is summarized in its own section of the same
report, and no Clifford cell enters a headline statistic or a headline cost
(Appendix~\ref{app:clifford}).

Instance counts in a split specification are per-role totals, divided across
family, width, and depth pools. The first campaign uses two family pools per
role. Its training totals are therefore 320 or 1280 circuits, with 640
validation and 320 test circuits, corresponding to the per-family counts in
Appendix~\ref{sec:results-help}. Random Clifford is outside this campaign.
The largest released legacy fixture has 48 training and 32 test instances.

Three observables ship, all Z-type Pauli strings: a single-site $Z$ at the middle
of the register, the neighboring $ZZ$ pair at the middle, and, for QAOA, the $ZZ$
term on one graph edge. No aggregate MaxCut cost and no whole-energy observable
is generated. This is also why the observable axis contributes no evaluation
setting. The generator requires every observable predicted in validation or test
to appear in training, and refuses to write a dataset that breaks that closure,
so a training-unseen physical observable cannot be produced at all. Observable
transfer is declared future work in Appendix~\ref{app:discussion-extended}.

The benchmark ships six noise families, each on a four-level strength grid from
weakest to strongest, and every level of every family carries a readout error
term. The registry itself marks no family as seen or unseen; each setting
declares its source and target families. Future transfer recipes use the two
trios below. The first campaign uses depolarizing plus readout alone.

\textbf{Training trio for future transfer experiments.} Depolarizing, dephasing,
and combined amplitude-and-phase damping supply the stochastic source channels
for those recipes.

\textbf{Held-out trio.} These are device-relevant mechanisms omitted by the
homogeneous, independent training channels. \textit{Coherent overrotations} from
imperfect calibration are a documented error class of their own; randomized
compiling exists precisely to convert them into stochastic noise
\citep{wallman2016noise, hashim2021randomized}. \textit{Correlated errors} are
real and nonlocal on hardware: simultaneous randomized benchmarking measures
addressability errors on coupled superconducting qubits
\citep{gambetta2012addressability}, crosstalk generates correlated multiqubit
errors \citep{sarovar2020detecting}, and a probabilistic-error-cancellation
experiment on IBM hardware learns sparse Pauli--Lindblad models to capture them
\citep{vandenberg2023probabilistic}. The benchmark implements a narrower version
of that mechanism: on each entangling gate it composes an independent
depolarizing pair, a correlated two-qubit depolarizing channel, and a coherent
$R_{ZZ}$ rotation, all supported on the two qubits the gate acts on. Spectator
error and error on qubits outside the gate stay out of scope.
\textit{Per-qubit heterogeneity} appears in calibration data from several qubits
on one processor \citep{yeteraydeniz2023stability}. The mixed model takes that
observation as motivation rather than fitting it. By deterministic seed, it
assigns every qubit and every directed two-qubit pair one of the other five
mechanisms and one scale factor. It therefore varies the mechanism as well as
its strength.

Whether methods trained on the stochastic trio survive these mechanisms is the
unseen-noise question, which this artifact can generate and the first campaign
does not run (Section~\ref{sec:results}). Two properties of the
strength grids bound what that question can currently answer. Within a family,
the four levels rise monotonically in every channel parameter, which the test
suite checks, so extrapolation in strength is well posed. Across families, the
levels are not matched. The shipped numbers are per-family placeholders. The
matching definition the protocol proposes, average gate infidelity per layer at a
reference circuit shape, is implemented and tested but inactive. Generation reads
the uncalibrated registry, and the cross-family agreement test is recorded as an
expected failure. Transfer to a held-out noise family is therefore transfer to a
different mechanism at an undeclared relative severity, and this release cannot
separate the two effects. Severity calibration was not applied to the shipped
registry before the campaign ran, and the cross-family agreement test remains a
standing expected failure rather than a resolved one.

\subsection{Artifact Capabilities: Six Evaluation Settings}
\label{app:benchmark-settings}

The six settings are predeclared, and each role uses a separately seeded circuit
pool. Physical-circuit disjointness is an additional campaign requirement,
checked on realized circuit hashes before fitting. A tuple component can move beyond training
support in two common ways. Either a categorical \emph{kind} is unseen, giving
an identity with no nearby training data, or a scalar \emph{magnitude} extends
beyond its trained range. Noise contributes one setting of each type, an unseen
family and stronger seen noise. Circuit does the same, through an unseen family
and greater depth. Keeping each pair separate distinguishes transfer across
families from range extrapolation within a family, two failures a regressor can
meet independently. A shot count is a single number with no kind, so it
contributes the lower-shot magnitude change alone. The observable contributes
nothing, for a reason given next. Five controlled changes and the familiar
reference make six settings.

\textbf{No observable setting.} An observable axis would ask the question the
closest published learned mitigator already poses, training on a small
fraction of the available Pauli observables and predicting the rest
\citep{liao2024practicalqem}. This release cannot ask it. The random-forest
candidate of the restricted-feature ablation fits one independent forest per
physical observable, and a fitted forest refuses any observable it did not see. Generation therefore
requires every observable appearing in validation or test to appear in training.
It refuses to write a dataset otherwise, and the same check runs again when an
artifact is loaded and when a run starts. A training-unseen physical observable
cannot be generated at all, so the axis is absent by construction rather than
merely unpopulated. Two further gaps sit behind it. The generator emits three
observable classes: a single middle site, the neighboring middle pair, and the
first graph edge for the variational family. There are no higher-locality
strings and no aggregate energies. A genuine shift also has to be defined on the
realized Pauli string and its locality, never on the class name. Two named
classes can resolve to the same string at the same locality. The middle-pair and
first-edge classes do exactly that on a three-qubit path graph, where both give
the same two-qubit $Z$ string. A split declared over class names alone can
therefore validate cleanly and move nothing physical. Observable transfer is
future work. It needs a model contract that predicts on an unseen observable,
and a generator for the observables to predict (Appendix~\ref{app:discussion-extended}).

Every setting specifies training and test domains and their declared difference.
The first campaign uses the familiar setting with the two families, one noise
family, and two strengths specified in Appendix~\ref{sec:results-help}.
The remaining recipes describe future transfer experiments. Each campaign
authors its pools in a split specification, and trained methods fit separately
within each dataset. The interface has three phases and the benchmark
enforces all three. Fitting sees training items with their labels. Model
selection sees validation items drawn from the training domain only. Prediction
sees test items whose exact labels have been stripped from the record the method
receives. Table~\ref{tab:settings} summarizes the six settings; the recipes
follow.

\textbf{Familiar conditions.} Motivation: the familiar-condition accuracy
comparison used by the closest prior study, reproduced under this benchmark's
budgets so the five changed settings have a baseline. Recipe: test items are new circuit instances (fresh parameters
and seeds) from the same families, noise models, strengths, and shot counts as
training.

\textbf{Unseen noise type.} Motivation: a method is calibrated against textbook
stochastic noise, and the device's measured error mechanisms include coherent
overrotation and correlated crosstalk (Section~\ref{sec:inventory}). Recipe:
training unchanged; test items use the held-out noise trio at the same strength
index. That index is a grid label rather than a calibrated physical scale, so
the mechanism and its severity move together. On average gate infidelity per
layer at a fixed reference circuit, the three held-out families sit below the
depolarizing anchor at every index, by 28\% to 99.8\%. Coherent overrotation at
its strongest setting is about ten times weaker than depolarizing at its
weakest. This setting therefore measures transfer to a different mechanism at a
different and generally milder severity. It does not isolate the mechanism, so a
degradation here cannot be attributed to the mechanism alone. A severity-matching
definition is drafted and tested against this metric, but the released grids do
not use it (Section~\ref{sec:inventory}). Fine-tuning on the new noise using
exact labels is impossible, because the prediction interface removes those
labels from the test items. Unsupervised adaptation to the unlabeled test batch
is forbidden by protocol rather than by the interface, which hands the method
the whole test batch at once.

\textbf{Stronger noise.} Motivation: device noise varies over time. IBM monitors
the relevant parameters approximately hourly and triggers recalibration when
needed \citep{ibm2026calibrations}. Relaxation times on superconducting qubits fluctuate
by up to an order of magnitude on 15-minute timescales
\citep{klimov2018fluctuations}, and drift appears across roughly nine months of
daily calibration data \citep{carroll2022dynamics}. For one IBM qubit over five
days, $T_1$ decreased by a factor of 2.8, $T_2$ decreased by a factor of 20, and
readout error increased by a factor of 3.6 \citep{yeteraydeniz2023stability}. A
model trained on last week's noise can face this week's. Recipe: training
unchanged; test items use the seen noise families at the held-out top strength.
Two qualifications belong with that recipe. The direction is a protocol
convention. Disjoint training and test strength sets are all the grammar checks;
unlike the depth and shot settings, it does not check which side is stronger.
Extrapolation also holds within a family rather than across the pooled training
trio, because the grids are uncalibrated. Dephasing at its top strength (0.0169
on the per-layer metric above) sits below depolarizing at its third (0.0170),
which training already covers.

\textbf{Unseen circuit family.} Motivation: data-driven mitigation deploys across
a training gap by design. CDR fits on efficiently simulable proxy circuits and
applies the fit to the target \citep{czarnik2021clifford}, and learned mitigation
is proposed as train-on-cheap, deploy-on-expensive \citep{liao2024practicalqem}.
Whether the fit survives a change of circuit family must be measured rather than
assumed. Recipe: leave one family out of training and test on it, rotating the
held-out family across the four families with continuous ideal values. The random
Clifford family stays out of this rotation: its ideal values are discrete, and
the runner refuses any dataset that mixes the two strata
(Section~\ref{sec:inventory}). Each fold must also keep its test observables
inside the set the training families realize, since observable closure holds here
as everywhere.

\textbf{Deeper circuits.} Motivation: the closest learned-QEM study tests
shallower-to-deeper transfer under simulated noise, while its hardware experiment
trains and tests at each Trotter depth \citep{liao2024practicalqem}. This setting
tests that depth gap under the benchmark's controlled noise models. Recipe:
training on the standard depth range; test items are family-native circuits at
extended depths, generated from fresh seeds. One native step is a different
physical gate count in each family, and nothing here equalizes that. The setting
varies circuit length at fixed noise and shots, rather than a measured quantity
of accumulated noise.

\textbf{Fewer shots.} Motivation: on-demand QPU use on Amazon Braket carries
both per-task and per-shot charges \citep{aws2026braketpricing}, while
probabilistic error cancellation requires extra samples set by its sampling
overhead \citep{vandenberg2023probabilistic}. A fixed cloud budget can therefore
make a reduced-shot test condition operationally relevant. Recipe: training
unchanged; test items are measured with a fraction of the training shot count,
raising the variance of the noisy estimate the method must correct. The grammar
enforces the direction here, refusing any specification whose test shot counts
are not all below the training counts.

Two directions stay outside the six settings. Circuit width cannot be written
into the grammar at all, because the training and test domains share one declared
width set, so a train-narrow, test-wide split is not expressible. Compound
changes along several axes at once are excluded by construction, since each
setting moves one declared tuple component. Both are recorded in
Appendix~\ref{app:limitations-full}.

In this version, the wiring to the results is narrow. The first campaign runs the
familiar setting alone (Section~\ref{sec:results}). The five changed settings are
specified, generated and validated by the artifact, and running them is future
work rather than a result reported here.

Three caveats bound the interpretation, stated once. Some changes move coupled
properties (a new circuit family also changes gate mix and depth statistics), so
each result is conditional on the declared construction rather than a
single-cause attribution. The five changed settings share no common severity
scale, so the benchmark never ranks them against each other; each answers its own
question. And the strength index is a grid label rather than a calibrated
physical scale. Two items carrying the same index under different noise families
are not at the same severity, so that gap sits inside the unseen-noise setting as
well as between settings.

\subsection{Methods, Budgets, and Scoring}
\label{app:benchmark-scoring}

Every method reaches an item through one narrow interface. During training it
receives items with their noisy estimates, their circuit and observable
structure, and their exact labels. At prediction time the runner hands it a
field-restricted copy of each test row with the exact label removed, and the
fit, select, and predict phases run in that order. This is a guarantee about
methods the runner executes. It is not a check on predictions produced elsewhere
and supplied as a file, and the paper does not claim otherwise.

The learned arms consume one versioned feature vector of thirty columns. It
holds the noisy estimate, the log shot count, the qubit count, family
indicators, family parameters, the two-qubit gate count, the compiled depth, and
the observable locality. That vector carries no noise family, no channel
parameter, and no strength index. A learned arm cannot read the generating noise
off its inputs. ZNE is different by construction. It
rebuilds and re-executes the circuit under the declared noise family and
strength, which is what makes it a simulator-side standard-QEM baseline rather
than a regression model over stored features.

Table~\ref{tab:methods} lists the eight registered methods. The raw estimate is
the uncorrected measurement. ZNE uses global digital
folding at the fixed scale factors 1, 3, and 5, with three-point Richardson
extrapolation. Scale one reuses the stored base measurement. Scales three and
five each cost one extra execution per measurement group, so sibling observables
of one circuit share that extra cost as they share the base measurement. The
scale set and the extrapolator are fixed in advance and are never selected on
validation data, so this arm reports one declared configuration. Two learned
arms follow: a ridge regressor with a prespecified three-value penalty grid selected on
source validation, and a restricted-feature ablation of the published learned
mitigator \citep{liao2024practicalqem}. Both consume the same feature vector,
the same source rows, and the same measurement charge.

The ablation is not a reproduction, and this paper does not read it as one. That
published feature construction uses a native-gate count vector, an angle-bin
histogram, and a sparse Pauli observable encoding. This benchmark's dataset
schema carries none of the three, and every run artifact records that omission
by name. A win or a loss for this arm bounds what this benchmark's feature
representation supports, and says nothing about the published method in either
direction. The direction is not even predictable in advance: the
parameter-bearing features \qemscore{} does carry may themselves help a model
emulate the simulator. Clifford data regression and its variable-noise extension
appear in Section~\ref{sec:related} as published methods, and are absent from
the roster. The release ships a closed-form cost model and a training-circuit
viability probe for them, and no executable mitigator.

Four surrogate controls fix what a learned score means. The feature-only control
receives every feature except the noisy estimate, and spends zero circuit
evaluations. Its counterpart, the noisy-only control, sees the noisy estimate
and the shot count and nothing else. A shrinkage control predicts the
source training mean for the item's family, using the pooled source mean for a
family absent from training. The shuffled-noisy control
permutes the noisy-estimate column of the training rows and leaves test inputs
intact. Its accuracy is a diagnostic of sensitivity to this training
intervention; it does not establish that the original model ignored the
measurement. Every fitted arm here, the controls included, trains on exact
simulator labels. The runner emits one coarse automatic
alarm, comparing the feature-only control to ridge at a ratio of 1.05 on a single
slice. Section~\ref{sec:results-controls} states the stricter criterion this
paper commits to in advance, which the released library implements.

Every method pays in one currency: circuit evaluations, meaning backend calls
times shots. Each bill has three buckets: training measurements (validation
measurements are charged here), extra measurements at prediction time, and the
base measurement of the test item itself. Compatible observables of one circuit,
noise setting, and shot count share a single execution, which is charged once.
Exact labels are logged separately, because they are simulator calls rather than
backend measurements, and burying them would flatter the supervised arms.

Budgets are accounted, and no method spends a cheap allowance down to a common
total. Feasibility runs at declared tiers of $2.5\times10^{6}$,
$2.5\times10^{7}$, and $2.5\times10^{8}$ circuit evaluations. A tier is a
combined cap on the three buckets, applied to each method inside each paired
budget cell, rather than a pot the roster shares. The runner refuses to execute
any configuration in which a registered method exceeds its cap. Every reported
run is therefore one in which every method was feasible. For methods outside the
executed roster, infeasibility is computed from the closed-form cost model and
labeled as modeled rather than measured. Realized totals differ across the
roster by construction. The raw estimate pays the base test measurement alone.
ZNE adds two folded executions per measurement group, so its realized
test-time bill is exactly three times the raw bill. The two learned arms add the
shared source measurement, so their totals scale with the size of the source
pool. The library's method-summary format supports nominal and realized
totals and test-time circuit evaluations per expectation. This version
supplies no per-arm campaign ledger and reports no such campaign counts,
training-inclusive amortization, or empirical comparison of realized
campaign costs. Training-inclusive amortization would divide total
realized cost by the number of test expectations
(Appendix~\ref{app:ledger}).
The reported campaign runs at feasibility tier H, a cap of
$2.5\times10^{8}$ circuit evaluations. That pairing is read from the released
code rather than from a result table: the campaign design declares
\texttt{budget\_tier="H"} and the budget module defines H as
250,000,000.

\begin{table}[t]
\caption{The eight registered methods, their roles, and the measurement
categories charged by the library. Takeaway: the accounting rule depends on
the method; this table lists cost categories rather than realized campaign
totals.}
\label{tab:methods}
\begin{center}
\small
\begin{tabular}{lll}
\toprule
\bf Method & \bf Role & \bf What it pays for \\
\midrule
Raw estimate & No mitigation & Base test measurement only \\
Zero-noise extrapolation & Standard QEM & Base test measurement, plus two \\
 & & folded executions per group \\
Ridge regression & Learned & Base test measurement, plus the \\
 & & shared source measurement, amortized \\
Restricted-feature ablation & Learned & Same as ridge \\
\midrule
Feature-only control & Diagnostic & Nothing \\
Noisy-only control & Diagnostic & Same as ridge \\
Shrinkage floor & Diagnostic & Nothing \\
Shuffled-noisy control & Diagnostic & Same as ridge \\
\bottomrule
\end{tabular}
\end{center}
\end{table}

Scoring uses two lenses. Accuracy is the macro mean of per-cell mean absolute
error against the exact label, so large easy cells cannot dominate. Intervals
come from a bootstrap that resamples whole physical circuits within their
stratum, keeping observables that share one circuit together. This campaign reports no harm endpoint. Model selection retains its
signed validation-improvement tie break, the signed total improvement over
the raw estimate on validation (Appendix~\ref{app:modelsel}).

The library supplies paired Wilcoxon signed-rank tests
\citep{wilcoxon1945individual}, Holm correction within a declared comparison
family, median paired differences, and rank-biserial effect sizes. Mean ranks
follow \citet{demsar2006statistical} as a secondary view. These support the
future comparison protocol of Appendix~\ref{app:stats}. The first campaign
uses the gain estimates and direct contrasts in Appendix~\ref{sec:results-help}.
Predictions are never clipped before scoring; leaving the physically valid
range is reported as its own rate.

\subsection{The Field Half: Reanalyzing Two Published Systems}
\label{app:benchmark-field}

Sections~\ref{sec:task} through~\ref{sec:scoring} describe a testbed this work
builds, in which every ideal value is exact and every measurement is charged
where it is spent. A second instrument reads data that other groups already
released, and applies the same ordered comparison to two published learned
mitigators: the Q-LEAR mitigator of \citet{muqeet2024qlear} and QRAFT
\citep{patel2021qraft}. We call this the field half. Its move is the one the
controlled design already uses. Sort a published method's inputs into an ordered
ladder of feature groups, refit at every rung under one declared recipe, and
score each rung by the score specified for it here. The Q-LEAR reproduction
panel uses the release's own evaluation routine; the Q-LEAR ladder uses the
standard Hellinger score specified here, and the QRAFT panels use the per-row
mean absolute error specified for them. One component
does not travel across. Nothing on the field half carries an exact ideal value,
so the exact-label machinery of Section~\ref{sec:task} governs the controlled
half alone.

\emph{Four rungs make the Q-LEAR ladder.} The first is an affine model over the
release's four circuit descriptors and the output state's weight. Its successor
reads those same inputs through a nonlinear model, so the step between the two
moves model capacity and leaves the inputs alone. That step is the capacity
control that Section~\ref{sec:results-controls} requires for a nonlinear arm.
Two statistics computed across the base executions enter at the third rung: a
percentile of the observed probability, and a maximum-over-minimum odds ratio.
The fourth rung is the release's full feature set, which adds three features
collected from three depth-cut copies of the circuit. Every rung is fitted from
scratch, so no arm here is a fitted model with a column removed at prediction
time. A fifth arm sits beside the ladder instead of on it, dropping the state
weight so that only the four circuit descriptors remain. That arm predicts one
value for every output state of a circuit, and it therefore returns a
distribution that is uniform within a circuit whenever its common prediction is
positive.

The affine control is fitted once and reused, because that fit is
deterministic. The nonlinear descriptor control, the base-statistics rung, the
full released method, and the off-ladder four-descriptor diagnostic are each
fitted at seeds 0 through 9, giving forty-one fits. Six applications run
on eight IBM backends: ground state, pricing call, pricing put, QAOA, routing
and TSP. That gives 48 paired application-and-backend results. A simulator companion runs the
same ladder on the release's simulator panel, and the two panels are reported
side by side wherever they disagree. Two gaps are the declared primary
comparisons for this half: the second rung against the fourth, and the third
rung against the fourth.

\emph{The rows are noisy observed states.} One row is one output state of one
application circuit on one backend, recorded from a noisy execution. The set of
states each arm predicts on is the support that the released base execution
produced, so that support was chosen by executing the circuit. No arm on this
half is therefore execution-free, and this paper describes none of them that
way. Supervision in the release is a 1,024-shot ideal simulation of the same
circuit, which is a sampled estimate rather than an exact statevector value.

\emph{The ladder uses the specified standard Hellinger score.} Each arm predicts one value per state on
that support, and those values are normalized into a distribution. Error is the
standard Hellinger distance from it to the release's target distribution on the
same support, one number per application and backend. Those numbers are averaged
equally over the eight backends and the six applications. Intervals resample
applications and are pointwise. Each holds one fit, the eight backends, and the
released labels fixed, and excludes retraining and model-selection uncertainty,
which the spread across the ten fits describes separately. Counts of intervals
above zero are
descriptive counts over those same six applications, and they are not
independent replication rates. One reading rule travels with them. A small
positive estimate whose interval crosses zero is reported as a pattern that is
not observed. That statement is weaker than absence, and it establishes no
equivalence.

\emph{A reproduction gate runs before any rung is read.} The evaluation protocol forbids
interpreting an ablation until a faithful replay of the release's own published
evaluation is verified. That replay therefore runs first, and the ladder is read
only after it passes. No reproduction tolerance was declared beforehand. Whether
the replay is faithful is therefore a judgement made after the numbers were
seen, with no bound fixed in advance to test it against.

\emph{Execution counts are read from the released source.} Per application
circuit, the hardware generator builds two copies of the transpiled circuit, one
full inverse, and the three depth-cut circuits, then runs every one of them at
1,024 shots. The first base batch supplies the support the descriptor rungs are
scored on, and it costs 1,024 shot-level circuit evaluations per circuit. Both
base batches are required by the third rung, at 2,048; the three depth-cut
batches bring the fourth rung to 5,120. Full released generation costs 6,144 per
circuit, because it also executes a full inverse that no feature reads. Ideal
supervision adds one 1,024-shot run per circuit, is disclosed separately, and is
excluded from the per-step figures. The simulator generator calls feature
collection inside its per-state loop, so the same feature groups cost per output
state there instead of once per circuit. Every rung is accounted at its own
declared total. This design has no shot-count sweep and no arm that reallocates
a fixed number of shots across feature groups, so it cannot say what one further
shot buys.

\begin{table}[t]
\caption{The field half at a glance. Both systems are read under an ordered
feature ladder. The Q-LEAR ladder uses the standard Hellinger score specified
here, and the QRAFT panels use the per-row mean absolute error specified for
them, which is not the objective QRAFT's own models were fitted for.
Takeaway: the two scores differ because the released QRAFT output cannot be
reassembled into per-circuit distributions, so they are reported separately and
never combined.}
\label{tab:field}
\begin{center}
\small
\begin{tabular}{lp{5.4cm}p{5.4cm}}
\toprule
 & \bf Q-LEAR half & \bf QRAFT panel \\
\midrule
What is read & Archived application data from the release & Released predictions,
and the release's processed inputs \\
Cohort & Six applications on eight IBM backends, 48 paired results & Five
machines; 1,524 released rows, 10,155 rows for retraining \\
Ladder & Four rungs, plus one arm beside it & Three arms, transcribed from the
release's training code \\
Fitting & A fitted once; C, O, F and C4 at ten seeds, 41 fits & None in one reading, ten seeds in the
other \\
Score & Standard Hellinger on the released observed-state support & Per-row mean
absolute error in probability points \\
Uncertainty & Application bootstrap, pointwise, conditional on each fit & None
reported \\
Split unit & Released validation built from the evaluation's own circuit
directory; no circuit identifier in the files & Rows, shuffled; no circuit
identifier in the released output \\
Specification & Fixed before the controlled campaign, with no pre-fit binding
artifact & Committed before the scoring script read any data \\
\bottomrule
\end{tabular}
\end{center}
\end{table}

\textbf{The QRAFT panel.} QRAFT supplies three arms of its own, transcribed from
the training code in its release. The first arm reads circuit descriptors alone:
width, depth, separate U1, U2, U3 and CX gate counts, and the output state's
Hamming weight, seven features together. Adding the forward observed probability
at three severities gives the second arm ten features. A third arm adds machine
identity and the reverse-execution statistics that are QRAFT's central
contribution, and reaches seventeen. In every arm the target is the
state's real probability, and it is never an input; the scoring script refuses
any arm whose feature set contains it.

Two readings run over those arms. One fits nothing, reading the three
predictions as shipped in the release's output file, 1,524 rows. The other
retrains all three arms from the release's processed input file, 10,155 rows,
under one fixed recipe with ten seeds. Rows are weighted equally within a
machine and the five machines equally. Its specification, including the anticipated
outcome, was committed before the scoring script was run against any
data.

\emph{The QRAFT panel is scored differently, for a stated reason.} Its score is
per-row mean absolute error in probability points. Hellinger needs a circuit's
output states assembled into one distribution, and the released QRAFT output
carries no circuit identifier and no state label, so those states cannot be
reassembled. The two scores are reported separately and are never compared or
combined. Neither reading carries an interval, and that was settled before the
run. Grouping the released rows by machine and descriptor tuple gives 735 groups
over 1,524 rows at a median size of two, which are partial circuits rather than
circuits. Five machine-level clusters are too few for a stable 95 percent
interval. The spread reported for the retrained reading is the range over ten
seeds, and it is not a confidence interval.

\textbf{What the field half does not have.} These absences belong here, because a
reader would otherwise assume the component exists. No row on this half carries
an exact ideal value. Nothing binds the Q-LEAR specification to a time before
fitting. The ladder, the inputs, the retraining recipe, the ten seeds, the
score, the aggregation, and the resample unit were all fixed before the
controlled campaign ran. Checksums taken before review show that the inputs did
not change afterwards. Neither fact establishes when the specification was
frozen, and no retrospective timestamp is invented for it. Both QRAFT readings
split rows; neither splits circuits, and the released split is shuffled by row,
so a circuit's output states appear on both sides of it. Neither QRAFT reading
says anything about unseen application circuits. The Q-LEAR half has the same
gap by another route. Its released generator builds the validation set from the
same circuit directory the evaluation reads. No circuit identifier survives in
its files, so a circuit-disjoint split cannot be built from the archive.

Three further absences bound how the field half can be read. Neither QRAFT
reading reproduces QRAFT's published results: one takes their released
predictions without checking them against the numbers printed in that paper, and
the other uses a different learner. QRAFT's own models minimize a weighted
classification cost, while this panel reads out mean absolute error. A result
here therefore bounds that score, and it says nothing about QRAFT's method under
the objective QRAFT chose. The Q-LEAR replay environment is a port. Python,
TensorFlow, Keras, ktrain, transformers, numpy, pandas, and scikit-learn
versions match the release's pins, while SciPy, h5py, and joblib differ, and no
matched original-environment counterfactual was run. Both systems also come from one
release, collected by the Q-LEAR authors on the same application circuits, so
the field half holds two feature designs and not two independent environments.

\emph{What the field half extends, and what it does not originate.} Sorting a
mitigator's inputs by whether they require executing a circuit is a move both
releases already make. Q-LEAR defines its features separately, retrains after
deleting individual features, and reports in its own RQ3 a 32 percent median
error increase from deleting the observed probability \citep{muqeet2024qlear}.
QRAFT compares static, static-plus-forward, and full forward-and-reverse feature
sets, and documents what executing each set requires \citep{patel2021qraft}.
What this half adds is the ordered group comparison under one matched retraining
recipe. The Q-LEAR reproduction panel uses the release's own evaluation routine.
The Q-LEAR ladder uses the standard Hellinger score specified here, and the
QRAFT panels use the per-row mean absolute error specified for them, which is
not the objective QRAFT's own models were fitted for. Execution counts travel with it,
derived from the released generator source instead of assumed.

%% file: A4_results.tex
\section{Campaign Results in Full}
\label{app:results-full}

The frozen design assigns the primary endpoint to the within-family
capacity-matched decomposition in Section~\ref{sec:results-breaks}.
Evolution-step and training-size contrasts are secondary and do not support
the primary claim. Section~\ref{sec:results-field} reports the field test
and collection requirements.

\textbf{Scope of this campaign.} The first campaign compares separately trained
experiments under familiar conditions. It does not evaluate transfer across the
five changed settings in Section~\ref{sec:settings}. Those settings remain part
of the benchmark specification. The auxiliary roster's untouched-test errors are reported in
Table~\ref{tab:roster-errors}; the primary comparison remains the
capacity-matched decomposition.
The validation gate is a diagnostic. Untouched-test estimates and their direct
contrasts determine the campaign's claims.

\textbf{How results are grouped.} Results are grouped per evaluation setting. A
reported cell is keyed by method, role, split identity, circuit family, noise
family, strength, and observable, so records from two settings cannot merge. The
continuous-regression stratum is the headline. Random Clifford circuits have
discrete ideal values, so that family is reported as a separate stratum, and the
runner refuses to put both strata in one dataset. The report keeps that
separation: it emits a headline section over the continuous stratum, a separate
section for the Clifford control, and one section per declared split, and no
Clifford cell enters a headline statistic. The default report also provides a
within-stratum aggregate across the supplied splits; we use the split-specific
sections for setting-specific claims. Split identity names the evaluation
setting rather than every campaign condition, so the campaign's own tables keep
circuit family, generator seed, training size, and step regime explicit.
Aggregation is the macro mean over cells, with circuit-blocked bootstrap
intervals.

\textbf{A future comparison protocol.} A shift-resolved reliability map needs a
declared win, tie, loss, and harm rule, and a frozen family of paired
method-versus-method comparisons with Holm correction. Appendix~\ref{app:stats}
states that protocol as future work on this artifact. It does not govern
this campaign, whose primary decomposition is specified in
Section~\ref{sec:results-breaks} and Appendix~\ref{app:protocol}.

\subsection{Do the Controls Leave Room for Mitigation?}
\label{app:results-full-1}

\textit{Does the noisy measurement carry predictive benefit for a learned
mitigator on this testbed, and under what conditions?}

This check bounds how later accuracy comparisons are interpreted. Every fitted
arm, including the controls, trains on exact simulator labels. A competitive
feature-only predictor shows that circuit features support accurate prediction
within that model class. Failure to beat it does not establish that the full
model ignores measurements or that positive benefit is absent.

The released library implements a source-validation diagnostic stricter than
the runner's coarse alarm. Passing requires the macro mean absolute error
to beat both the feature-only and noisy-only controls by a factor of 1.05
in at least two circuit families with continuous ideal values. For each
qualifying family, both paired circuit-blocked intervals on the
control-minus-full difference must lie strictly above zero. The margin was
fixed before this diagnostic was applied. Its outcome assigns no attribution
decision.

Three limits travel with that diagnostic, and the paper states them wherever
it reports a pass or a failure. First, the interval establishes a positive difference rather
than a population benefit above the margin. Second, failure is not an equivalence
result. A predictor that returns the measured value alone can fail by tying the
noisy-only control. A claim that no practically relevant positive benefit exists
therefore needs its own upper bound, below $1 - 1/1.05$ on the relative gain, and
such a bound is compatible with substantial harm. Third, for a
nonlinear arm the affine controls do not separate measurement benefit from model
capacity. The restricted-feature ablation is therefore read against a
capacity-matched feature-only counterpart, fitted from the same candidate family
under the same budget and selection rule, rather than against the linear controls
alone.

The source-validation gate is a diagnostic and assigns no attribution
decision. Untouched-test claims use the six primary decompositions of
Section~\ref{sec:results-breaks}. The secondary evolution-step and
training-size endpoints are direct contrasts of absolute capacity-matched
gaps, each with its own pointwise interval
(Appendix~\ref{sec:results-help}). A change in a diagnostic pass/fail
category is not an estimate of either contrast.

Continuous integration ships six small single-domain fixtures, and the smallest
carry ten training circuits and six test circuits in one family.
The coarse alarm fires on three of the six. On one of them
the feature-only control reaches macro mean absolute error 0.0576 against the
ridge arm's 0.0580, and the shuffled-noisy control reaches 0.0558, so the coarse
alarm fires. Those fixtures are far too small to carry evidence, and they are not
generated by the setting grammar of Section~\ref{sec:settings}. They are reported
here so that a reader knows this subsection was written before the frozen
campaign rather than in response to it.

The criterion has now been applied to the frozen campaign, and one arm
fails it. For the restricted-feature ablation the gate passes on all twelve
campaign datasets. Ridge passes on the six large-step datasets,
and fails on all six at the shipped step sizes, because fewer than two
families met both comparisons. Across the twelve family results at the
shipped step sizes and both training sizes, the affine feature-only
control's error is 1.0004 to 1.0145 times ridge's, inside the 1.05 margin.
The noisy-only control's error is 1.60 to 4.52 times ridge's on those same
rows, so the feature-only comparison is the one that falls inside the
margin. These are diagnostics on rows used in model selection. They are
neither an independent replication nor a test-set estimate of measurement
benefit, and the frozen rule assigns no attribution decision to the gate or
its margin.

Figure~\ref{fig:share-ladder} carries the four arms of that decomposition and no
measurement totals. The detailed record behind it, with all four error levels
printed and the rounded pointwise share intervals beside them, is
Table~\ref{tab:controls} in Appendix~\ref{app:figure-tables}. Ridge is not used to support the primary claim, so its untouched-test errors
are reported here rather than in either place. At the shipped step sizes with
640 training circuits per family,
ridge reaches macro mean absolute error 0.09735, 0.08935 and 0.10219 on
Heisenberg at seeds 101, 211 and 307. On transverse-field Ising the same three
seeds give 0.04413, 0.04527 and 0.04661. These are the same six primary
evaluations Figure~\ref{fig:share-ladder} reports.

For each of these six evaluations, the campaign report retains a legacy ridge
relative-gain diagnostic against the affine feature-only control, with a
pointwise interval on that ratio. The absolute errors printed here were
recomputed from the campaign's retained per-setting records after the run.
Those intervals attach to the relative gain, so this version reports no
interval on the six absolute errors themselves. The report retains its own ridge macro error for
all six, and every one of the six printed here agrees with it to within
$2.8\times10^{-17}$.

Two of the six evaluations reverse. On transverse-field Ising, ridge is worse
than the affine feature-only control at seed 101, 0.044131 against 0.044124,
and at seed 211, 0.045268 against 0.045091. At seed 101 the stored interval on
that relative gain runs from $-0.010667$ to $0.008976$ and spans zero. These
six points therefore support no ranking of ridge against that control, and on
two of them the retained evidence is reversed or spans zero.

The campaign's archived roster artifacts record untouched-test macro MAE for
all eight rostered methods across every evaluation setting. Table~\ref{tab:roster-errors}
reports the raw reference and four auxiliary controls for the six primary evaluations (regime \texttt{shipped},
640 training circuits per family). Each entry is the equally weighted mean over
that family's four strength-by-observable cells (160 test circuits per seed).

\begin{table}[t]
\caption{Untouched-test macro MAE for auxiliary controls beside the raw reference, at shipped step sizes and 640 training circuits per family. Noisy, Shrink, and Shuf denote the linear noisy-only, shrinkage, and shuffled-noisy controls. Each value equally weights four strength-by-observable cells. These are descriptive point estimates, without intervals.}
\label{tab:roster-errors}
\begin{center}
\small
\setlength{\tabcolsep}{4.5pt}
\begin{tabular}{llrrrrr}
\toprule
\bf Seed & \bf Family & \bf Raw & \bf ZNE & \bf Noisy & \bf Shrink & \bf Shuf \\
\midrule
101 & Heisenberg & 0.60258 & 0.46773 & 0.16762 & 0.09674 & 0.09851 \\
101 & TFI        & 0.26010 & 0.16589 & 0.21832 & 0.21793 & 0.04414 \\
211 & Heisenberg & 0.60916 & 0.47121 & 0.17295 & 0.08866 & 0.09046 \\
211 & TFI        & 0.28133 & 0.17210 & 0.18865 & 0.22150 & 0.04512 \\
307 & Heisenberg & 0.59649 & 0.45952 & 0.16277 & 0.10272 & 0.10348 \\
307 & TFI        & 0.25090 & 0.15539 & 0.22800 & 0.21445 & 0.04711 \\
\bottomrule
\end{tabular}
\end{center}
\end{table}

ZNE reduces raw error to $0.46$--$0.47$ on Heisenberg and
$0.16$--$0.17$ on transverse-field Ising, but remains less accurate than
the feature-based estimators. The shuffled-noisy control matches the affine
feature-only control $A$ within $0.0001$ in these six fitted linear comparisons.
These controls contextualize the primary result; they are not lower bounds
on the error achievable with measurements.

\subsection{Secondary: Measurement Benefit across Evolution-Step Regimes}
\label{sec:results-help}
\label{app:results-secondary}

\textit{Does the larger evolution step change the capacity-matched gap at 640
training circuits per family?}

This is a secondary contrast motivated by exploratory runs.
The campaign uses transverse-field Ising and Heisenberg circuits at ten qubits
and three family-native steps. The shipped step sizes are 0.2 and 0.15,
respectively; the larger step size is 0.6 for both families. Each regime is a
separately trained familiar-condition experiment. This comparison does not
measure transfer from one step size to another.

The generator seeds are 101, 211, and 307. Each family has 160 or 640 training
circuits, 320 source-validation circuits, and 160 untouched test circuits.
Every experiment uses depolarizing-plus-readout noise at the first and third
strength levels, the middle-site and middle-pair observables, and 2048 shots.
These choices give twelve datasets. At fixed regime and seed, the smaller
training set is a prefix of the larger one; validation and test identities
are shared across sizes.

For each family, seed, and size, we report the full and feature-only test MAEs,
their absolute difference, and the relative gain
$g=1-\mathrm{MAE}_{\mathrm{full}}/\mathrm{MAE}_{\mathrm{feature\mbox{-}only}}$.
Each MAE equally weights the four strength-by-observable cells. The secondary evolution-step endpoint is
$D_{\mathrm{large}}-D_{\mathrm{shipped}}$ at 640 training circuits per family,
where $D$ is the capacity-matched feature-only error minus the full
nonlinear method's error. Its six family-and-seed comparisons are reported
separately. The paired training-size endpoint is
$D_{640}-D_{160}$, reported separately for each regime, family, and seed.
Neither secondary endpoint is used to support the primary claim.

Intervals use 10,000 physical-circuit bootstrap draws with root seed
20260904 and the frozen stream mapping. Each draw keeps all strength and
observable rows of a sampled circuit together. Model arms share each draw
within a dataset, while the two regimes use independent streams. Each draw
recomputes the absolute gaps and their evolution-step contrast. The paired
training-size contrast shares physical-circuit draws across the two sizes.
The 2.5th and 97.5th percentiles define the pointwise intervals, conditional
on the fitted and selected models.

These secondary comparisons are reported one row at a time, with their
pointwise intervals and every dissenting seed. They are not used to support the primary claim. A positive contrast can occur
between two negative gains, so a sign here is not by itself a benefit claim.

The step change alters gate angles and total evolution time. The estimates
describe these learners under the declared parameter regime and feature access.
They do not establish an information-theoretic boundary for measurement.
Every estimate reported for this question comes from the frozen campaign
run.

\begin{table}[t]
\caption{The secondary evolution-step contrast, $D$ at the larger step minus $D$
at the shipped step, at 640 training circuits per family. $D$ is the
capacity-matched feature-only error minus the full nonlinear error. Intervals
are pointwise and conditional on the fitted and selected models. Takeaway: three
intervals lie below zero, two above, and one spans it, and both positive rows
belong to seed 211.}
\label{tab:secondary-step}
\begin{center}
\small
\begin{tabular}{llrl}
\toprule
\bf Seed & \bf Family & \bf Contrast & \bf Pointwise 95\% CI \\
\midrule
101 & Heisenberg & $-0.001320$ & $[-0.003162,\ 0.000607]$ \\
101 & Transverse-field Ising & $-0.001661$ & $[-0.002458,\ -0.000890]$ \\
211 & Heisenberg & $0.002744$ & $[0.000622,\ 0.004842]$ \\
211 & Transverse-field Ising & $0.002734$ & $[0.001806,\ 0.003635]$ \\
307 & Heisenberg & $-0.003787$ & $[-0.006939,\ -0.000822]$ \\
307 & Transverse-field Ising & $-0.003961$ & $[-0.004905,\ -0.003009]$ \\
\bottomrule
\end{tabular}
\end{center}
\end{table}

The larger evolution step changes the capacity-matched gap in both
directions: three of six intervals lie entirely below zero, two entirely above,
and one spans zero. Table~\ref{tab:secondary-step} gives the six rows with
their intervals, and both positive contrasts belong to seed 211. For the paired training-size
contrast, Table~\ref{tab:secondary-size} gives all twelve rows: ten of twelve
point differences are positive, while five intervals lie above zero, one below
zero, and six span zero. These descriptive counts do
not establish a general evolution-step or training-size trend and are not used
to support the primary claim.

\begin{table}[t]
\caption{The secondary paired training-size contrast,
$D_{640}-D_{160}$, for every regime, seed, and family. Differences use
unrounded $D$ estimates. Intervals come from 10,000 paired physical-circuit
bootstrap draws shared across the two training sizes, and are pointwise and
conditional on the fitted and selected models. Takeaway: ten of twelve point
differences are positive, while five intervals lie above zero, one below, and
six span it.}
\label{tab:secondary-size}
\begin{center}
\footnotesize
\setlength{\tabcolsep}{4pt}
\begin{tabular}{lllrrrl}
\toprule
\bf Regime & \bf Seed & \bf Family & \bf $D_{160}$ & \bf $D_{640}$ &
\bf Difference & \bf Paired 95\% CI \\
\midrule
Shipped & 101 & Heisenberg & $0.001320$ & $-0.000385$ & $-0.001705$ & $[-0.002585,\ -0.000853]$ \\
Shipped & 101 & Transverse-field Ising & $0.002399$ & $0.002461$ & $0.000062$ & $[-0.000626,\ 0.000764]$ \\
Shipped & 211 & Heisenberg & $-0.003497$ & $0.001005$ & $0.004501$ & $[0.003488,\ 0.005484]$ \\
Shipped & 211 & Transverse-field Ising & $-0.002258$ & $0.001156$ & $0.003414$ & $[0.002978,\ 0.003855]$ \\
Shipped & 307 & Heisenberg & $0.001301$ & $0.000800$ & $-0.000500$ & $[-0.001510,\ 0.000458]$ \\
Shipped & 307 & Transverse-field Ising & $0.002017$ & $0.005547$ & $0.003529$ & $[0.002635,\ 0.004414]$ \\
Larger & 101 & Heisenberg & $-0.005566$ & $-0.001705$ & $0.003861$ & $[-0.000784,\ 0.008649]$ \\
Larger & 101 & Transverse-field Ising & $-0.005895$ & $0.000801$ & $0.006696$ & $[0.005068,\ 0.008301]$ \\
Larger & 211 & Heisenberg & $-0.000227$ & $0.003748$ & $0.003975$ & $[-0.000193,\ 0.007948]$ \\
Larger & 211 & Transverse-field Ising & $0.002335$ & $0.003890$ & $0.001555$ & $[-0.000487,\ 0.003655]$ \\
Larger & 307 & Heisenberg & $-0.003416$ & $-0.002987$ & $0.000429$ & $[-0.003958,\ 0.004692]$ \\
Larger & 307 & Transverse-field Ising & $-0.000997$ & $0.001586$ & $0.002582$ & $[0.000836,\ 0.004284]$ \\
\bottomrule
\end{tabular}
\end{center}
\end{table}

\subsection{RQ1: The Capacity-Matched Feature-Only Control}
\label{app:results-full-2}

\textit{What gain does the nonlinear arm show against a feature-only control with
comparable model capacity?}

The additional control uses the ablation's existing RF and MLP candidates,
with the noisy-estimate column physically deleted at fitting, validation, and
prediction. The other 29 columns retain their order. Forests remain grouped
by register width and physical observable; the MLP remains global. Full and
feature-only pipelines select independently under the existing one-standard-error
rule and tie breaks, using the dataset seed and the same candidate searches,
preprocessing, and training schedules. The feature-only counterpart has zero
measurement cost. This additional campaign control is outside the eight-method
registry summarized in Table~\ref{tab:methods}.

The affine controls remain diagnostics. Training-shuffle refits and fixed-model
input shuffles are reported separately, with their permutation groups and seeds.
Neither shuffle substitutes for feature removal. We report endpoint MAEs,
signed gains, absolute reductions, and intervals for the capacity-matched
comparison, whatever their direction. That comparison carries the campaign's primary endpoint, and the shuffle
diagnostics beside it remain secondary. The campaign
also reports the same nonlinear arm against the affine feature-only control on
these untouched test rows. That comparison holds the fitted learner fixed and
moves only the control. We report it whichever way the two arms fall, and it
carries no promotion verdict. If the arms do diverge, it is what separates a
change in the control from a change in the learner.

The primary result comprises the six within-family decompositions at the
shipped step sizes and 640 training circuits per family. Let $A$, $C$,
and $F$ denote the affine feature-only, capacity-matched feature-only,
and full nonlinear errors. We report $T=A-F$, $K=A-C$, $D=C-F$, and
$S=K/T$ separately for every family-and-seed row, with pointwise
circuit-bootstrap intervals conditional on the fitted and selected models.
No share is selected or averaged. Counts of interval signs are descriptive;
no seed vote determines the primary claim. Evolution-step and training-size
contrasts are secondary and are not used to support the primary claim.

The campaign record retains the source-validation training-shuffle intervals
and fixed-model shuffle diagnostics for every condition, together with
item-keyed predictions, selected configurations, permutation groups, and seeds.
Those diagnostics are held in the run artifacts and are not displayed here. These validation diagnostics reuse
rows seen during selection. Untouched-test intervals condition on the selected
models and omit training variation. Three generator seeds do not estimate a
population replication rate. The raw estimate itself is in Figure~\ref{fig:share-ladder}. Failed conditions remain visible and
cannot count as successful replications. Synthetic checks verify gains and
contrasts, predictor pairing, independent regime streams, circuit blocks that
span both severities and observables, and unavailable denominators. A separate
check confirms that changing a test label changes neither the selected
configurations nor the predictions. These checks are implemented and pass.

The capacity-matched control, the test-gain analysis, and the campaign
audit are implemented and verified against small artifacts. Twelve of
twelve campaign datasets were then audited in the frozen campaign, with no
failed dataset, no missing roster, and no missing zero-noise replay. Every one of
the twenty-four grid cells returned an estimate, and no draw was discarded.
The additive identity held on every family, with a largest absolute
residual of $2.8\times10^{-17}$ against a tolerance of $7.1\times10^{-15}$.
Two limits belong with that record rather than after it. The frozen
environment record lists package versions and leaves the
continuous-integration lock unverified. Ridge also fails the
source-validation gate on all six shipped-step datasets, which
Section~\ref{sec:results-controls} reports with its ratios.

This appendix expands Finding~1 of Section~\ref{sec:results-breaks} with
per-row estimates.

\begin{figure}[htbp]
\centering
\includegraphics[width=5.2in]{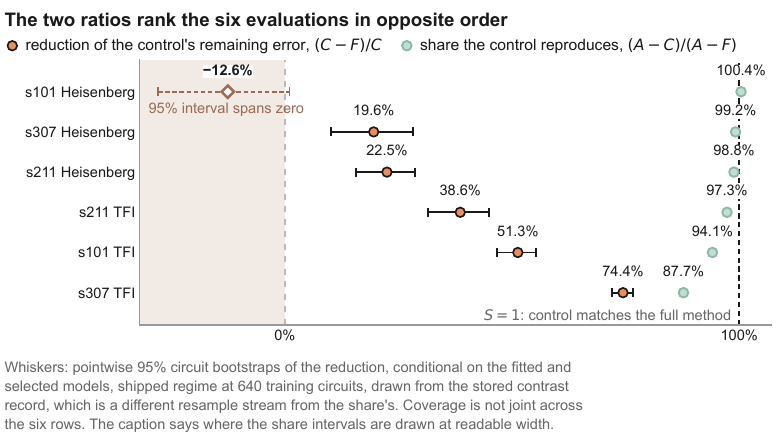}
\caption{\textbf{Sorted by the share it reproduces, the six primary
evaluations come out in increasing order of remaining-error reduction.}
Rows run by share, descending, which is not the fixed evaluation order of
Figure~\ref{fig:gap-and-share}; each row names its seed and family. The share
$S=(A-C)/(A-F)$ divides the part the capacity-matched control reproduces by the
ablation's improvement over the affine control. The remaining-error reduction
$D/C=(C-F)/C$ divides the gap $D=C-F$ by the control's own error. Numerator and
denominator both differ, so the second does not follow from the first. Whiskers are
pointwise 95 percent circuit-bootstrap intervals of the reduction, conditional
on the fitted and selected models, at 640 training circuits. Coverage is not
joint across rows, and the shares keep their own intervals in
Figure~\ref{fig:gap-and-share}, which is why none is drawn here. Every arm was
trained on exact labels, so a reduction compares two fitted arms and is no
evidence that reading the noisy estimate earns its cost: the degree-five
polynomial in the couplings alone (Table~\ref{tab:poly-degree-5}) beats the full
method on all six rows while reading no measurement. A share near one therefore
leaves the remaining-error reduction open, and on these six rows the two
readings order the evaluations oppositely.}
\label{fig:share-vs-reduction}
\end{figure}

\noindent\textbf{In the six primary evaluations, the capacity-matched
feature-only control reproduces 87.7 to 100.5 percent of the improvement the
restricted-feature ablation shows over the affine feature-only control.}
Five of the six shares have intervals entirely below one; the sixth, seed
101 with Heisenberg, runs from 0.9997 to 1.0084.
Figure~\ref{fig:gap-and-share} gives the six gaps and the six shares with their
intervals. Table~\ref{tab:primary-gap} in Appendix~\ref{app:figure-tables}
prints the same six gaps beside the gap as a fraction of the total improvement
$T=A-F$. The comparison against the
affine control moves learner capacity and measurement access together, so
it does not isolate deletion of that measurement. Holding the learner class
fixed, the ablation cuts the control's remaining error by 51.3, 22.5, 38.6,
19.6 and 74.4 percent on the five rows whose gap excludes zero. The range is
19.5 to 74.5 percent, with its endpoints rounded outward to contain all five
estimates. On seed 101
with Heisenberg that reduction is $-12.6$ percent, and the gap's own
interval spans zero.
Figure~\ref{fig:share-vs-reduction} pairs the two ratios row by row, with the
rows sorted by share. Under that order the six reductions come out increasing:
the row with the highest share carries the only negative reduction, and the row
with the lowest share carries the largest one.

\subsection{Controlled Diagnostics: Signal and Polynomial Baselines}
\label{app:controlled-diagnostics}

Table~\ref{tab:signal-difficulty} describes the noisy and ideal expectations
in the shipped-step testbed. Each cell contains 160 physical test circuits
per seed. Statistics are computed separately for seeds 101, 211, and 307,
then averaged with equal weights. All circuits have ten qubits and three
family-native steps.

\begin{table}[t]
\caption{Signal diagnostics at shipped step sizes, averaged over three seeds.
The centered OLS slope $\beta=\mathrm{Cov}(y,r)/\mathrm{Var}(y)$ regresses
the noisy expectation $r$ on the ideal expectation $y$ with an intercept;
Res.\ SD is the residual standard deviation. These descriptive statistics
do not bound recoverable information or explain the reproduced share $S$.}
\label{tab:signal-difficulty}
\begin{center}
\small
\setlength{\tabcolsep}{4.5pt}
\begin{tabular}{lllrrrrr}
\toprule
\bf Family & \bf Sev. & \bf Obs. & \bf Mean $|y|$ & \bf SD($y$) & \bf Mean $|r|$ & \bf $\beta$ & \bf Res.\ SD \\
\midrule
Heisenberg & $L_1$ & $z_{\mathrm{mid}}$  & 0.9037 & 0.1118 & 0.6083 & 0.6613 & 0.0176 \\
Heisenberg & $L_1$ & $zz_{\mathrm{mid}}$ & 0.8854 & 0.1282 & 0.4651 & 0.4955 & 0.0204 \\
Heisenberg & $L_3$ & $z_{\mathrm{mid}}$  & 0.9037 & 0.1118 & 0.0766 & 0.0692 & 0.0221 \\
Heisenberg & $L_3$ & $zz_{\mathrm{mid}}$ & 0.8854 & 0.1282 & 0.0228 & 0.0064 & 0.0227 \\
\midrule
TFI        & $L_1$ & $z_{\mathrm{mid}}$  & 0.6834 & 0.2208 & 0.5880 & 0.8633 & 0.0172 \\
TFI        & $L_1$ & $zz_{\mathrm{mid}}$ & 0.5515 & 0.2595 & 0.4362 & 0.7905 & 0.0195 \\
TFI        & $L_3$ & $z_{\mathrm{mid}}$  & 0.6834 & 0.2208 & 0.2642 & 0.4047 & 0.0209 \\
TFI        & $L_3$ & $zz_{\mathrm{mid}}$ & 0.5515 & 0.2595 & 0.1259 & 0.2270 & 0.0217 \\
\bottomrule
\end{tabular}
\end{center}
\end{table}

Attenuation differs by family, severity, and observable. For TFI the slopes
range from $0.79$ to $0.86$ at $L_1$ and from $0.23$ to $0.40$ at $L_3$.
Heisenberg's $L_3$ slopes are $0.0692$ and $0.0064$, with mean $|r|$
of $0.0766$ and $0.0228$, respectively. This identifies severe attenuation
in part of the testbed, not a uniform absence of signal.

\paragraph{Post-Hoc Polynomial Diagnostic on Coupling Descriptors.}
A post-campaign diagnostic fitted ordinary least squares polynomials in
$(J,h)$ for TFI and $(J_x,J_y,J_z)$ for Heisenberg, separately per observable.
Inputs were centered and scaled using training rows only. We report the
fixed degree-five arm of this exploratory diagnostic, with 21 coefficients
per observable for TFI and 56 for Heisenberg; the degree was not selected
using test error. The exploratory script fitted degrees one through eight,
reported fixed degrees three and five, and also selected a separate arm by
minimum validation MAE, with ties favoring the lower degree and no refitting.
The comparison uses 640 training circuits per family and
the same four-cell macro weighting as the primary results.

\begin{table}[t]
\caption{Recorded post-hoc degree-five polynomial errors beside recomputed
capacity-matched control $C$ and full-method $F$ errors for the six primary
evaluations. The polynomial uses coupling descriptors alone and has lower
recorded error on every row. Entries are descriptive point comparisons,
not paired interval estimates.}
\label{tab:poly-degree-5}
\begin{center}
\small
\setlength{\tabcolsep}{5.5pt}
\begin{tabular}{llrrrr}
\toprule
\bf Seed & \bf Family & \bf Control $C$ & \bf Full $F$ & \bf Poly-5 & \bf Poly / $F$ \\
\midrule
101 & Heisenberg & 0.00306 & 0.00345 & 0.000256 & 7.4\% \\
101 & TFI        & 0.00480 & 0.00234 & 0.000042 & 1.8\% \\
211 & Heisenberg & 0.00447 & 0.00347 & 0.000292 & 8.4\% \\
211 & TFI        & 0.00300 & 0.00184 & 0.000051 & 2.8\% \\
307 & Heisenberg & 0.00409 & 0.00329 & 0.000266 & 8.1\% \\
307 & TFI        & 0.00745 & 0.00191 & 0.000051 & 2.7\% \\
\bottomrule
\end{tabular}
\end{center}
\end{table}

The polynomial entries reproduce the recorded post-campaign diagnostic
summary; $C$ and $F$ were recomputed from archived test predictions.
The compact diagnostic package behind these tables, which is separate from
\texttt{campaign-archive-v1}, does not contain the polynomial fit script, training
rows, or per-item polynomial predictions, so it supports reproduction of the
recorded comparison, not an independent refit or a paired interval.
The recorded polynomial errors are about 12 to 55 times lower than $F$.
This challenges the adequacy of the evaluated learners on the smooth,
low-dimensional coupling parameterization, rather than establishing an
information-theoretic measurement requirement.

One planned analysis is absent from this paper. An earlier design scored a
routed system that sent low-confidence predictions to a conservative fallback,
and that layer has been removed from the release. The reason is a verification
limit rather than a negative result. No local check on a decoded prediction file
can show that the prediction was produced without access to the test label. A
routed number could therefore not be trusted from the artifact alone. The layer
returns only with a trusted fitting stage inside the runner, or with a
verifiable record of who produced the predictions. Appendix~\ref{app:discussion-extended}
carries the fuller statement.

%% file: A5_field.tex
\section{RQ2 and RQ3 in Full: Two Published Mitigators}
\label{app:field-full}
\textit{Under each released protocol, which inputs carry the ordered predictive
gap, and what collection requirements accompany Q-LEAR's gains?}

The controlled half shows what an evaluation of this shape can reward. It cannot
show how often that shape occurs. We therefore ran ordered input comparisons on the
released data of two published learned mitigators, Q-LEAR
\citep{muqeet2024qlear} and QRAFT \citep{patel2021qraft}. Q-LEAR's archive covers
six application circuits on eight IBM backends, with a simulator companion.
QRAFT's archive covers five machines and ships both its processed inputs and its
predictions.

Commitments and absences frame what follows. For Q-LEAR, the ladder of arms, the
inputs, the retraining recipe, the ten seeds, the score, the aggregation and the
resample unit were fixed before the controlled campaign ran. The QRAFT panel
specification was committed before its scoring script was run against any data.
No pre-fit artifact binds the Q-LEAR specification to a date. Checksums taken
before review establish that the inputs did not change afterwards; they cannot
establish when the specification was frozen. The replay also runs on a Linux
aarch64 port rather than a replica of the released Windows environment. Pinned
versions of Python, TensorFlow, Keras, ktrain, transformers, numpy, pandas and
scikit-learn match the release; SciPy, h5py and joblib differ. Neither archive
supports a claim about unseen application circuits. The Q-LEAR generator builds
its validation set from the same circuit directory the evaluation reads, and both
QRAFT panels split rows rather than circuits.

Scores on Q-LEAR are Hellinger distances on the released support, one per
application and backend, averaged equally over eight backends and six
applications. Intervals are pointwise application bootstraps conditional on each
fit, the eight fixed backends and the released labels. They exclude retraining
and model-selection uncertainty, which the ten-fit spread describes separately.
The scorer restarts its draw stream at a fixed seed inside the panel loop, so the
hardware and simulator panels resample the same application indices. Their
intervals are therefore dependent by construction, and the two panels agreeing is
not an independent replication.
Counts of intervals above zero are descriptive counts over the same six
applications rather than independent replication rates.
Appendix~\ref{app:field-stats} states the full protocol.

One absence governs every number below. The design carries no shot-count sweep
and no arm that reallocates a fixed number of evaluations across methods, so
nothing here says what one further evaluation would buy.

\subsection{Does the Replay Reproduce the Published Numbers?}
\label{sec:field-repro}
\label{app:field-replay}

\textit{Before any ablation is read, does our retrained copy of the released
method land where the published table puts it?}

The evaluation protocol forbids interpreting any ablation until a faithful reproduction
is verified, so this panel came first.

\textbf{All six published application values fall inside our ten-fit range, with
a largest application mean deviation of 0.0134.} Table~\ref{tab:field-repro}
gives the six rows. Coverage of a published point by a ten-fit range is weaker
than an equivalence test, and no reproduction tolerance was declared before the
run. Calling this replay faithful is therefore a judgement made after seeing it.
Coverage also fails at backend resolution: on the Lagos backend the published
value is 0.44 against our range of 0.4446 to 0.4595.

\begin{table}[t]
\caption{Q-LEAR reproduction on hardware. The published column is the value the
release reports for its neural model on each application; the remaining columns
are ten retrained fits scored by the release's own routine. Takeaway: every
published value lies inside our range at application resolution, which supports
the port and is not an equivalence test.}
\label{tab:field-repro}
\begin{center}
\small
\begin{tabular}{lrrrr}
\toprule
\bf Application & \bf Published & \bf Our ten-fit mean & \bf Min & \bf Max \\
\midrule
Ground state & 0.45 & 0.451 & 0.443 & 0.457 \\
Pricing call & 0.60 & 0.613 & 0.557 & 0.661 \\
Pricing put  & 0.59 & 0.593 & 0.572 & 0.632 \\
QAOA         & 0.57 & 0.577 & 0.565 & 0.587 \\
Routing      & 0.10 & 0.102 & 0.095 & 0.111 \\
TSP          & 0.36 & 0.364 & 0.348 & 0.377 \\
\bottomrule
\end{tabular}
\end{center}
\end{table}

\subsection{Does the Descriptor Control Carry the Improvement on Q-LEAR?}
\label{app:field-qlear}

\textit{Under the released Q-LEAR protocol, how much of the ordered improvement
does a nonlinear descriptor control reproduce, and how much needs the
measurement-derived inputs?}

The first step changes affine to nonlinear capacity at fixed descriptors; the
remaining two steps retain the nonlinear recipe and add measurement-derived
feature groups. The ladder begins with an affine descriptor control over
the four released circuit descriptors and the state weight. A nonlinear
descriptor control over those same inputs follows, then the base-statistics rung,
then the statistics from three depth-cut circuits, which completes the full
released method. The affine control is fitted once and reused. C, O, F, and the
off-ladder C4 diagnostic are each fitted at seeds 0 through 9, giving forty-one
fits.

\textbf{The nonlinear descriptor control adds a ten-fit hardware macro mean of
$+0.000055$ over the affine one, and every per-fit interval spans zero.} Its
ten-fit range runs from $-0.003115$ to $+0.004225$. On the simulator companion
the same step gives a ten-fit macro mean of $+0.000091$ and a range from
$-0.003124$ to $+0.004333$, with every per-fit interval spanning zero. The
anticipated large capacity contribution is therefore not observed, which is
weaker than absence and is the strongest statement these fits support. A small
positive estimate whose interval crosses zero does not demonstrate equivalence,
and the affine control already receives the same descriptors. Both arms are
separately fitted predictors with approximately equal macro errors and differing
per-application errors, so they are not the same control.

\textbf{Measurement-derived inputs carry the ordered gap on hardware, where the
full released method improves on the nonlinear descriptor control by a ten-fit
macro mean of 0.2191.} The per-fit intervals lie above zero on all ten fits. Split in
the declared order, the base-statistics rung opens 0.1969 of that gap and the
three depth-cut circuits 0.0222, whose interval is above zero on nine of ten
fits. Both are descriptive splits of an ordered comparison rather than unique
causal contributions, and no interval is reported for either fraction. On the
simulator companion the same two steps give 0.4407 and 0.0099, and seven of its
ten depth-cut intervals cross zero. Figure~\ref{fig:qlear-endpoints} reports the
two measurement steps on both panels, where disagreement between them about the
depth-cut features is the finding rather than something to average away.
Table~\ref{tab:field-ladder} in Appendix~\ref{app:figure-tables} prints the whole
ordered ladder, including the capacity step that opens it and the count of
per-fit intervals above zero on every rung.

\textbf{The near-zero average is partly cancellation: the nonlinear descriptor
control improves on the affine one for both pricing applications on all ten
fits, by hardware means of $+0.020113$ and $+0.017292$.} It worsens the other
four applications on all ten fits, and the simulator carries the same signs.
These capacity gains are stable in sign and are not negligible, and they do not
rescue the anticipated macro pattern. The gap from the nonlinear descriptor
control to the full released method varies by an order of magnitude across
applications, from 0.053 on QAOA to 0.565 on routing. It stays positive on
every one of the six. Figure~\ref{fig:qlear-cancellation} gives the six
capacity contrasts on both panels, and Figure~\ref{fig:qlear-endpoints} gives
the per-application step that the depth-cut circuits add.
Table~\ref{tab:field-apps} in Appendix~\ref{app:figure-tables} prints the
hardware rows, including the per-application gap of the full method over the
descriptor control that neither figure draws.

The descriptors are not devoid of query information. Q-LEAR's four circuit
descriptors are constant across every output state of a circuit, so only the
state weight can separate states within a circuit. That weight does separate
states of unequal ideal probability in at least some released files. The
limitation is real and partial, and it does not license a claim that these
descriptors carry no query information.

\subsection{Does a Richer Descriptor Set Carry It on QRAFT?}
\label{app:field-qraft}

\textit{Does a descriptor set richer than Q-LEAR's four circuit numbers and a
state weight close the gap in a second published system?}

QRAFT records circuit width, depth, separate single-qubit and two-qubit gate
counts, and the output state's Hamming weight. Three arms come transcribed from
its released training code. Descriptors alone give seven features. Adding the
forward observed probability at three severities gives ten, and adding the
reverse-execution statistics and machine identity gives seventeen. The target is
the true probability of the output state in every arm and never an input. One
reading scores the predictions QRAFT shipped and fits nothing; the other retrains
all three arms under one recipe at ten seeds. Both are scored by per-row mean
absolute error in probability points, with rows weighted equally within a machine
and the five machines equally.

\textbf{Descriptors alone score roughly six times worse than descriptors plus the
forward observed probability: 12.4838 against 2.1074 in the released predictions,
and 13.5903 against 2.2685 under matched retraining.} Per machine the released
ratios are 8.96, 6.07, 6.16, 3.73 and 6.50. The retrained ratios by seed are
6.36, 5.80, 6.55, 6.29, 5.93, 5.59, 6.38, 5.87, 5.44 and 5.88. Conditions favor
the descriptor-only arm, because the released split is shuffled by row and a
circuit's output states therefore appear in both training and test. It loses
anyway, on every machine and every seed.

\textbf{The final feature group, reverse-execution statistics and machine
identity, improves matched retraining by about six
percent, from 2.2685 to 2.1246, and loses in the released predictions, 2.4467
against 2.1074.} Its retrained direction holds on all ten seeds, and its
released direction is worse on four of the five machines. Two differences account
for the disagreement without a third. The panels use different learners, and
QRAFT's own models minimize a weighted classification cost that penalizes errors
on high-probability states, while this readout is mean absolute error. So the
bounded statement concerns this score, and it is not a claim that QRAFT's
reverse-circuit method fails under the objective QRAFT chose, which this work
does not evaluate.

\begin{table}[t]
\caption{\textbf{Descriptors alone score roughly six times worse than descriptors plus the forward observed probability in both readings: 12.4838 against 2.1074 (released) and 13.5903 against 2.2685 (retrained).}
Each cell is a per-row mean absolute error in probability points, with rows weighted equally within a machine and the five machines equally. The released-prediction panel reads 1,524 shipped predictions with no fitting; the matched-retraining panel refits all three arms over 10,155 processed rows under one recipe. The ten-seed range is a spread across seeds, not a confidence interval, and no interval is reported for either panel. Takeaway: the final feature group, reverse-execution statistics and machine identity, changes which arm wins between them.}
\label{tab:field-qraft}
\begin{center}
\small
\setlength{\tabcolsep}{3pt}
\begin{tabular}{llrrl}
\toprule
\bf Panel & \bf Arm & \bf Features & \bf Macro MAE & \bf Ten-seed range \\
\midrule
Released predictions & Descriptors only & 7 & 12.4838 & not applicable \\
Released predictions & Adds forward observed probability & 10 & 2.1074 & not applicable \\
Released predictions & Adds reverse statistics and machine identity & 17 & 2.4467 & not applicable \\
\midrule
Matched retraining & Descriptors only & 7 & 13.5903 & [13.1060, 13.9975] \\
Matched retraining & Adds forward observed probability & 10 & 2.2685 & [2.0663, 2.5100] \\
Matched retraining & Adds reverse statistics and machine identity & 17 & 2.1246 & [1.9439, 2.3343] \\
\bottomrule
\end{tabular}
\end{center}
\end{table}

Three limits travel with this panel, and Appendix~\ref{app:field-stats} gives the
reasoning for each. Neither reading carries an interval, and that was decided
before the run. The released output has no circuit identifier and no state label,
and five machine-level clusters are too few for a stable one. Neither reading
reproduces QRAFT's published results. One reads their shipped predictions without
checking them against their printed numbers; the other substitutes a histogram
gradient-boosting regressor for their MATLAB classification ensembles. Absolute
errors from the retrained panel therefore carry no weight, and only the ordering
across arms does. This score also differs from the Hellinger distance used on
Q-LEAR, which that archive cannot support without a circuit grouping, so the two
are never compared or combined.

\subsection{What Two Instances Establish}
\label{app:field-establish}

\textit{Taken together, what do the two reanalyses support, and what stays
outside them?}

The Q-LEAR ladder estimates the capacity contrast the controlled campaign
anticipated, and does not show that pattern occurring. The QRAFT panel compares
three feature sets instead, with no affine arm, no reproduced share and no
interval, so its error ratio does not test the controlled share.
Two instances are not prevalence, and
this design cannot support prevalence in either direction. The QRAFT data were
collected by the Q-LEAR authors on the same application circuits in the same
release, which makes this a second feature design rather than a second
environment. A prevalence claim needs a design built for prevalence, and this is
not one.

Neither reanalysis claims execution-free prediction. The rows are noisy observed
states, and the released support was chosen by executing the circuit. What these
panels estimate is the contribution of the retained numerical features
conditional on that support.

The reported quantities are also not a fraction of Q-LEAR's published figure.
Naming such a fraction would need controls on the same evaluation contract and a
separately named denominator, which this scope does not build.

Grouping information sources by origin predates this work. The Q-LEAR release
already defines its features separately, retrains after individual feature
deletions, and reports a 32 percent median error increase from deleting the
observed probability \citep{muqeet2024qlear}. QRAFT already compares static,
static-plus-forward, and full forward-and-reverse feature sets, and documents
their execution requirements \citep{patel2021qraft}. What these panels add is the
ordered group comparison, under one retraining recipe and a single score per
panel, with the
controls this paper specifies. Appendix~\ref{app:field-cost} places the released
collection counts beside each step.

The controlled instrument still demonstrates that an evaluation of its shape can
reward interpolation over circuit descriptors alone.
Appendix~\ref{sec:results-third} reports separately documented comparisons
on three other published releases under a scoring rule committed before
fitting. Its screen was informed by prior knowledge, and its timing is
supported by contemporaneous working records rather than
independent attestation. The readings establish their stated comparisons,
with the reported split, budget and target restrictions; they establish
neither an unqualified extension to published mitigation practice nor its
absence. A prevalence claim would require a design that supports prevalence.

%% file: A6_three_systems.tex
\section{An Additional Question: Three Screened Published Systems}
\label{sec:results-third}
\label{app:three-systems}

\textit{On published systems admitted by a structural screen, how much of the
affine-to-full improvement does a capacity-matched descriptor control reproduce,
and what does a reproduced share measure?}

The comparisons introduced in Section~\ref{sec:results-third-brief} are separately
documented and governed by a scoring rule committed before fitting.
The screening criteria were recorded before the search; their timing
rests on contemporaneous working records rather than
independent attestation. This question is additional
to the paper's three, is not used to support the primary claims, and declares no finding
slot. A screen over six structural admission criteria admitted three systems,
none sharing authors, application set or release with Q-LEAR or QRAFT. They are
ML-QEM, the learned mitigator of \citet{liao2024practicalqem} that this paper
already enters as a restricted-feature ablation; the convolutional mitigator of
\citet{cantori2024synergy}, called Synergy CNN below; and Q-Cluster
\citep{patil2025qcluster}.

A scoring rule committed before any arm was fitted on any of them governs what
follows, in a commit that carries no scores. The arms rebuild the campaign's
ladder on each system's own released rows. Here $A$ is an affine descriptor
control over that system's execution-independent descriptors. Its
capacity-matched counterpart $C$ reads the same descriptors under the system's
own model family and selection rule, and $F$ is the released full feature set.
The reproduced share is $S=(A-C)/(A-F)$, as in Section~\ref{sec:results-breaks},
read against 0.5 on the interval declared for that system before fitting. An
interval entirely above 0.5 reads present, entirely below reads absent, and one
spanning 0.5 reads indeterminate. Device calibration is a declared third group.
A system carrying it is scored once with calibration available to $A$ and $C$
and once with it withheld, and both numbers are reported. Every admitted system
appears in Figure~\ref{fig:three-systems}, and no share is averaged or pooled
across them. Table~\ref{tab:third-instance} in Appendix~\ref{app:figure-tables}
states each reading beside the sensitivities and scope limits it carries.
QRAFT is not a fourth reading, having been scored with no affine
arm, no share and no interval, by a decision recorded before its run
\citep{patel2021qraft}.

\begin{figure}[p]
\centering
\includegraphics[width=6.5in]{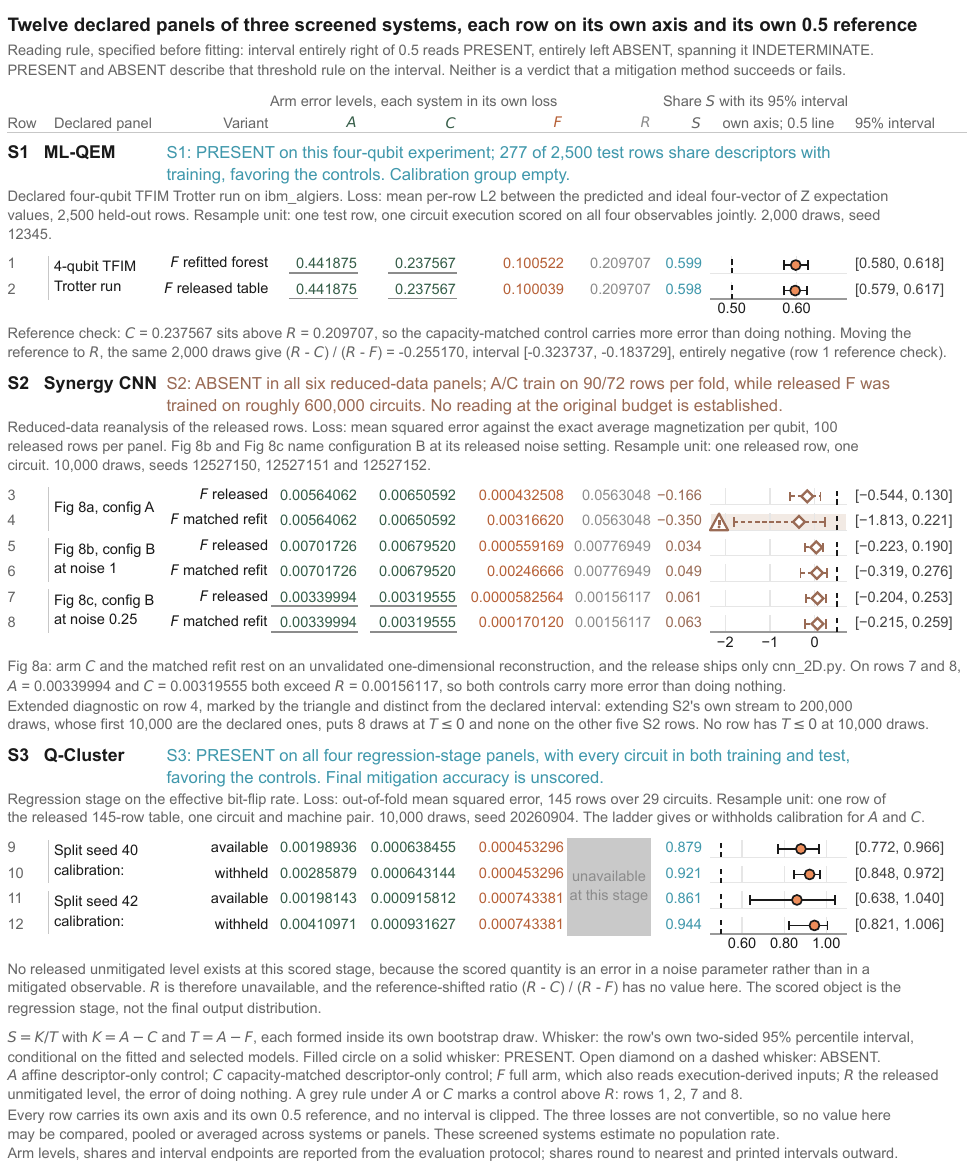}
\caption{\textbf{The three readings disagree, and each is bounded by a property
of the release it was read from.}
The twelve declared panels of the three admitted systems, one axis and
one $0.5$ reference per row. The reproduced share is $S=(A-C)/(A-F)$, for $A$
the affine descriptor-only control, $C$ its capacity-matched counterpart, $F$
the full arm in the indicated released or refitted variant and $R$ the unmitigated level. Whiskers are that row's own
95 percent percentile interval, conditional on the fitted and selected models.
Each system is scored in its own loss, so nothing here may be pooled or compared
across systems.}
\label{fig:three-systems}
\end{figure}

\textbf{ML-QEM reads present at $S = 0.5985$ $[0.5808, 0.6172]$, from a
conditional row bootstrap of 2,000 draws on its declared four-qubit experiment,
whose calibration group is empty.} Of the 2,500 test rows, 277 carry a descriptor
block that also appears in training, an overlap that favors the descriptor
control and travels with this reading. None of the single substitutions tested
reverses it, though joint ones can. Ridge as the affine control together with the
separately released saved model, over all 2,500 rows, gives $0.4721$ $[0.4529,
0.4928]$. Those are sensitivities reported beside the declared reading, and not
evidence that it lacks an answer.

\textbf{Synergy CNN reads absent, with all six upper interval limits at or below
$+0.2755$ across its three panels and both full-arm variants, at 10,000 draws
each.} Its declared full arm is the released prediction, and a predeclared
matched-refit supplement reads absent as well. The reading is a reduced-data
reanalysis: the affine control fits on 90 rows per fold and the capacity-matched
control on 72, against a released model trained on roughly 600,000 circuits. One
panel also uses a reconstructed one-dimensional architecture. Applicability at
the original published budget is therefore not established, and no reading at
that budget is claimed in either direction.

\textbf{Q-Cluster reads present under its declared primary-seed row bootstrap, at
$S = 0.8795$ $[0.7723, 0.9658]$ with calibration available to the controls and
$0.9211$ $[0.8480, 0.9716]$ with it withheld, at 10,000 draws.} Both declared
split seeds give that reading in both calibration ladders, and the released split
places every circuit in both training and test, which favors the descriptor
control. The scored object is the effective bit-flip regression stage rather than
the final output distribution the scoring plan named, so it decomposes no final
mitigation accuracy. An exploratory circuit bootstrap is borderline at declared
secondary seed 42 with calibration, where 20 of 40 replicate seeds put the lower
limit above 0.5. Unseen-circuit robustness is not established.

The descriptor-only diagnostics use different selection procedures.
On ML-QEM, the best tested predictor attains a mean per-observable held-out
$R^2$ of 0.6805, selected on the same test rows on which it is reported. On
Q-Cluster, nested selection within the outer training folds gives at least
0.9348 under either calibration ladder. Both use the released
circuit-overlapping splits, and neither supplies a universal bound on descriptor
prediction. Synergy CNN's descriptors determine its target exactly: an
independently written state-vector simulator agrees to $1.46\times10^{-13}$
across all 300 rows. The controlled-generator explained-variance comparison
required by the scoring plan was not performed. We therefore report these
diagnostics as an incomplete feasibility audit and make no controlled-reference
high-or-low saturation classification or claim of discriminative validity. No absent reading here is therefore an artifact of
descriptors carrying no information. The controlled-reference half of that
diagnostic, which the scoring plan also requires, was computed for none of the
three.

\textbf{A large affine-referenced share need not represent improvement over the
raw estimate: on ML-QEM, $S = 0.5985$ coexists with $(R-C)/(R-F) = -0.2552$.}
Here $R$ is the raw estimate's error, and the unit is mean per-row distance from
the predicted four-vector of expectation values to the ideal one. On that scale
$A = 0.44188$, $C = 0.23757$, $R = 0.20971$ and $F = 0.10052$. The affine control
is therefore 2.107 times worse than doing nothing and the descriptor control
1.133 times worse, while the full arm improves on both. A paired bootstrap over
the same resample unit puts $C-R$ at $+0.02786$ $[+0.02006, +0.03545]$, and $C$
minus the system's own zero-noise comparator at an observed paired mean
of $+0.05075$, with interval $[+0.04209, +0.05910]$.

The controlled campaign does not sit in that position. There $A$, $C$ and $F$ all
improve on $R$ on all six primary rows. On those rows $R$ runs between 5.3 and
6.8 times the affine control's error and between 111 and 182 times the
ablation's (Table~\ref{tab:controls} in Appendix~\ref{app:figure-tables}). Where
both ratios are defined, they are
related by
\begin{equation}
\frac{R-C}{R-F} \;=\; 1-(1-S)\,\frac{A-F}{R-F},
\label{eq:rereference}
\end{equation}
so a weak affine control raises $S$ while the descriptor control still loses to
the raw estimate.

This is a post-result diagnostic, prompted by the ML-QEM numbers after that
reading was fixed. It changes no classification above and fixes no eligibility
condition on $S$. These comparisons do not establish that $A$ beating $R$ is
necessary or sufficient for interpreting $S$: at $R=1$, $A=2$, $C=0.2$ and
$F=0.1$, the share is 0.947 while $(R-C)/(R-F) = 0.889$. What follows is a
reporting rule. A reproduced share travels with same-unit $A$, $C$, $F$ and $R$
errors, its reference, the uncertainty in its $A-F$ denominator, and the circuit
overlap its split allows.

Three limits on the screen bound every reading above. Its criteria were recorded
before the search and committed together with the screen's result, so this is no
independently timestamped preregistration. Prior knowledge reaches the criteria
as well as the pool. ML-QEM was already named in this project two
months before the screen ran, and four of the six criteria name properties its
release was known to have. This paper's restricted-feature ablation is built from
that same published architecture, and no neutral literature search produced this
set. Adjudication is incomplete as well, because the surviving screening record
carries favorable and conflicting verdicts with no named disposition. These three
systems therefore do not exhaust the screen's eligible candidates. Candidates
blocked by unavailable material or an unconfirmed training recipe are recorded
separately, and non-admission is not evidence that a release is insufficient.

Three admitted systems estimate no rate in any population of published
mitigators, in either direction. Nothing here supports a claim about what the
field's released material can decide, and ML-QEM's own release supplied enough
material to expose the sensitivities above. No quantity in this appendix prices a
further measurement. What the three support is narrower: a qualified reading on
one four-qubit experiment, a reduced-data reading at a declared budget, and an
intermediate regression-stage reading on a circuit-overlapping split. None of
them establishes a clean independent occurrence of the controlled pattern in a
published system's final mitigation accuracy, and none establishes its absence.

%% file: A7_discussion.tex
\section{Extended Discussion}
\label{app:discussion-extended}

This appendix expands the interpretation in Section~\ref{sec:discussion}.

The controls bound the predictive-benefit claims supported by a low error here. Four surrogate controls run
beside the mitigation methods in every campaign dataset
(Appendix~\ref{app:modelsel}). One predictor sees the circuit and observable
description but never the noisy measurement. A second sees only the noisy
measurement. The third predicts the training mean, and the fourth is a copy of
the learned mitigator whose training noisy values have been shuffled. One
versioned feature contract feeds the learned methods, and it is
parameter-bearing: it carries the sampled couplings, angles, and evolution
step size alongside compiled structure. A model given those parameters can
approximate the ideal target from circuit features. The campaign measures the
gain from including the noisy input in specified learning pipelines.

The restricted-feature ablation needs the same care. It selects between the
published random-forest and multilayer-perceptron architectures of
\citet{liao2024practicalqem} on source validation, using \qemscore{}'s own
feature vector. That vector carries no native-gate count vector, no angle-bin
histogram, and no sparse Pauli encoding of the observable, and the run
artifact declares the omission in its own configuration record. A win or a
loss for this arm therefore describes one restricted representation trained
under one protocol. It establishes nothing about the published method.
Appendix~\ref{sec:results-third} also scores that same published system, under a
screen whose criteria were not independent of this paper's prior use of it. The
ablation's direction is not predictable in advance, because parameter-bearing
features may support accurate prediction without a noisy input. The
capacity-matched feature-only comparison measures the nonlinear arm's gain
under that feature access (Section~\ref{sec:results-breaks}).

Four pieces of work would change what this benchmark can settle. First, the
strength grids need a calibration to a common severity. Only then can the
unseen-noise setting be read as isolating a change of mechanism. The solver
and a candidate definition are already released, so adopting them is a
decision rather than an implementation (Appendix~\ref{app:noise}). Second, an
executable Clifford data regression and its variable-noise variant would
restore the standard data-driven comparators. An allocator that spends a cheap
method's remaining allowance on extra base shots would then turn the accounted
comparison of Appendix~\ref{app:ledger} into an equal-spend one. Third, a
training-unseen observable axis needs a model that can predict an observable
it never trained on; the fixed per-observable forest cannot, which is why
generation refuses that split today. Fourth, a pre-label reject-and-fallback
layer would answer whether harmful corrections are identifiable before the
ideal value is known. That layer was built and then removed from the codebase.
No local check on a decoded run artifact can establish that a prediction was
produced without access to the test labels, so a reject score fitted from one
cannot be shown to be label-free. It returns when fitting runs inside a
trusted stage before labels are exposed, or when artifacts carry a verifiable
producer signature.

\paragraph{Further Evaluation Priorities.}
The artifact can generate and validate the five changed settings, S1 to S4
and S6 in Table~\ref{tab:settings}, but this campaign has not evaluated them.
Q-LEAR's depth-cut step adds 3,072 hardware evaluations for a 0.0222 gap,
and seven of its ten simulator intervals cross zero. A shot-count sweep with
an arm that reallocates evaluations would clarify the return on this spend.
None of the three screened systems establishes a clean instance of the
controlled pattern or its absence. Prevalence needs a representative sample of
releases scored on final mitigation accuracy, with full training data and
circuit-disjoint training and test sets. Hardware transfer needs controlled
comparisons on physical processors. These evaluations remain future work.

\section{Reproducibility and Release in Full}
\label{app:repro-full}
\label{sec:artifact}

This appendix expands the Reproducibility Statement.
Appendix~\ref{app:repro} provides additional reproducibility details.

Every dataset is generated deterministically from a specification. One master
seed derives named per-item streams, and every per-item seed is recoverable
from the manifest. Items are hashed in a canonical order, and split artifacts
store circuits, sampled counts, observables, and noise models as
content-addressed sidecars. Generation refuses to write into a path that
already exists, so a dataset directory is written once and never edited. Run
artifacts never modify a dataset, and reports are views over result manifests.

The dataset validator is a Python interface rather than a command. For split
artifacts it rebuilds circuits from their canonical descriptors and recomputes
exact labels, then checks the specification hash, the resolved pools and cells,
the per-item and whole-stream hashes, and the sidecar references. It also checks
row-derived counts and the generation ledger against the manifest, and refuses
a test-role spend above the declared cap. A second entry point checks a run
result's identities and ledger conservation. These checks cover the split
artifacts the evaluation settings use. The six legacy micro fixtures predate
the format, carry no sidecar hashes, and have no validation role.

The command surface is two console subcommands and one module entry point.
\texttt{qemscore generate} takes a named preset, an output directory, and an
optional master seed. \texttt{qemscore run} takes a dataset directory, an
output directory, and a budget tier, and executes the whole registered roster.
The tier flag applies to split-v2 artifacts and is rejected for legacy-v1.
Reporting is \texttt{python -m qemscore.reports} with a result manifest and
an output directory. A split specification other than a shipped preset is
authored in Python; no entry point reads an arbitrary specification file.
Additional methods are registered through a phased fit, select, and predict
interface that is tested but not declared stable across releases.

The digital zero-noise-extrapolation implementation is cross-checked against
Mitiq \citep{larose2022mitiq} in a dedicated continuous-integration job. It is
the only external cross-check in the suite; the other methods have no
comparable reference implementation to check against. The suite collects 828
cases. Continuous integration runs it under the pinned dependency lock with the
Mitiq cross-check routed to its own job, together with one expected failure
asserting that the shipped severity grids do not agree across noise families
(Appendix~\ref{app:noise}). A separate workflow regenerates and checks the benchmark
dataset hashes under the baseline and candidate locks. It also cross-checks
zero-noise extrapolation against Mitiq under the candidate lock.

The release ships deterministic generators, small fixtures, and one worked
example in the repository. That example holds six runs, a table, a figure, and
a numerical trace naming the source artifacts and cells behind every number.
Two benchmark dataset hashes regenerate exactly on a clean
continuous-integration machine under the recorded dependency lock. Every
manifest records the package versions, the seed derivation, the active
severity grids, and the feature specification. No single
command reproduces a table or a figure. The example bundle needs a
hand-authored report manifest, and its table needs an external LaTeX run that
no shipped entry point performs. The completed campaign's outputs are archived
at \releaselink{https://github.com/yzhao062/QEMScore/releases/tag/campaign-archive-v1}%
{a public release tagged \texttt{campaign-archive-v1}, whose host is withheld for
double-blind review}:
67 files under a per-file checksum list that the repository carries as well, so
the download can be checked against a copy that arrived by another route. Generated
item streams are excluded and regenerate from the recorded seed and
dependency lock against the twelve dataset hashes the campaign wrote. The reference suite is
publicly released at
\releaselink{https://github.com/yzhao062/QEMScore}{the same host, withheld here for
double-blind review}, under the BSD 2-Clause License.

\section{Limitations in Full}
\label{app:limitations-full}

This appendix expands the limitations in Section~\ref{sec:limits}.

Exact simulator labels define the controlled benchmark's coverage. Its
conclusions are conditional on the tested simulator regimes, circuit
families, and budgets, and do not establish hardware transfer. The field
conclusions are conditional on archived application data from eight IBM
backends, the simulator companion, and the released protocols. The field
reanalysis uses sampled ideal-simulator supervision and establishes no
generalization to unseen application circuits. It performs no new quantum
executions and makes no quantum-advantage or hardware-speedup claim. The
additional question's three readings are conditional on each screened system's
released material, declared evaluation unit, and released split
(Appendix~\ref{sec:results-third}).
Label cost also bounds the controlled benchmark's width. Four of the five samplers
enforce a cap in code: Heisenberg and QAOA at 12 qubits, near-Clifford at 14,
random Clifford at 20. The transverse-field Ising sampler enforces none, so
its cap is a choice the campaign manifest imposes.

The method roster is smaller than the field's. Clifford data regression and
its variable-noise variant are not comparators in this paper. The release
carries a cost formula for each and a construction viability probe, and no
executable mitigator. Lasso and averaged or stacked ensembles are absent as well. Gradient
boosting is absent from the controlled roster; the QRAFT panel's matched
retraining uses it, under the recipe that panel's specification fixes. The learned mitigator of
\citet{liao2024practicalqem} enters as a restricted-feature ablation rather than
a reproduction, so the controlled campaign reports no comparison against the
published method (Section~\ref{sec:discussion}). Appendix~\ref{sec:results-third}
scores descriptor controls against that system's released full feature set under
a separate protocol, which is no reproduction of the published method either.
Controlled method comparisons concern this roster and testbed; the field
comparisons concern their separately specified released-data protocols.

Cost is accounted, and it is not equalized. The library records each method's
three buckets and refuses a method whose total exceeds the tier cap. This
paper reports no per-method bucket totals. No allocator spends a cheap method's remaining allowance down to a
common total. On the shipped in-distribution micro preset at the lowest tier,
the realized totals were 1024 circuit evaluations for the raw estimate, 3072 for
zero-noise extrapolation, and 4608 each for ridge and the restricted-feature
ablation. The two zero-cost controls pay nothing, and the noisy-only and
shuffled-noisy diagnostics pay the same 4608 as ridge. Equal tiers therefore do
not mean equal spending, as the micro-preset totals illustrate. The
campaign's roster artifacts do record realized per-arm totals, as each method's
declared counts rather than audited measurements. Refusal is blunt as well: one infeasible method aborts the whole run,
with its shortfall reported in the error, instead of leaving an infeasible row
in the table. A partial table carrying reported infeasible cells is future work.

The completed campaign evaluates the familiar setting alone, and two further
limits bound the setting coverage the artifact specifies. The four
strength levels are monotone placeholders that no calibration has matched to a
common severity across noise families. That defines the unseen-noise recipe as
transfer to a different channel at a different and undeclared
strength, so a result there could not be read as isolating noise-family transfer
(Appendix~\ref{app:noise}). An observable axis is absent. The split grammar
reserves one, and generation refuses to emit a dataset whose prediction
observables never appear in training, because the fixed per-observable forest
cannot predict an unseen observable key. Six settings therefore remain: one
familiar reference and five single changes. Circuit width and compound
multi-axis changes are also omitted. The pre-label reject-and-fallback layer is not part of this
release, for the reason given in Appendix~\ref{app:discussion-extended}. Several changes
also move coupled properties, so a result under one of them would be
conditional on the declared construction rather than attributable to a single
cause.

What the artifact verifies is narrower than what it records. Validation checks
schema, hashes, roles, seeds, observables, exact labels, and ledger
conservation, and refuses artifacts whose rows disagree with their manifest.
It does not replay the sampled measurement counts from the circuit, the noise
model, and the seed. Nor does it recompute the stored transpilation structure
features that enter every learned model. Consistent but wrong counts or
structure fields would pass. An independent audit matched the recorded
transverse-field Ising row statistics, exact labels, parameter draws, and
structure features to fresh seeded executions. The legacy fixtures retain no
historical full count histograms, so the audit cannot recover those
retrospectively. The campaign's stratified audit completed on all
twelve of its datasets, with no failed dataset, missing roster, or missing
zero-noise replay. Appendix~\ref{app:protocol} states its coverage; it is
not an independent replay of every campaign execution. The quantum objects (noise construction, baseline
implementations, evolution-step choices) are gated by executable cross-checks:
stabilizer versus statevector label agreement, structured-circuit unitary
checks, noise-channel oracles, and the Mitiq cross-check of
Appendix~\ref{sec:artifact}. The intended user is a researcher or toolchain
builder deciding whether a learned mitigator can be trusted under a named change
of regime and at what measurement budget.

Stabilizer simulation keeps exact Clifford labels tractable beyond statevector
reach \citep{gidney2021stim}. Reliable mitigation remains worth measuring on
classical simulators regardless of any quantum-advantage claim
\citep{schuld2022advantage}.